\documentclass[preprint,aps,amsmath,nofootinbib,tightenlines,superscriptaddress]{revtex4}
\usepackage{graphicx}
\usepackage{bm}
\usepackage{epsfig}

\newif\ifpdf
\ifx\pdfoutput\undefined
\pdffalse % we are not running PDFLaTeX
\else
\pdfoutput=1 % we are running PDFLaTeX
\pdftrue
\fi

\def\Bslash{B\!\!\!\!\slash}
\def\Dslash{D\!\!\!\!\slash}

\def\nslash{n\!\!\!\slash}
\def\bnslash{\bar n\!\!\!\slash}

\def\xslash{x\!\!\!\slash}

\def\vslash{v\!\!\!\slash}
\def\OMIT#1{}

\newcommand{\nn}{\nonumber} 

\newcommand{\bn}{{\bar n}}
\newcommand{\bea}{\begin{eqnarray}}
\newcommand{\eea}{\end{eqnarray}}

\newcommand{\bnP}{\bar {\cal P}}

\newcommand{\cP}{{\cal P}}

\newcommand{\mcdot}{\!\cdot\!}

\newcommand{\SCETa}{\mbox{${\rm SCET}_{\rm I}$ }}
\newcommand{\SCETb}{\mbox{${\rm SCET}_{\rm II}$ }}

\newcommand{\DSppP}{{{D\!\!\!\!\hspace{0.04cm}\slash}}_c^\perp}

\begin{document}
%%%%%%%%%%%%%%%%%%%%%%%%%%%%%%%%%%%%%%%%%%
%Some more stuff to get graphics to work
\ifpdf
\DeclareGraphicsExtensions{.pdf, .jpg}
\else
\DeclareGraphicsExtensions{.eps, .jpg}
\fi
%%%%%%%%%%%%%%%%%%%%%%%%%%%%%%%%%%%%%%%%%%

%%%%%%%%%%%%%%%%%%%%%%%%%%%%%%%%%%%%%%%%%%
%Define Title, Author, Address, Preprint#

\preprint{ \vbox{ \hbox{hep-ph/0306254} \hbox{MIT-CTP-3370} }}

\title{\phantom{x}
\vspace{0.5cm}
Strong Phases and Factorization for\\[3pt]
Color Suppressed Decays
%Strong Phases and Heavy Quark Symmetry\\[3pt]in
%$\bar B^0\to D^{(*)0}\pi^0$ Decays
\vspace{0.6cm}
}

\author{Sonny Mantry}
\affiliation{Center for Theoretical Physics, Massachusetts Institute for
Technology,\\ Cambridge, MA 02139\footnote{Electronic address: mantry@mit.edu, 
iains@mit.edu}
\vspace{0.2cm}}
\author{Dan Pirjol}
\affiliation{Department of Physics and Astronomy, The Johns Hopkins University,\\
        Baltimore, MD 21218\footnote{Electronic address: dpirjol@pha.jhu.edu}
\vspace{0.4cm}}
\author{Iain W. Stewart\vspace{0.4cm}}
\affiliation{Center for Theoretical Physics, Massachusetts Institute for
Technology,\\ Cambridge, MA 02139\footnote{Electronic address: mantry@mit.edu, 
iains@mit.edu}
\vspace{0.2cm}}

%\vspace{0.2cm}
%\date{\today\\ \vspace{1cm} }
%\vspace{0.3cm}

%%%%%%%%%%%%%%%%%%%%%%%%%%%%%%%%%%%%%%%%%%
\begin{abstract}
\vspace{0.3cm}

We prove a factorization theorem in QCD for the color suppressed decays $\bar
B^0\to D^0 M^0$ and $\bar B^0\to D^{*0}M^0$ where $M$ is a light meson. Both the
color-suppressed and $W$-exchange/annihilation amplitudes contribute at lowest
order in $\Lambda_{\rm QCD}/Q$ where $Q=\{m_b, m_c, E_\pi\}$, so no power
suppression of annihilation contributions is found.  A new mechanism is given
for generating non-perturbative strong phases in the factorization framework.
Model independent predictions that follow from our results include the equality
of the $\bar B^0\to D^0 M^0$ and $\bar B^0\to D^{*0}M^0$ rates, and equality of
non-perturbative strong phases between isospin amplitudes, $\delta^{(DM)} =
\delta^{(D^*M)}$.  Relations between amplitudes and phases for $M=\pi,\rho$ are
also derived. These results do not follow from large $N_c$ factorization with
heavy quark symmetry.

\end{abstract}

\maketitle

\section{Introduction} \label{sect_intro}

Many of the most frequent hadronic decay channels of $B$ mesons are mediated by
the quark level transition $b\to c d \bar u$. The same hadronic dynamics also
governs the Cabibbo suppressed $b\to cs\bar u$ decays. Typical decays of this
kind are $\bar B\to D\pi$, $\bar B\to D^*\pi$, $\bar B\to D\rho$, $\bar B\to
D^*\rho$, $\bar B\to D K$, $\bar B\to D^*K$, $\bar B\to DK^*$, $\bar B\to
D^*K^*$, $\bar B\to D_s K^-$, $\bar B\to D_s K^{*-}$, $\ldots$ and will be
generically referred to as $\bar B\to D\pi$ decays.  Since these decays are the
simplest of a complicated array of hadronic channels a great deal of theoretical
work has been devoted to their
understanding~\cite{BSW,DG,phen,BlSh,BuSi,PW,bbns,Ligeti,bps,xing0,NePe,Ben,Rosner,xing,Li}.

\begin{figure}[!t]
\vskip0.1cm
 \centerline{
  \mbox{\epsfxsize=4.5truecm \hbox{\epsfbox{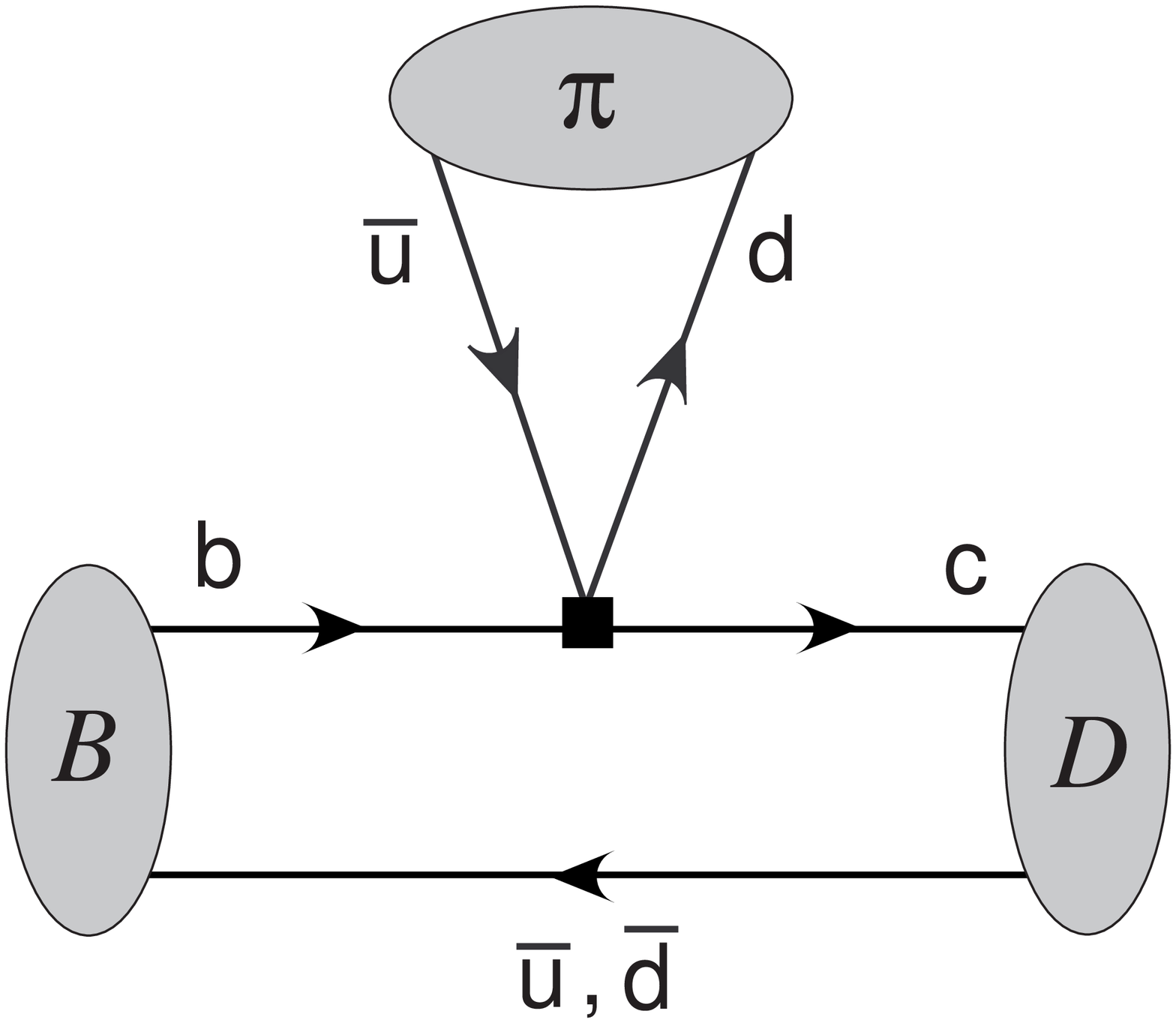}} }
   \hspace{0.7cm}
  \mbox{\epsfxsize=4.5truecm \hbox{\epsfbox{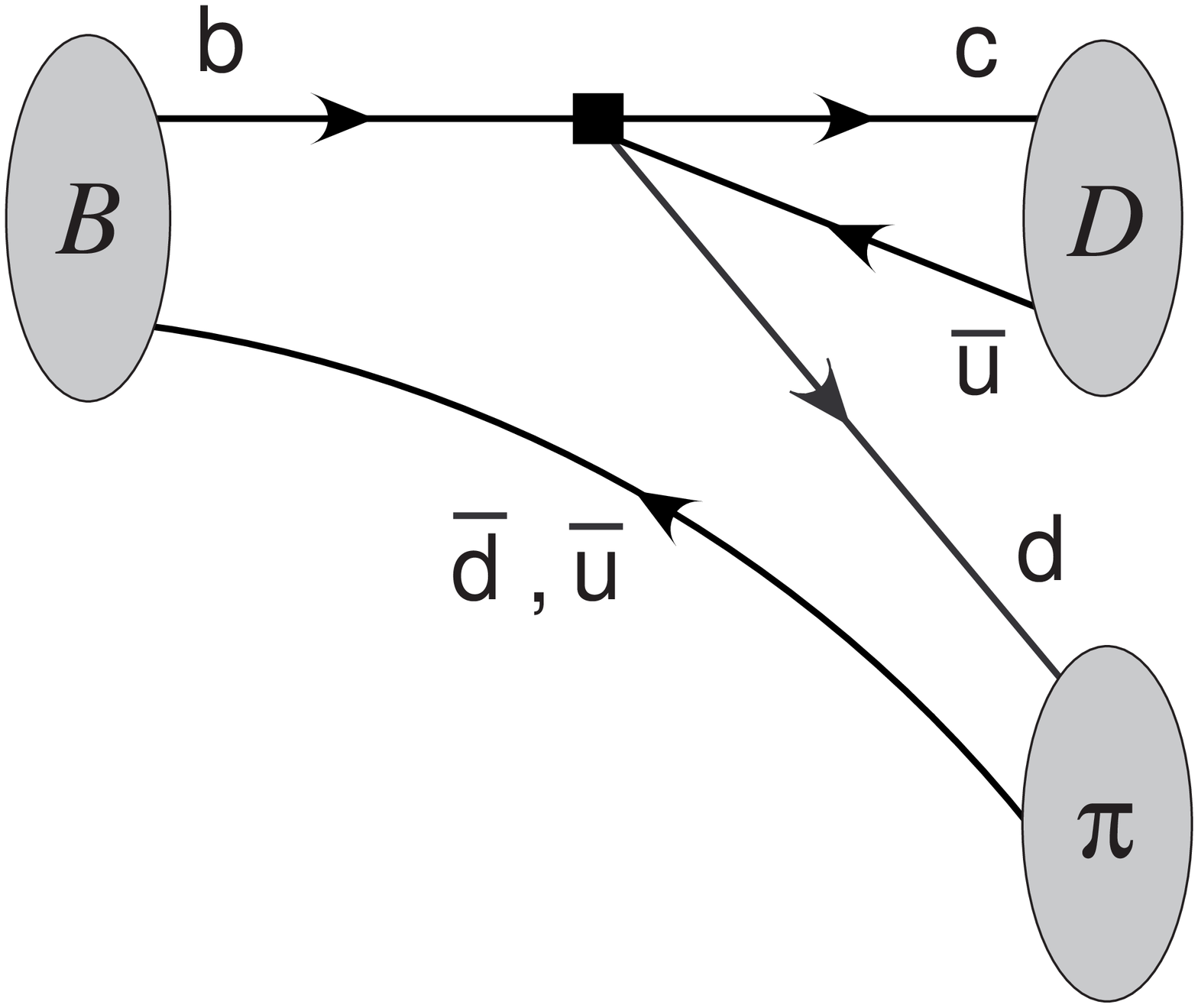}} }
   \hspace{0.7cm}
  \mbox{\epsfxsize=4.5truecm \hbox{\epsfbox{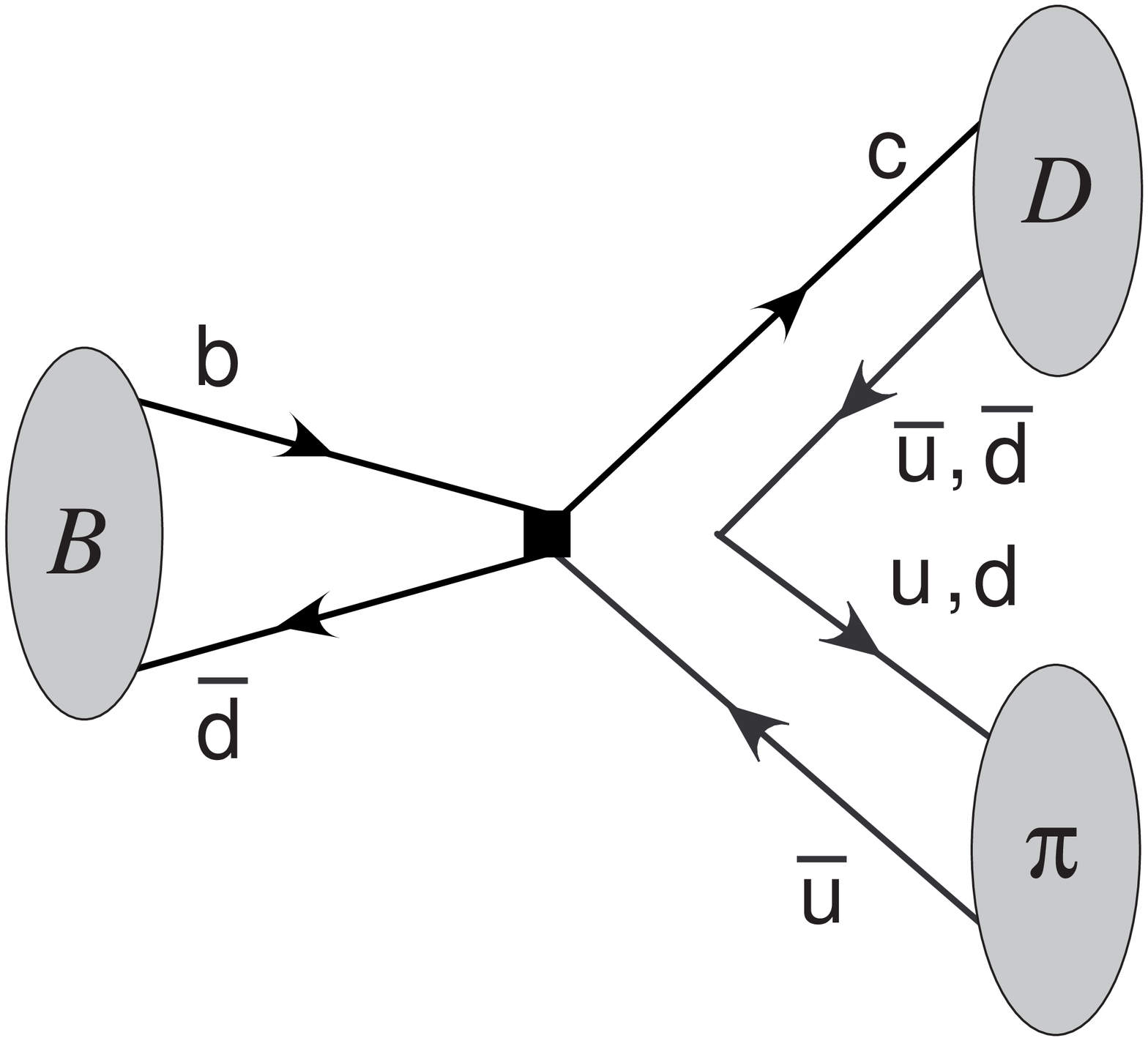}} }
  } 
 \vspace{0.2cm}
 \centerline{ 
  \vbox{\raisebox{-0.15cm}{T:}\hspace{0.3cm}
        $\bar B^0\to D^+\pi^-$ \hspace{2.5cm} 
        \raisebox{-0.15cm}{C:}\hspace{0.3cm}
        $ B^-\to D^0\pi^-$ \hspace{2.1cm}
        \raisebox{-0.15cm}{E:}\hspace{0.3cm}
        $\bar B^0\to D^+\pi^-$  \nn \\ 
        \hspace{0.55cm}
        $B^-\to D^0\pi^-$ \hspace{3.4cm}
        $\bar B^0\, \to D^0\pi^0$ \hspace{3.05cm}
        $\bar B^0 \to D^0\pi^0$ \hspace{0.1cm}}
 } 
\vskip-0.3cm
\caption[1]{Decay topologies referred to as tree (T), color-suppressed
(C), and $W$-exchange (E) and the corresponding hadronic channels to which they 
contribute. }
\label{fig_qcd} 
\vskip0cm
\end{figure}

After integrating out the $W$-boson the weak Hamiltonian for $\bar B\to
D\pi$ decays is 
\begin{eqnarray}\label{Hw}
 {\cal H}_W = \frac{G_F}{\sqrt2} V_{cb} V_{ud}^* [ C_1(\mu)
 (\bar c b)_{V-A} (\bar d u)_{V-A} + C_2(\mu)
 (\bar c_i b_j)_{V-A} (\bar d_j u_i)_{V-A} ]\,,
\end{eqnarray}
where $i,j$ are color indices, and for $\mu_b=5\,{\rm GeV}$, $C_1(\mu_b)= 1.072$
and $C_2(\mu_b)=-0.169$ at NLL order in the NDR scheme~\cite{Buras}. For the
Cabibbo suppressed ${\cal H}_W$ we replace $\bar d\to \bar s$ and $V_{ud}^*\to
V_{us}^*$. It is convenient to categorize the decays into three classes
\cite{BSW}, depending on the role played by the spectator in the B meson (where
``spectator'' is a generic term for the flavor structure carried by the light
degrees of freedom in the $B$). Class I decays receive contributions from graphs
where the pion is emitted at the weak vertex (Fig.~\ref{fig_qcd}T), while in
class II decays the spectator quark ends up in the pion
(Figs.~\ref{fig_qcd}C,\ref{fig_qcd}E).  Finally, class III decays receive both
types of contributions. Many of these channels have been well studied
experimentally~\cite{PDG,CLEO,CLEOdata,BELLEdata,CLEOhelicity,Belle}, see Table
\ref{table_data}. Another method to categorize these decays makes use of
amplitudes corresponding to the different Wick contractions of flavor
topologies.  These can be read off from Fig.~\ref{fig_qcd} and are denoted as
$T$ (tree), $C$ (color-suppressed), and $E$ ($W$-exchange or weak annihilation).

Long ago, it was observed that approximating the matrix elements by the factorized
product $\langle D | (\bar cb)_{V-A}| B\rangle \langle \pi | (\bar
du)_{V-A}|0\rangle $ gives an accurate prediction for the branching fractions of
type-I decays, and a fair prediction for type-III decays. For all class-I and
-II amplitudes a similar procedure was proposed~\cite{BSW}. In terms of two
phenomenological parameters $a_{1,2}$,
\begin{eqnarray} \label{naive}
 &&i A(\bar B^0\to D^+\pi^-) = \frac{G_F}{\sqrt{2}}{V_{cb}^{\phantom{*}}V_{ud}^*}
  \ a_1(D\pi)\:
 \langle D^+ |(\bar cb)_{V-A}|\bar B^0 \rangle
 \langle \pi^-|(\bar du)_{V-A}|0\rangle\,, \quad\\
 && i A(\bar B^0\to D^0\pi^0) 
  = \frac{G_F }{\sqrt{2}}   { V_{cb}^{\phantom{*}}V_{ud}^*} 
  \ a_2(D\pi) \:
  \langle \pi^0 |(\bar db)_{V-A}|\bar B^0\rangle
  \langle D^0|(\bar cu)_{V-A}|0\rangle\nn \,.
\end{eqnarray}
Type-III amplitudes are related by isospin to linear combinations of type-I and
II decays.  Naive factorization\footnote{In this paper we will use the phrase
  naive factorization to refer to factoring matrix elements of four quark
  operators even though this may not be a justified procedure, and will use the
  phrase factorization for results which follow from a well-defined limit of
  QCD.} predicts the universal values $a_1 = C_1 + C_2/N_c$ and $a_2 = C_2 +
C_1/N_c$. Phenomenological analyses testing the validity of the factorization
hypothesis have been presented in \cite{phen}, where typically contributions
from $E$ are not included. These contributions can be modeled using the vacuum
insertion approximation which gives the $D\to \pi$ form factor at a large
time-like momentum transfer $q^2 = m_B^2$. For this reason, they are often
estimated to be suppressed relative to the $T$ amplitudes by $\Lambda_{\rm
  QCD}^2/m_b^2$ \cite{bbns}.  

One rigorous method for investigating factorization in these decays is based on
the large $N_c$ limit of QCD. In this limit the amplitudes for type-I decays
start at $O(N_c^{1/2})$ while type-II decays are suppressed by $1/N_c$ (whence
the name color-suppressed). The type-I amplitudes have a form similar to
Eq.~(\ref{naive}) since non-factorizable diagrams are suppressed, while type-II
decays simultaneously receive contributions from factorized and
non-factorizable diagrams. For a typical class-II decay, a Fierz transformation
puts the amplitude into the form
\begin{eqnarray}\label{fierzHw}
 i A(\bar B^0 \to D^{0} \pi^0) &=& \frac{G_F}{\sqrt2} V_{cb} V_{ud}^* 
 \Big\{ \Big(C_2 + \frac{C_1}{N_c}\Big) 
 \langle D^0\pi^0| (\bar d b)(\bar c u)|\bar B^0\rangle \\ 
 && \qquad\qquad\quad
  + 2C_1  \langle D^0\pi^0| (\bar d T^a b)(\bar c T^a u)|\bar B^0\rangle
  \Big\} \,. \nonumber
\end{eqnarray}
where the $(V-A)\otimes (V-A)$ structure is implicit.  The two matrix elements
have expansions in $1/N_c$ which start with terms of order $N_c^{1/2}$ and
$N_c^{-1/2}$, respectively
\begin{eqnarray}
 && \frac{1}{N_c^{1/2}}
    \langle D^{(*)0} \pi^0| (\bar d b)(\bar c u)|\bar B^0\rangle 
    = F_0^{(*)} + \frac{1}{N_c^2} F_2^{(*)} + \cdots\\
 && \frac{1}{N_c^{1/2}}
    \langle D^{(*)0} \pi^0| (\bar d T^a b)(\bar c T^au)|\bar B^0\rangle =
    \frac{1}{N_c} G_1^{(*)} + \frac{1}{N_c^3} G_3^{(*)} + \cdots\,,\nn
\end{eqnarray}
where $F_i^{(*)}\sim N_c^0$, $G_i^{(*)}\sim N_c^0$.  The Wilson coefficients in
Eq.~(\ref{Hw}) can be assigned scalings with $N_c$ following from their
perturbative expansions $C_1 \sim O(1)$, $C_2 \sim N_c^{-1}$, which roughly
corresponds to the hierarchy in their numerical values at $\mu_b$. The leading
terms are the matrix elements $F_0^{(*)}$, which factor in terms of large
$N_c$ form factors and decay constants
\begin{eqnarray}\label{F0}
  N_c^{1/2} F_0^{(*)} \sim 
  \langle D^{(*)0} |\bar c u|0\rangle \langle \pi^0 | \bar d b |\bar B^0\rangle 
 +\langle D^{(*)0} \pi^0|\bar c u|0\rangle \langle 0 |\bar d b|\bar B^0\rangle \,,
\end{eqnarray}
plus the matrix elements $G_1^{(*)}$ which are nonfactorizable.  The naive
factorization assumption would keep only $F_0^{(*)}$ and neglect $G_1^{(*)}$.
This approximation is not justified in the $1/N_c$ expansion since $G_1^{(*)}$
is enhanced by the large Wilson coefficient $C_1$.  In either case, no
prediction is obtained for the ratio of the $\bar B\to D\pi$ and $\bar B\to
D^*\pi$ amplitudes,
\begin{eqnarray} \label{R0lgN}
  R_0^\pi \equiv \frac{A(\bar B^0 \to D^{*0} \pi^0)}
  {A(\bar B^0 \to D^{0}\pi^0)} 
 = \frac{(C_2+C_1/N_c) F_0^* + (2C_1/N_c) G_1^*}{(C_2+C_1/N_c) F_0 + 
   (2C_1/N_c)  G_1}\,.
\end{eqnarray}
Heavy quark symmetry does not operate with large $N_c$ factorization because for
$C$ and $E$ it is broken by the allowed exchange of energetic hard gluons
between the heavy quarks and the quarks in the pion.  In contrast, we will show
in this paper that expanding about the limit $E_\pi\gg\Lambda$ this ratio is
predicted to be 1 at leading order in $\Lambda/Q$. Here $\Lambda\sim
\Lambda_{\rm QCD}$ is a typical hadronic scale.

Another rigorous approach to factorization becomes possible in the limit $E_\pi
\gg \Lambda_{\rm QCD}$ which corresponds to having an energetic light hadron in
the final state. In this paper we analyze type-II decays using QCD and an
expansion in $\Lambda_{\rm QCD}/m_b$, $\Lambda_{\rm QCD}/m_c$, and $\Lambda_{\rm
  QCD}/E_\pi$ (or generically $\Lambda_{\rm QCD}/Q$ where $Q=\{m_b,m_c,m_b\!-\!
m_c\}$). We derive a factorization theorem and show that $E$ and $C$ appear at
the same order in the power counting, and are suppressed by $\Lambda_{\rm
  QCD}/Q$ relative to $T$.  Arguments for the suppression of $C$ by
$(\Lambda_{\rm QCD}/Q)^1$ and $E$ by $(\Lambda_{\rm QCD}/Q)^{1,2}$ appear in the
literature~\cite{bbns}, but we are unaware of a derivation that is model
independent.  Our leading order result disagrees with the $a_2$-factorization
result. Instead the amplitudes for $\bar B^0\to D^{(*)0}\pi^0$ and $\bar B^0\to
D^{(*)0}\rho^0$ are determined by the leading light-cone wavefunctions
$\phi_{\pi,\rho}$, and two new universal $\bar B\to D^{(*)}$ distribution
functions. Long distance contributions also occur at this order in $\Lambda_{\rm
  QCD}/Q$, but are shown to be suppressed relative to the short distance
contributions by an additional $\alpha_s(Q)/\pi$.

For type-I decays a color transparency~\cite{color} argument given by Bjorken
suggested $A(\bar B^0 \to D^+ \pi^-) \simeq (C_1+C_2/N_c) f_\pi
F_0^{BD}(m_\pi^2)+ O(\alpha_s(Q))$.  In Ref.~\cite{DG} it was argued that this
factorization is the leading order prediction in the large energy limit $E_\pi
\gg \Lambda_{\rm QCD}$, and in Refs.~\cite{PW,bbns} that $\alpha_s$ corrections
can be rigorously included.  This factorization was extended to all orders in
$\alpha_s$ with the proof of a factorization theorem using the soft-collinear
effective theory~\cite{bps}
\begin{eqnarray} \label{factLO}
  A(B\to D^{(*)}\pi) = N^{(*)}\: \xi(w_0,\mu) \int_0^1\!\!dx\: 
   T^{(*)}(x,m_c/m_b,\mu)\: \phi_\pi(x,\mu) + \ldots \,,
\end{eqnarray}
where the ellipses denote power suppressed terms. This result is similar to
predictions obtained from the hard exclusive scattering formalism of
Brodsky-Lepage~\cite{BL}, except for the presence of the Isgur-Wise function,
$\xi(\omega_0,\mu)$. The normalization factor is given by~\footnote{Note for
  longitudinal $D^*$, $n\mcdot \varepsilon_{D^*}=n\mcdot v'$. Production of
  transverse $\rho$'s is suppressed by $\Lambda/Q$.}
\begin{eqnarray} \label{N}
  N^{(*)}= \frac{G_F V_{cb}^{\phantom{*}}V_{ud}^*}{\sqrt{2}}\:
   { E_\pi f_\pi} \sqrt{m_{D^{(*)}} m_B} \: 
    \Big( 1+\frac{m_B}{m_{D^{(*)}}} \Big) \,.
\end{eqnarray}
The proof of Eq.~(\ref{factLO}) uses the heavy quark limit, so $m_D=m_{D^*}$ and
$N=N^*$.  In Eq.~(\ref{factLO}), $\phi_\pi(x,\mu)$ is the non-perturbative pion
light-cone wave function, and $\xi(w_0,\mu)$ is evaluated at maximum recoil
$v\mcdot v'\to w_0 = (m_B^2+m_{D^{(*)}}^2)/(2m_B m_{D^{(*)}})$. The hard
coefficient $T^{(*)}(x,\mu)=C^{(0)}_{L\pm R}((4x-2)E_\pi,\mu,m_b)$, where the
$\pm$ correspond to the $D$ and $D^*$ respectively, and $C^{(0)}_{L\pm R}=
C^{(0)}_{L} \pm C^{(0)}_{R}$ is the calculable Wilson coefficient of the
operators defined in Eq.~(\ref{QV}) below.  The renormalization scale dependence
of the hard scattering function $T(x,\mu)$ cancels the $\mu$ dependence in the
Isgur-Wise function and pion wave function.  In this framework~\cite{bbns} there
is no longer a need to identify by hand a factorization scale.\footnote{In naive
  factorization the hadronic matrix elements in Eq.~(\ref{naive}) are
  independent of the scale that separates hard and soft physics.  The scale
  dependence in $a_1$ and $a_2$ then causes the physical amplitudes to become
  scale dependent. The parameters $a_1$ and $a_2$ were therefore assumed to be
  evaluated at a specific scale called the "factorization scale".  In other
  words, the non-factorizable effects were accounted for by allowing $a_1$ and
  $a_2$ to be free parameters that are fit to data. The factorization scale can
  then be extracted from the scale dependence of $a_1$ and $a_2$ \cite{phen}.}
In the language of SCET~\cite{bps}, the scale dependence is understood from the
matching and running procedure.

Eq.~(\ref{factLO}) implies equal rates for $\bar B^0\to D^+\pi^-$ and $\bar
B^0\to D^{*+} \pi^-$ up to the $\alpha_s(m_b)$ corrections in $T^{(*)}$ and
power corrections.  This prediction is in good agreement with the observed data
for type-I and III decays to $\pi$, $\rho$, $K$ and $K^*$ as shown in Tables I
and II.  For two-body type-I decays both the large $N_c$ and large energy
mechanisms make similar phenomenological predictions. However, these mechanisms
can be distinguished with $B\to DX$ decays where $X$ is a multi-hadron
state~\cite{Ben}.

So far, no results of comparable theoretical rigor exist for the
color-suppressed type-II decays.  In fact existing results in $B\to D\pi$ and
$B\to \psi K^{(*)}$ do not support naive factorization with a universal
coefficient $a_2$~\cite{NePe}. Furthermore, it has been argued that in general
factorization will not hold for type-II decays~\cite{bbns}.

Using the soft-collinear effective theory (SCET)~\cite{SCET,bps2}, we prove
in this paper a factorization theorem for color-suppressed (type-II) $\bar B\to
DM$ decays, $M=\{\pi^0,\rho^0,\ldots\}$.  These decays are power suppressed
relative to the type I decays, and our results are valid at leading nonvanishing
order in $\Lambda/Q$.  The main results of our paper are
\begin{itemize}
  
\item The color suppressed (C) and exchange (E) contributions to $B^0\to
  D^{(*)0}\pi^0$ are both suppressed by $\Lambda/Q$ relative to the 
  amplitude (T). The $C$ and $E$ amplitudes are found to be of comparable
  size since the factorization theorem relates them to the same perturbative and
  non-perturbative quantities.  Our result is incompatible with the naive $a_2$
  type factorization.
  
\item When our result is combined with heavy quark symmetry it predicts the
  equality of the amplitudes for $\bar B^0\to D^0\pi^0$ and $\bar B^0\to
  D^{*0}\pi^0$ (in fact for any $DM$ and $D^*M$).  This prediction is in good
  agreement with existing data and will be tested by future measurements.
  
\item Our result gives a new mechanism for generating non-perturbative strong
  phases for exclusive decays within the framework of factorization. For $D M$
  and $D^* M$ it implies the equality of the strong phases $\delta$ between
  isospin amplitudes.  Furthermore, certain cases with different light mesons
  $M$ are predicted to also have a universal non-perturbative strong phase
  $\phi$ in their isospin triangle.
  
\item The power suppressed amplitudes for all color suppressed $\bar B\to
  D^{(*)}M$ decays are factorizable into two types of terms, which we refer to
  as short distance ($\mu^2\sim E_M\Lambda$) and long distance ($\mu^2\sim
  \Lambda^2$) contributions. The short distance contributions depend on complex
  soft $B^0\to D^{(*)0}$ distribution functions, $S^{(0,8)}_{L,R}(k_+,\ell_+)$,
  which depend only on the direction of $M$ (the superscripts indicate that
  two color structures contribute). For $M=\pi,\rho$ the long distance
  contributions vanish at lowest order in $\alpha_s(Q)/\pi$.

\end{itemize}
Combined with Eq.~(\ref{factLO}) the results here give a complete leading order
description of the $B\to D\pi$ isospin triangles.  

In Section~\ref{sect_data} we review the current data for $B\to D\pi$ decays.
The derivation of a factorization theorem for the color suppressed channels
$\bar B^0\to D^{(*)0}\pi^0$ and $\bar B^0\to D^{(*)0}\rho^0$ is carried out in
section~\ref{sect_SCET} using SCET. Then in section \ref{sect_Kaon} the
formalism is applied to decays with kaons, $\bar B^0\to D^{(*)0}K^0$, $\bar
B^0\to D^{(*)0}K^{*0}$, $\bar B^0\to D^{(*)}_s K^-$, and $\bar B^0\to D^{(*)}_s
K^{*-}$.  In section \ref{sect_lNc} we contrast our results with the large $N_c$
limit of QCD and prior theoretical expectations. Readers only interested in
final results can safely skip sections~\ref{sect_SCET}, \ref{sect_Kaon}, and
\ref{sect_lNc}.  In section \ref{sect_results} we discuss the phenomenological
predictions that follow from our new formalism for color suppressed channels.
Conclusions are given in \ref{sect_concl}. In Appendix~\ref{app_G} we prove that for
$\pi^0$ and $\rho^0$ the long distance contributions are suppressed. Finally in
Appendices B and C we elaborate on the properties of the jet functions and our
new soft $B\to D^{(*)}$ distribution functions respectively.

%%%%%%%%%%%%%%%%%%%%%%%%%%%%%%%%%%%%%%%%%%%%%%%%%%%%%%%%%%%%%%%%%%%%%%%%%%%%%%%

\section{Data} \label{sect_data}

We start by reviewing existing data on the $\bar B\to D^{(*)}\pi$ decays. 
The branching ratios for most of these modes have been measured and the
existing results are collected in Table~\ref{table_data}. 
Taking into account that the $D^{(*)}\pi$ final state can have isospin 
$I=1/2, 3/2$, these decays can be parameterized by 2 isospin amplitudes 
$A_{1/2}$, $A_{3/2}$:
\begin{eqnarray} \label{As}
  A_{+-} = A(\bar B^0\, \to D^+ \pi^-) &=& \frac{1}{\sqrt3} A_{3/2} + 
  \sqrt{\frac23} A_{1/2} = T+E   \,,\nn   \\
  A_{\,0-} = A( B^- \to D^0\, \pi^-) &=& \sqrt3 A_{3/2} = T+C  \,,\nn  \\
  A_{\,00\,} = A(\bar B^0\, \to D^0\, \pi^0\,) &=& \sqrt{\frac23} A_{3/2} -
  \frac{1}{\sqrt3} A_{1/2} = \frac{1}{\sqrt2}(C-E)\,.
\end{eqnarray}
Similar expressions can be written for the decay amplitudes of $B\to D^*\pi$,
$B\to D\rho$, $B\to D^*\rho$ with well defined helicity of the final state
vector mesons. Eq.~(\ref{As}) also gives the alternative parameterization of
these amplitudes in terms of the amplitudes $T,C,E$ discussed in
Sec.~\ref{sect_intro}.  

Using the data in Table~\ref{table_data}, the individual isospin amplitudes
$A_I$ and their relative phase $\delta={\rm arg}(A_{1/2} A_{3/2}^*)$ can be
extracted using
\begin{eqnarray}
 {\rm Br}(\bar B\to D^{(*)}M) = \tau_B \Gamma(\bar B\to D^{(*)} M) &=&  
   \frac{\tau_B |{\bf p}|}{8\pi m_B^2} \: 
   \sum_{\rm pol} \big|A(\bar B\to D^{(*)} M)\big|^2 \,.
\end{eqnarray}  
with $\tau_{\bar B^0}=2.343\times 10^{12}\,{\rm GeV}^{-1}$ and
$\tau_{B^-}=2.543\times 10^{12}\,{\rm GeV}^{-1}$. We find
\begin{eqnarray}
|A_{1/2}^D | &=& (4.33 \pm 0.47)\times 10^{-7} \mbox{ GeV}\,, \qquad
   \delta^{D\pi} = 30.5^{\circ+7.8}_{-13.8}\,, \\
|A_{3/2}^D| &=&  (4.45 \pm 0.17)\times 10^{-7} \mbox{ GeV}\,, \nn\\[5pt]
|A_{1/2}^{D^*} | &=& (4.60\pm 0.36)\times 10^{-7} \mbox{ GeV}\,, \qquad
   \delta^{D^*\pi} = 30.2 \pm 6.6^\circ \,,\nn\\
    % ^{\circ+7.8}_{-13.0} \,,\nn\\
|A_{3/2}^{D^*}| &=&  (4.33 \pm 0.19)\times 10^{-7} \mbox{ GeV} \,.\nn
\end{eqnarray}
The ranges for $\delta$ correspond to $1\sigma$ uncertainties for the
experimental branching ratios. A graphical representation of these results is
given in Fig.~\ref{fig_isospin}, where we show contour plots for the ratios of
isospin amplitudes $R_I = A_{1/2}/(\sqrt2 A_{3/2})$ for both $D\pi$ and $D^*\pi$
final states.  For $\bar B\to D\pi$ an isospin analysis was performed recently by
CLEO \cite{CLEO} including error correlations among the decay modes; we used their
analysis in quoting errors on $\delta^{D\pi}$.

\begin{table}[t!]
\begin{center}
\begin{tabular}{|c|c|c|c|c|c|c|}
\hline
Type & Decay & Br$(10^{-3})$ & $|A|$ ($10^{-7}$ GeV) 
 & Decay & Br$(10^{-3})$ & $|A|$ ($10^{-7}$ GeV) 
 \\
\hline\hline
I & $\bar B^0 \to D^+ \pi^-$ & $2.68\pm 0.29$ ${}^a$ & $5.89\pm 0.32$ 
 & $\bar B^0 \to D^{*+} \pi^-$ & $ 2.76 \pm 0.21 $  & $6.05\pm 0.23$
 \\ 
III & $B^- \to D^0 \pi^-$ &  $4.97\pm 0.38$ ${}^a$ % $5.3\pm 0.5$ 
 & $7.70\pm 0.29$
 & $B^- \to D^{*0} \pi^-$ & $ 4.6 \pm 0.4 $ & $7.49 \pm 0.33$
 \\ 
II & $\bar B^0 \to D^0 \pi^0$ & $0.292\pm 0.045$ ${}^b$ & 
$1.94\pm 0.15$ & $\bar B^0 \to D^{*0} \pi^0$ ${}^b$ 
 & $0.25 \pm 0.07$ & $1.82\pm 0.25$
  \\
\hline
I & $\bar B^0 \to D^+ \rho^-$ & $7.8\pm 1.4$  & $10.2 \pm 0.9$ 
 & $\bar B^0 \to D^{*+} \rho^-$ & $6.8 \pm 1.0$ ${}^c$ & $9.08\pm 0.68$ ${}^\dagger$ 
 \\ 
III & $B^- \to D^0 \rho^-$ & $13.4\pm 1.8$  & $12.8\pm 0.9$
 & $B^- \to D^{*0} \rho^-$ & $9.8 \pm 1.8$ ${}^c$ & $10.5\pm 0.97$ ${}^\dagger$
 \\ 
II & $\bar B^0 \to D^0 \rho^0$ & $0.29 \pm 0.11$  ${}^d$  & $1.97 \pm 0.37$
 & $\bar B^0 \to D^{*0} \rho^0$ & $< 0.56$  & $<2.77$
  \\ \hline\hline
\end{tabular}
\end{center}
{\caption{Data on $B\to D^{(*)}\pi$ and $B\to D^{(*)}\rho$ decays from
    ${}^a$Ref.\cite{CLEO}, ${}^b$Ref.\cite{CLEOdata,BELLEdata},
    ${}^c$Ref.\cite{CLEOhelicity}, ${}^d$Ref.\cite{Belle}, or if not 
  otherwise indicated from Ref.\cite{PDG}. ${}^\dagger$For $\bar B\to D^*\rho$ 
the amplitudes for longitudinally polarized $\rho$'s are displayed.}
\label{table_data} }
\end{table}

For later convenience we define the amplitude ratios 
\begin{eqnarray}  \label{ratios}
 && R^M_0 \equiv \frac{A(\bar B^0 \to D^{*0}M^0)}  
  {A(\bar B^0 \to D^0 M^0)} \,,   \qquad\quad
  R^{M/M'}_0 \equiv \frac{A(\bar B^0 \to D^{(*)0}M^0)}  
    {A(\bar B^0 \to D^{(*)0} M^{\prime\, 0})} \,, \\
  R_I &\equiv& \frac{A_{1/2}}{\sqrt2 A_{3/2}} 
    = 1 - \frac32\ \frac{C-E}{T+C}\,, \qquad
  R_c \equiv \frac{A(\bar B^0 \to D^{(*)+} M^-)}{A(B^- \to D^{(*)0}
    M^-)} = 1 - \frac{C-E}{T+C}\,, \nn 
\end{eqnarray} 
where the ratios $R_I$ and $R_c$ are defined for each $D^{(*)}M$ mode.
Predictions are obtained for the ratios in Eq.~(\ref{ratios}), including the
leading power corrections to $R_I$ and $R_c$.  The relation $R_I= 1 + {\cal
  O}(\Lambda/Q)$ can be represented graphically by a triangle with base
normalized to 1 (see Fig.~\ref{fig_isospin} in section~\ref{sect_results}). The
two angles adjacent to the base are the strong isospin phase $\delta$, and
another strong phase $\phi$. The usual prediction is that $\delta \sim
1/Q^k$~\cite{bbns,NePe}, and that there is no constraint on the strong phase
$\phi$ which can be large.  In section~\ref{sect_results} we show that at lowest
order the angle $\phi$ is predicted to be the same for all channels in
Table~\ref{table_data}, and that $\delta$ can be dominated by a constrained
non-perturbative strong phase. From $R_I$ in Eq.~(\ref{ratios}) we note that for
a leading order prediction of $\delta$ it is not necessary to know the power
corrections to the $T$ amplitude.

%\subsection{$B \rightarrow DK $ Data Analysis}

\begin{table}[t!]
\begin{center}
\begin{tabular}{|c|c|c|c|c|c|c|}
\hline
Type & Decay & Br$(10^{-5})$ & $|A|$ ($10^{-7}$ GeV) 
  & Decay & Br$(10^{-5})$ & $|A|$ ($10^{-7}$ GeV) \\
\hline\hline
I & $\bar B^0 \to D^+ K^-$ & $20.0\pm 6.0$  & $1.62\pm 0.24$ 
  & $\bar B^0 \to D^{*+} K^-$ & $20\pm 5$ & $1.64\pm 0.20$  \\ 
III & $B^- \to D^0 K^-$ &  $37.0\pm 6.0$ & $2.11\pm 0.17$ 
  & $B^- \to D^{*0} K^-$ &  $36\pm 10$ & $2.11\pm 0.29$\\
II & $\bar B^0 \to D^0 \bar K^0$ & $5.0^{+1.1}_{-1.2}\pm 0.6$ \cite{D0K0} 
  & $0.81\pm 0.11$
  &  $\bar B^0 \to D^{*0} \bar K^0$ & -- & -- \\ 
\hline
I & $\bar B^0 \to D^+ K^{*-}$ & $37.0\pm 18.0$  & $2.24\pm 0.54$ 
  &  $\bar B^0 \to  D^{*+} K^{*-} $ & $38\pm 15$  & $2.30\pm 0.45$${}^\dagger$\\ 
III & $B^- \to D^0 K^{*-}$ &  $61.0\pm 23.0$ & $2.76\pm 0.52$  
  & $B^- \to D^{*0} K^{*-}$ & $72\pm 34 $ & $3.03\pm 0.72$${}^\dagger$ \\
II & $\bar B^0 \to D^0 \bar K^{*0}$ & $4.8^{+1.1}_{-1.0}\pm 0.5$ \cite{D0K0} 
& $0.81\pm 0.11$
  &  $\bar B^0 \to D^{*0} \bar K^{*0}$ & -- & -- \\
\hline\hline
\end{tabular}
\end{center}
{\caption{Data on Cabibbo suppressed $\bar B\to DK^{(*)}$ decays. Unless 
otherwise indicated, the data is taken from Ref.\cite{PDG}. ${}^\dagger$Since
no helicity measurements for $D^*K^*$ are available we show effective amplitudes
which include contributions from all three helicities.}
\label{table_data2} }
\end{table}
A similar analysis can be given for the Cabibbo suppressed $ \bar B \rightarrow
DK^{(*)}$ decays. Although several of these modes had been seen for some time,
it is only recently that some of the corresponding class-II decays have been
seen by the Belle Collaboration \cite{D0K0} (see Table II).  For this case the
final $D^{(*)}K^{(*)}$ states can have isospins $I=0,1$, so these
decays are parameterized in terms of 2 isospin amplitudes $A_{I=0,1}$ (for given
spins of the final particles)
\begin{eqnarray}\label{AsDK}
  A_{+-} = A(\bar B^0\, \to D^+ K^-) &=& \frac{1}{2} A_{0} +
  \frac{1}{2} A_{1} = T \,,\nonumber   \\
  A_{\,0-} = A( B^- \to D^0\, K^-) &=& A_{1} = T+C \,,\nonumber \\
  A_{\,00\,} = A(\bar B^0\, \to D^0\, K^0\,) &=& \frac{1}{2} A_{1} -
  \frac{1}{2} A_{0} = C\,.
\end{eqnarray}
Isospin symmetry implies the amplitude relation among these modes
$A_{+-} + A_{00} = A_{0-}$, 
which can be used to extract the isospin amplitudes $A_{0,1}$ and their
relative phase $\delta={\rm arg}(A_{0} A_{1}^*)$. Using Gaussian error
propagation we obtain
\begin{eqnarray}
|A_{0}^{DK} | &=& (1.45 \pm 0.62)\times 10^{-7} \mbox{ GeV}, \qquad
   \delta^{DK} = 49.9\pm 9.5^\circ \\
|A_{1}^{DK}| &=&  (2.10 \pm 0.17)\times 10^{-7} \mbox{ GeV}, \nonumber \\
\nonumber \\
|A_{0}^{DK^*} | &=& (1.93\pm 1.49)\times 10^{-7} \mbox{ GeV}, \qquad
   \delta^{DK^*} = 34.9 \pm 19.4^\circ\\
|A_{1}^{DK^*}| &=&  (2.76 \pm 0.52)\times 10^{-7} \mbox{ GeV}.\nonumber
\end{eqnarray}
However, note that scanning the amplitudes $A_{+-}$, $A_{00}$, $A_{0-}$ in their
$1$$\sigma$ allowed regions still allows a flat isospin triangle~\cite{Rosner}.

%%%%%%%%%%%%%%%%%%%%%%%%%%%%%%%%%%%%%%%%%%%%%%%%%%%%%%%%%%%%%%%%%%%%%%%%%%%

\section{Soft-collinear effective theory analysis} \label{sect_SCET}

The key idea of the soft-collinear effective theory~\cite{SCET,bps2} is to
separate perturbative and non-perturbative scales directly at the level of
operators. The relevant scales have virtualities $p^2\sim m_b^2,m_c^2,E_M^2$
(hard), $p^2\sim E_M\Lambda$ (intermediate), and $p^2\simeq \Lambda^2$ (soft).
The $p^2\sim \Lambda^2$ scales are described by soft $(p^+,p^-,p^\perp)\sim
(\Lambda,\Lambda,\Lambda)$ and collinear $(p^+,p^-,p^\perp)\sim
(\Lambda^2/E_M,E_M,\Lambda)$ degrees of freedom. We follow the notation in
Refs.~\cite{bps,bfprs}.

The weak Hamiltonian $\mathcal{H}_W $ in Eq.~(\ref{Hw}) is matched onto
effective operators containing soft and collinear fields. In exclusive processes
this matching can be simplified by a two-stage procedure~\cite{bps4}.  We first
match QCD onto a theory \SCETa with ultrasoft fields with $p^2\sim\Lambda^2$,
but intermediate-collinear fields with $p^2\sim E_M\Lambda$. This theory gives a
simplified description of the $p^2\sim E_M\Lambda$ exchanges that necessarily
mediate interactions between soft and collinear particles. Then at a scale
$\mu^2\sim E_M\Lambda$ we match \SCETa onto the final theory \SCETb which has
only the propagating long distance soft and collinear particles.  This procedure
determines which factors of $\alpha_s(\mu)$ belong\footnote{A more accurate
  statement is that the scale dependence is determined by anomalous dimensions
  of operators in \SCETa and \SCETb.} at the hard scale $\mu^2=Q^2$, which
belong at the intermediate scale $\mu^2=E_M \Lambda$, and what non-perturbative
matrix elements appear.

Since the collinear fields do not interact with soft fields at lowest order, if
one can rearrange the fields in the \SCETb operator to express it as a product
of collinear fields and soft fields, the factorization of matrix elements is
achieved. This is precisely what happens in type-I decays, and as we will see
also type-II decays, with operators of the form
\begin{eqnarray}
 &\mbox{Type-I:}&\qquad   \big[ \bar h_{v'} S\Gamma S^\dagger h_{v}\big]
 \ \big[(\bar\xi _n W)\Gamma'  (W^{\dagger } \xi _n) \big] \nonumber \\
 &\mbox{Type-II:}&\qquad  \big[ (\bar h_{v'} S) \Gamma (S^\dagger 
  h_{v})\: (\bar q S) \Gamma'' (S^\dagger q)\big] \
  \big[(\bar \xi _n W)\Gamma' (W^{\dagger }\xi _n) \big] \,, \\
& & \hspace{-0.5cm} \int\!\! d^4x \: T\,  \big[
   (\bar h_{v'} S) \Gamma (S^\dagger h_{v})\:  
  (\bar \xi _n W)\Gamma' (W^{\dagger }\xi _n) \big] (0)\,
  \big[ (\bar q S) \Gamma'' (S^\dagger q) \:
  (\bar \xi _n W)\Gamma' (W^{\dagger }\xi _n) \big](x)   \,. \nn
\end{eqnarray}   
In type-II decays the first and second \SCETb operators give short and long
distance contributions respectively.  We use here the notation in
Ref.~\cite{bps2} so that $h_v$ are HQET fields, $\xi_n$ are collinear quark
fields, $q$ are soft quark fields, and $S, W$ are soft and collinear Wilson
lines.  Since collinear particles do not connect with the heavy meson states and
soft particles do not connect with the collinear light meson state, the matrix
elements of these operators factor into the product of a soft $B\rightarrow
D^{(*)}$ matrix element and a collinear matrix element involving $M$.

We start by reviewing type-I decays. Using SCET, the factorization of the leading
amplitude for type-I decays has been proven in Ref.~\cite{bps} at leading order
in $1/Q$ (and non-perturbatively to all orders in $\alpha_s$).  The operators in
Eq.~(\ref{Hw}) are matched onto effective operators at a scale $\mu_Q\simeq Q$
\begin{eqnarray} \label{QVI0}
  \sum_{1,2} C_i O_i \to 4 \sum_{j=L,R} \int\!\! d\tau_1 d\tau_2 \big[
  C^{(0)}_{j} (\tau_1,\tau_2)
  {\cal Q}^{(0)}_{j}(\tau_1,\tau_2) + C^{(8)}_{j} (\tau_1,\tau_2)
  {\cal Q}^{(8)}_{j}(\tau_1,\tau_2) \big]\,.
\end{eqnarray}
At leading order in \SCETa there are four operators [$j=L,R$]
\begin{eqnarray}\label{QVI}
 {\cal Q}^{(0)}_{j}(\tau_1,\tau_2) &=& 
\big[\bar h_{v'}^{(c)} \Gamma^h_j  h_v^{(b)} \big] 
 \big[(\bar\xi_n^{(d)} W)_{\tau_1} \Gamma_n 
  (W^\dagger \xi_n^{(u)})_{\tau_2} \big] \,,\\
 {\cal Q}^{(8)}_{j}(\tau_1,\tau_2) &=&
\big[\bar h_{v'}^{(c)} Y \Gamma^h_j T^a Y^\dagger h_v^{(b)} \big] 
 \big[(\bar\xi_n^{(d)} W)_{\tau_1} \Gamma_n T^a 
 (W^\dagger \xi_n^{(u)})_{\tau_2} \big]\nn \,.
\end{eqnarray}
The superscript $(0,8)$ denotes the $1\otimes 1$ and $T^a\otimes T^a$ color
structures. The Dirac structures on the heavy side are \hbox{$\Gamma^h_{L,R}=
  \nslash P_{L,R}$} with $P_{R,L} = \frac12(1\pm \gamma_5)$, while on the
collinear side we have $\Gamma_n = \bnslash P_L/2$. The momenta labels are
defined by $(W^\dagger \xi_n)_{\omega_2} = [\delta(\omega_2\!-\!\bnP)\:
W^\dagger \xi_n]$.

The matching conditions for the Wilson coefficients at tree level at
$\mu=E_{\pi}$ are
\begin{eqnarray}
  C_L^{(0)}(\tau_i) &=& C_1 + \frac{C_2}{N_c} \,,\qquad
  C_L^{(8)}(\tau_i) = 2 C_2 \,,\qquad C_R^{(0,8)}(\tau_i) = 0\,.
\end{eqnarray}
Matching corrections of order ${\cal O}(\alpha_s)$ can be found in 
Ref.~\cite{bbns}.  

The operators in Eq.~(\ref{QVI}) are written in terms of collinear fields which
do not couple to usoft particles at leading order. This was achieved by a
decoupling field redefinition~\cite{bps2} on the collinear fields $\xi_n\to
Y\xi_n$ etc.  The operators in Eq.~(\ref{QVI}) are then matched onto \SCETb to
give [$\omega_i=\tau_i$]
\begin{eqnarray}\label{QV}
 {\cal Q}^{(0)}_{j}(\omega_1,\omega_2) &=& 
\big[\bar h_{v'}^{(c)} \Gamma^h_j  h_v^{(b)} \big] 
 \big[(\bar\xi_n^{(d)} W)_{\omega_1} \Gamma_n 
  (W^\dagger \xi_n^{(u)})_{\omega_2} \big] \,,\\
 {\cal Q}^{(8)}_{j}(\omega_1,\omega_2) &=&
\big[\bar h_{v'}^{(c)} S \Gamma^h_j T^a S^\dagger h_v^{(b)} \big] 
 \big[(\bar\xi_n^{(d)} W)_{\omega_1} \Gamma_n T^a 
 (W^\dagger \xi_n^{(u)})_{\omega_2} \big]\nn \,,
\end{eqnarray}
where the collinear and soft Wilson lines $W$ and $S$ are defined in
Eq.~(\ref{WS}) of Appendix~\ref{app_S}.  At leading order in $1/Q$ only the
operators ${\cal Q}^{(0)}_{L, R}$ and the leading order collinear and soft
Lagrangians (${\cal L}_{c}^{(0)}$, ${\cal L}_s^{(0)}$), contribute to the
$B^-\to D^{(*)0}\pi^-$ and $\bar B^0\to D^{(*)+}\pi^-$ matrix elements.  The
matrix elements of ${\cal Q}_{L,R}^{(8)}$ vanish because they factorize into a
product of bilinear matrix elements and the octet currents give vanishing
contribution between color singlet states~\cite{bps}. 

Note that we take the pion state or interpolating field to be purely collinear
and the $B$ and $D^{(*)}$ states to be purely soft. Power corrections to these
states are included as time ordered products. This includes asymmetric
configurations containing one soft and one collinear quark which involve
$T$-products with subleading Lagrangians~\cite{bps4}.

Next we consider type-II decays. The matrix elements of the leading order
operators vanish, $\langle D^0\pi^0| Q_j^{(0,8)}| \bar B^0 \rangle =0$. This
occurs due to a mismatch between the type of quarks produced by ${\cal Q}_j^{(0,8)}$
and those required for the light meson state, where we need two collinear quarks
of the same flavor.  The operator ${\cal Q}_{j}^{(0,8)}$ produces collinear quarks
with $(d u)$ flavor.  Therefore it can not produce a $\pi^0$ since the leading
order SCET Lagrangian only produces or annihilates collinear quark pairs of the
same flavor. For this reason the leading contributions to $\bar B^0\to
D^{(*)0}\pi^0$ are power suppressed.

In SCET$_{\rm I}$ there are several sources of power suppressed contributions
obtained by including higher order four quark operators, higher order
contributions from the Lagrangians, or both.  However, there is only a {\em
  single type of SCET$_{I}$ operator} which contributes to $\bar B^0\to D^{(*)0}
M^0$ decays at leading order.  They are given by $T$-ordered products of the
leading operators in Eq.~(\ref{QVI}) with two insertions of the usoft-collinear
Lagrangian ${\cal L}_{\xi q}^{(1)}$:
\begin{eqnarray}\label{Tprod}
  T_j^{(0,8)} &=& \frac12 
  \int\!\! d^4\!x\, d^4\!y\ T\big\{ {\cal Q}_j^{(0,8)}(0)\,, 
   i{\cal L}_{\xi q}^{(1)}(x)\,, i{\cal L}_{\xi q}^{(1)}(y)\big\}\,.
\end{eqnarray}
Here the subleading Lagrangian is~\cite{bcdf,bps4}
\begin{eqnarray}\label{Lxiq}
  {\cal L}^{(1)}_{\xi q} &=&   
 (\bar\xi_n W) \: \Big(\frac{1}{\bnP}\: W^\dagger ig\, \Bslash_\perp^c W\Big)
  q_{us} - \bar q_{us} \Big( W^\dagger ig\, \Bslash_\perp^c W 
  \frac{1}{\bnP^\dagger} \Big) (W^\dagger \xi_n) \,,
\end{eqnarray}
where $ig \Bslash_\perp^c =[i\bn\mcdot D^c,i\Dslash_\perp^{\,c}]$. The two
factors of $i{\cal L}_{\xi q}^{(1)}$ in Eq.~(\ref{Tprod}) are necessary to swap
one $u$ quark and one $d$ quark from ultrasoft to collinear.  In contrast to the
tree amplitude, for this case both the ${\cal Q}^{(0)}_j$ and ${\cal Q}^{(8)}_j$
operators can contribute. By power counting, the $T_j^{(0,8)}$'s are suppressed by
$\lambda^2 = \Lambda/Q$ relative to the leading operators.  They will give order
$\Lambda/Q$ contributions in \SCETb, in agreement with our earlier statements.

\begin{figure}[!t]
\vskip-0.3cm
 \centerline{\hspace{0.6cm}
  \mbox{\epsfxsize=5.5truecm \hbox{\epsfbox{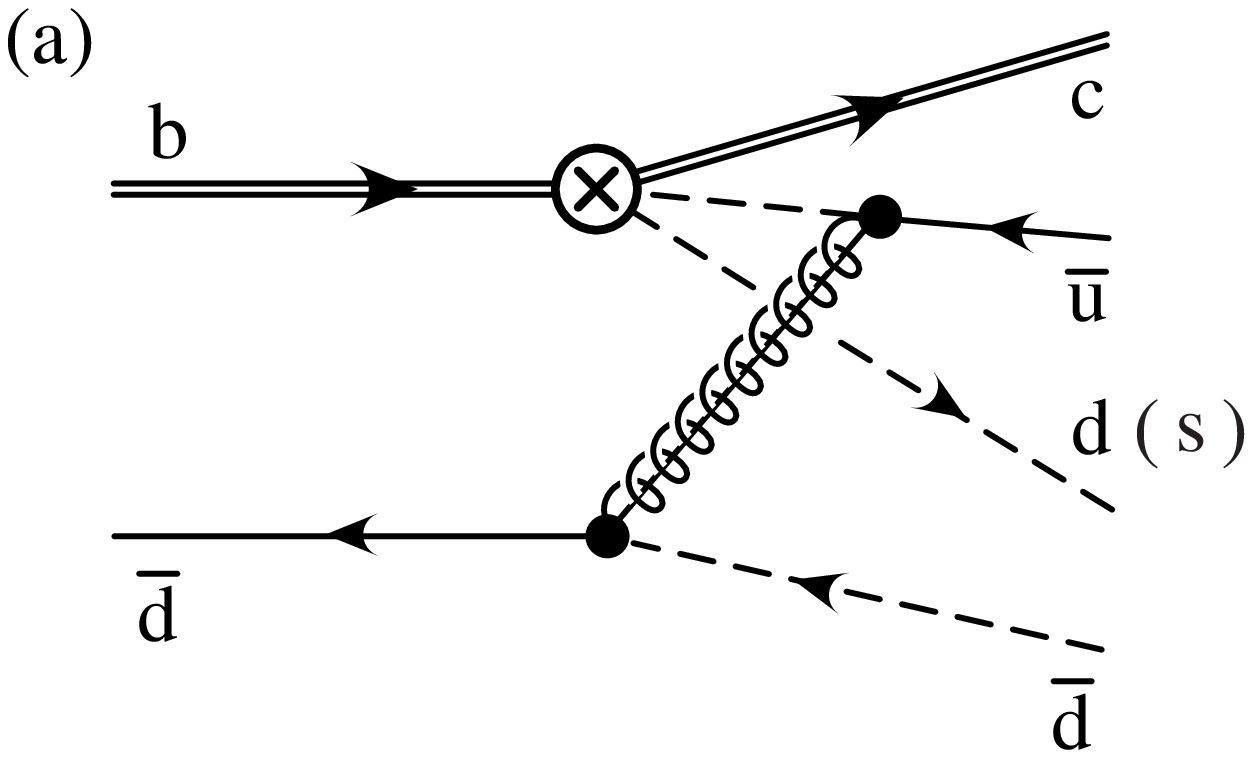}} }
  \hspace{0.cm}
  \raisebox{0.4cm}{\mbox{\epsfxsize=5.7truecm
      \hbox{\epsfbox{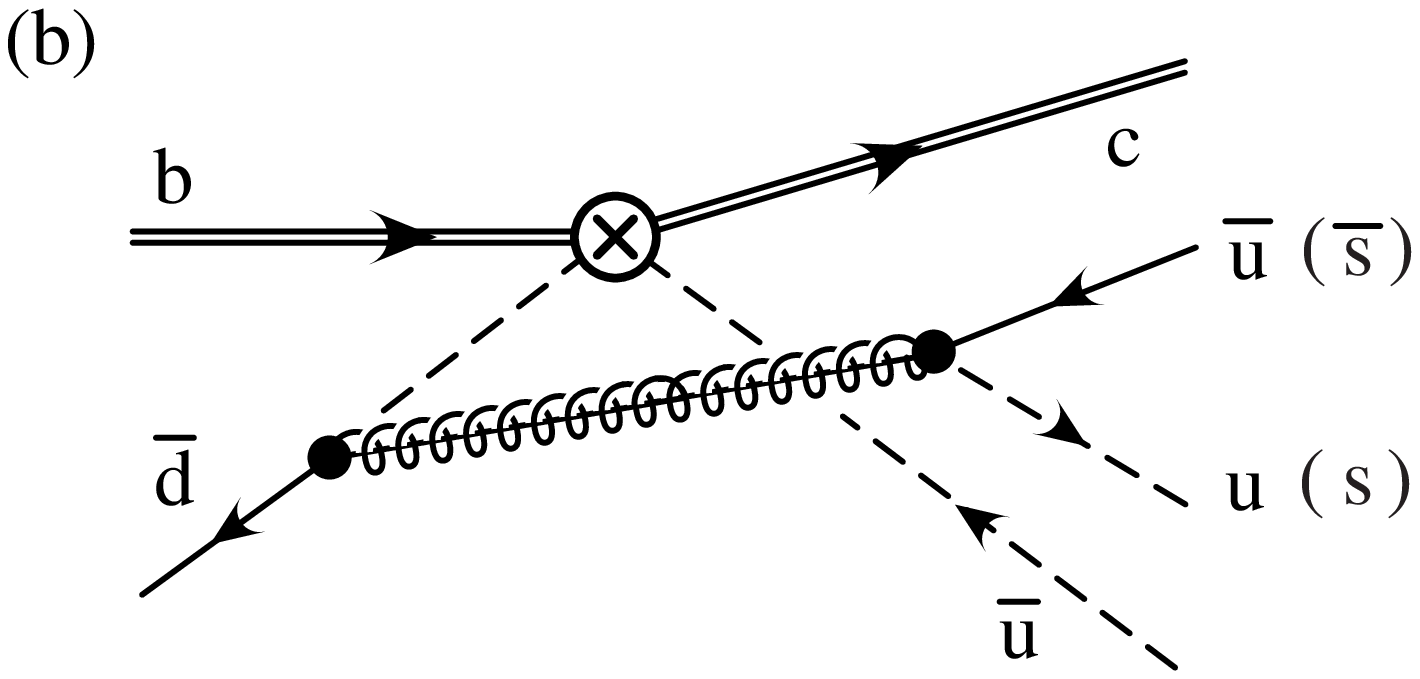}}} }
  }
\vskip-0.2cm
  \centerline{\hspace{0.6cm}
  \mbox{\epsfxsize=5.5truecm \hbox{\epsfbox{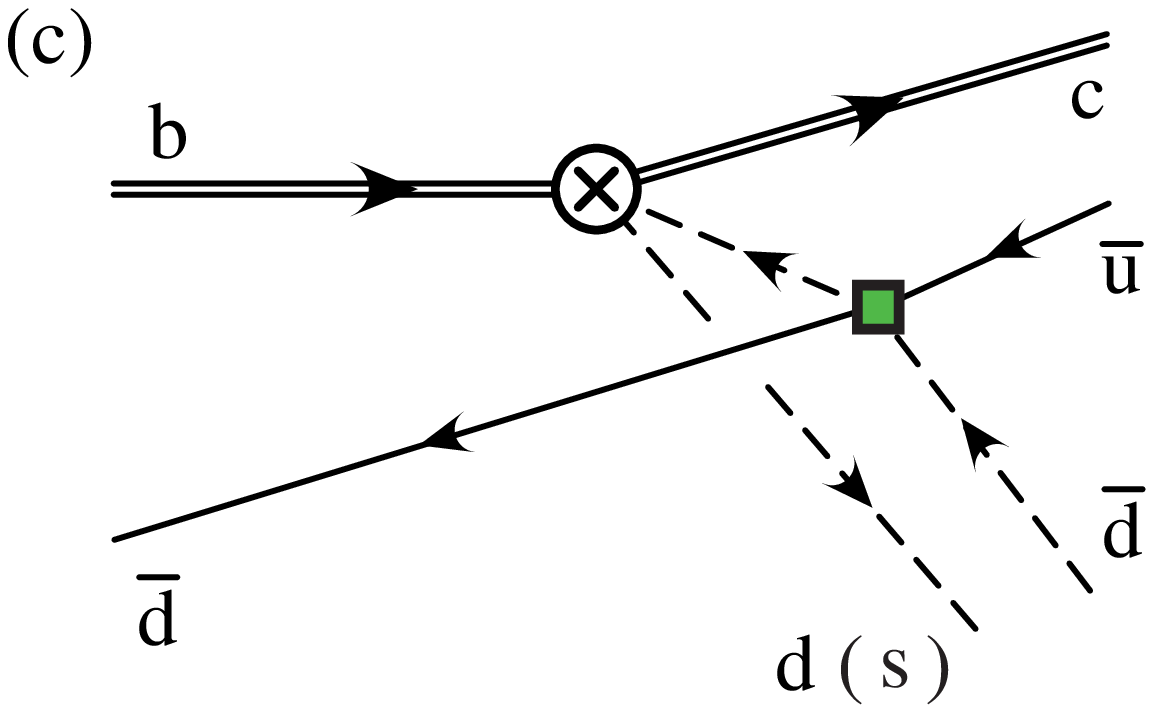}} }
  \hspace{0.cm}
  \raisebox{0.cm}{\mbox{\epsfxsize=5.7truecm
      \hbox{\epsfbox{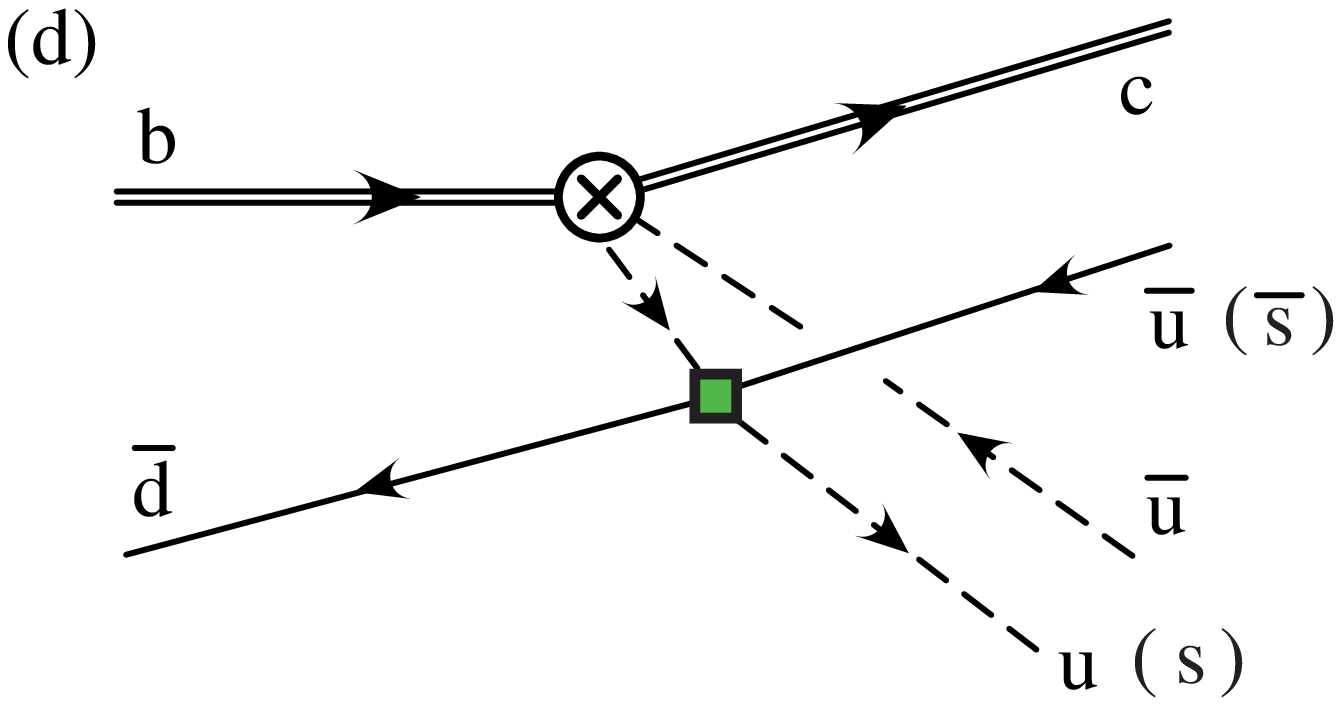}}} }
  \hspace{0.5cm}
  \raisebox{0.2cm}{\mbox{\epsfxsize=5.3truecm
      \hbox{\epsfbox{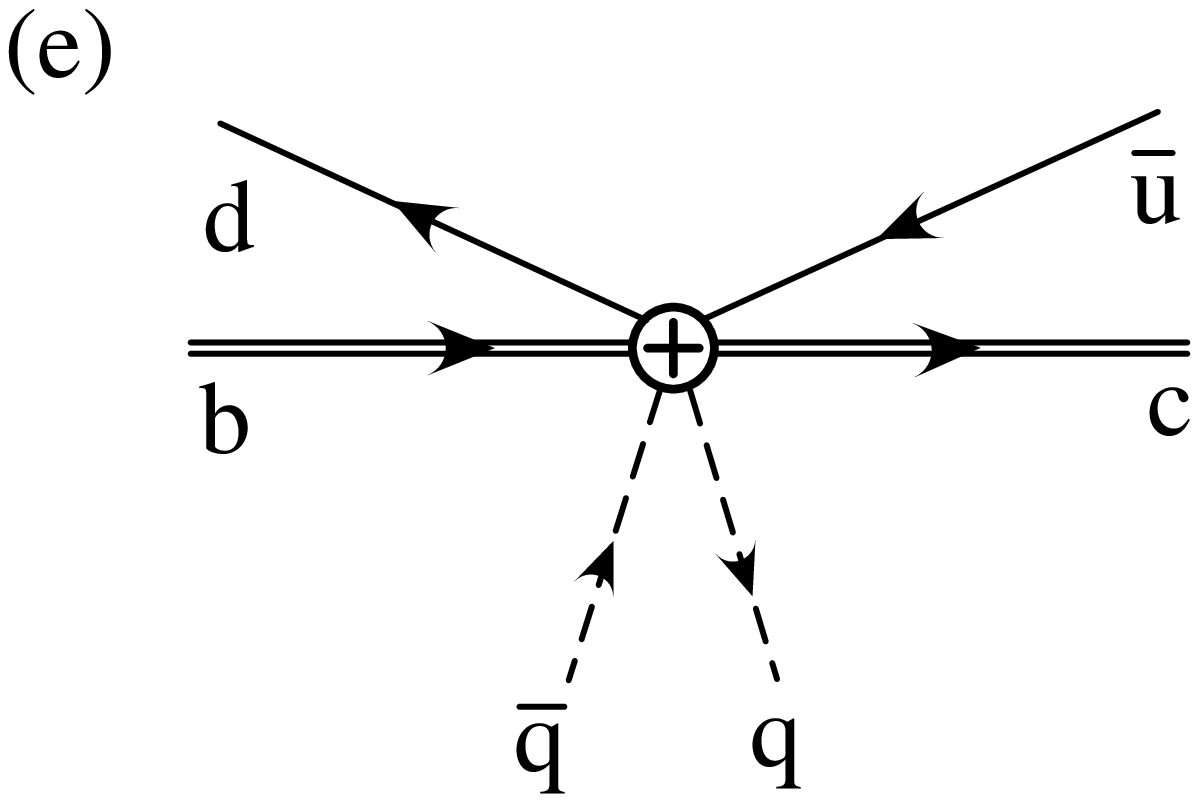}}} }
  }
\vskip-0.3cm
\caption[1]{Graphs for the tree level matching calculation from
  \SCETa (a,b) onto \SCETb (c,d,e). The dashed lines are collinear quark
  propagators and the spring with a line is a collinear gluon. Solid lines in
  (a,b) are ultrasoft and those in (c,d,e) are soft. The $\otimes$ denotes an
  insertion of the weak operator, given in Eq.~(\ref{QVI}) for (a,b) and in
  Eq.~(\ref{QV}) in (c,d). The $\oplus$ in (e) is a 6-quark operator from
  Eq.~(\ref{OV}).  The two solid dots in (a,b) denote insertions of the mixed
  usoft-collinear quark action ${\cal L}_{\xi q}^{(1)}$.  The boxes denote the
  \SCETb operator ${\cal L}_{\xi\xi qq}^{(1)}$ in Eq.~(\ref{L4q1}).}
\label{fig_scet1} 
\vskip0cm
\end{figure}

In Fig.~\ref{fig_scet1} we show graphs contributing to the matching of \SCETa
operators (a,b) onto operators in \SCETb (c,d,e).  In Figs.~\ref{fig_scet1}a,b
the gluon always has offshellness $p^2\sim E_M\Lambda$ due to momentum
conservation, and is shrunk to a point in \SCETb.  However, the collinear quark
propagator in (a,b) can either have $p^2\sim E_M\Lambda$ giving rise to the
short distance \SCETb contribution in Fig.~\ref{fig_scet1}e, or it can have
$p^2\sim \Lambda^2$ which gives the long distance \SCETb contribution in
Figs.~\ref{fig_scet1}c,d. To match onto the short distance contribution in
Fig.~\ref{fig_scet1}e we subtract the \SCETb diagrams (c,d):
\begin{eqnarray}
 (a) + (b) - (c) - (d) = (e) \,.
\end{eqnarray} 
The operators in Figs.~\ref{fig_scet1}a,b are from the $T$-products
$T_j^{(0,8)}$ in Eq.~(\ref{Tprod}), while Figs.~\ref{fig_scet1}c,d involve the
\SCETb $T$-products $\overline O_j^{(i)}$ in Eq.~(\ref{Obar}), and
Fig.~\ref{fig_scet1}e involves $O_j^{(i)}$ in Eq.~(\ref{OV}).

To generate connected \SCETa diagrams from the time-ordered product in
Eq.~(\ref{Tprod}) requires at least two contractions, of which the minimum basic
possibilities can be grouped as follows:
\begin{itemize}
 \item[1)] Contraction of $\xi_n^{(u)}\, \bar\xi_n^{(u)}$ and the $\perp$ gluon
   in $B_{\perp}^\mu B_{\perp}^\nu$ (C-topology, Fig.~\ref{fig_scet1}a),
 \item[2)] Contraction of $\xi_n^{(d)}\, \bar\xi_n^{(d)}$ and the $\perp$ gluon
   in $B_{\perp}^\mu B_{\perp}^\nu$ (E-topology, Fig.~\ref{fig_scet1}b),
 \item[3)] Contraction of $\xi_n^{(u)}\, \bar\xi_n^{(u)}$ and 
   $\xi_n^{(d)}\, \bar\xi_n^{(d)}$ (topology with two external collinear gluons
   and no external collinear quarks, not shown).
\end{itemize} 
All more complicated contractions have one of these three as a root. Case 3)
only contributes for light mesons with an isosinglet component ($\eta$, $\eta'$,
$\omega$, $\phi$), which we will not consider here.

Each of the \SCETa $T$-products is matched onto \SCETb operators at scale
$\mu=\mu_0$, and
\begin{eqnarray}\label{jets}
  &&\hspace{-3cm} \int d\tau_1\:d\tau_2\: C_j^{(0,8)}\, T_{j}^{(0,8)} 
   \to \big[ T_{j}^{(0,8)} \big]_{\rm short} 
      + \big[ T_{j}^{(0,8)} \big]_{\rm long} \,,\\
\big[ T_{L,R}^{(0,8)} \big]_{\rm short} 
 &=&
   \int\!\!  \mbox{d}\tau_i \: \mbox{d}k^+_\ell\:\mbox{d}\omega_k \: 
  C^{(0,8)}_{L,R}(\tau_i,\mu_0)\:
  J^{(0,8)}(\tau_i,k^+_\ell,\omega_k,\mu_0,\mu)\:
  O^{(0,8)}_{L,R}(k^+_\ell,\omega_k,\mu)\,, \nn\\
\big[ T_{L,R}^{(0,8)} \big]_{\rm long} 
 &=&
   \int\!\!\mbox{d}k^+\,
  \mbox{d}\omega_1\:  \mbox{d}\omega_2\: \mbox{d}\omega\:  
  C^{(0,8)}_{L,R}(\omega_i,\mu_0)\:
  \overline J^{(0,8)}(k^+\omega,\mu_0,\mu)\:
  \overline O^{(0,8)}_{L,R}(\omega_i,k^+\! ,\omega,\mu)  
 \,, \nn
\end{eqnarray}
where the subscripts $i,\ell,k$ run over values $1,2$.  Here $J$, $\overline J$
are jet functions containing effects at the $p^2\sim E_M\Lambda$ scale and are
Wilson coefficients for the \SCETb operators $O$ and $\overline O$. The
$[T_{L,R}^{(0,8)}]_{\rm short}$ and $[T_{L,R}^{(0,8)}]_{\rm long}$ terms are
respectively Fig.~\ref{fig_scet1}e and Fig.~\ref{fig_scet1}c,d (after they are
dressed with all possible gluons).  The $\mu_0$ and $\mu$ dependence in
Eq.~(\ref{jets}) signifies the scale dependence in \SCETa and \SCETb
respectively. The jet functions are generated by the contraction of intermediate
collinear fields with couplings $\alpha_s(\mu_0)$ (where $\mu_0^2\sim E_\pi
\Lambda$).  In general the jet functions depend on the large light-cone momenta
$\tau_i$ coming out of the hard vertex, the large light-cone momenta $\omega_k$
of the external collinear \SCETb fields, and the $k_j^+$ momenta of the external
soft \SCETb fields. No other soft momentum dependence is possible since the
leading \SCETa collinear Lagrangian depends on only $n\mcdot\partial_{us}$.

The difference between the time-ordered products $T^{(0,8)}_{L,R}$ and the
time-ordered products $\overline O_{L,R}^{(0,8)}$ gives the six quark \SCETb
operator $O_{L,R}^{(0,8)}$, whose coefficients are the jet functions
$J^{(0,8)}$. In this \SCETa\!--\SCETb matching calculation the $\overline
O_{L,R}^{(0,8)}$ graphs subtract long distance contributions from the
$T^{(0,8)}_{L,R}$ graphs so that $J^{(0,8)}$ are free of infrared singularities.
In general the matrix elements for color suppressed decays then include both
short and long distance contributions as displayed in Eq.~(\ref{jets}). However,
for the isotriplet $\pi$ and $\rho$ a dramatic simplification occurs at leading
order in $C_J^{(i)}$. In this case it can be proven that the long distance
contributions $[T_j^{(i)}]_{\rm long}$ {\em vanish} to all orders in the
$\alpha_s$ couplings in \SCETa, and with the $\alpha_s$ couplings in \SCETb
treated non-perturbatively. The proof of this fact uses the $G$-parity
invariance of QCD and is carried out in Appendix~\ref{app_G}. At leading order
in the coefficients $C_{L,R}^{(0,8)}$ the $M=\pi,\rho$ factorization theorem is
therefore more predictive since possible long-distance contribution from
$\overline O_{j}^{(i)}$ are absent.  Most of the following discussion will focus
on $O_{j}^{(i)}$, but $\overline O_{j}^{(i)}$ is fully included in the final
factorization theorem.

\begin{figure}[!t]
\vskip-0.3cm
 \centerline{\hspace{0.6cm}
  \mbox{\epsfxsize=5.5truecm \hbox{\epsfbox{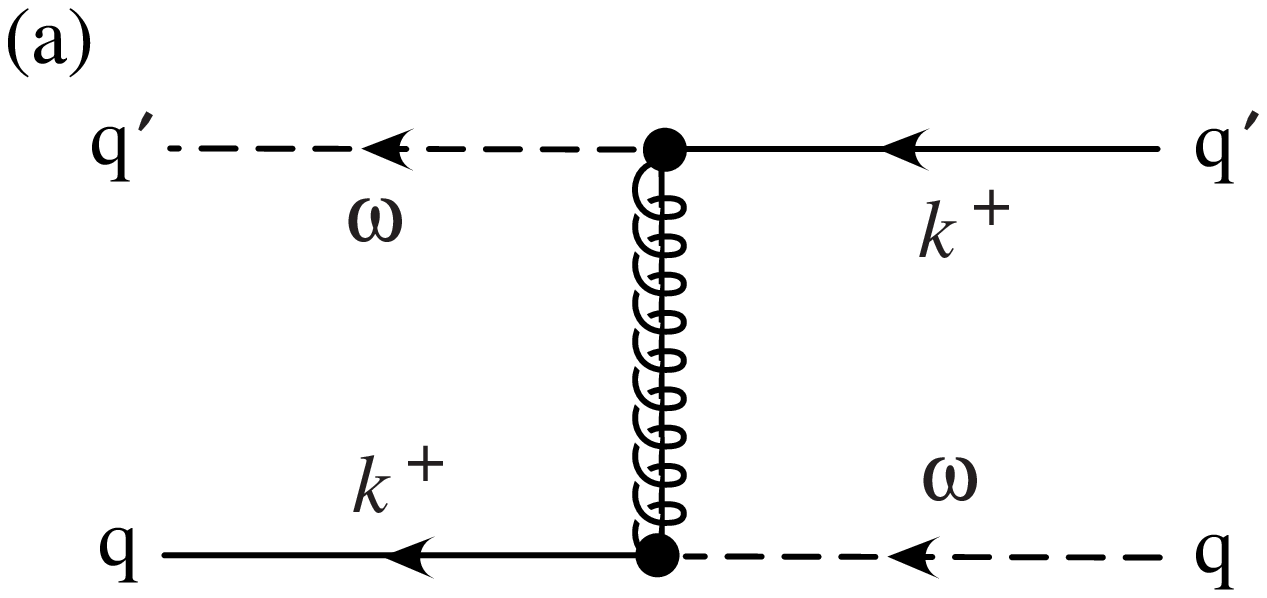}} }
  \hspace{2.cm}
  \raisebox{0.0cm}{\mbox{\epsfxsize=5.5truecm
      \hbox{\epsfbox{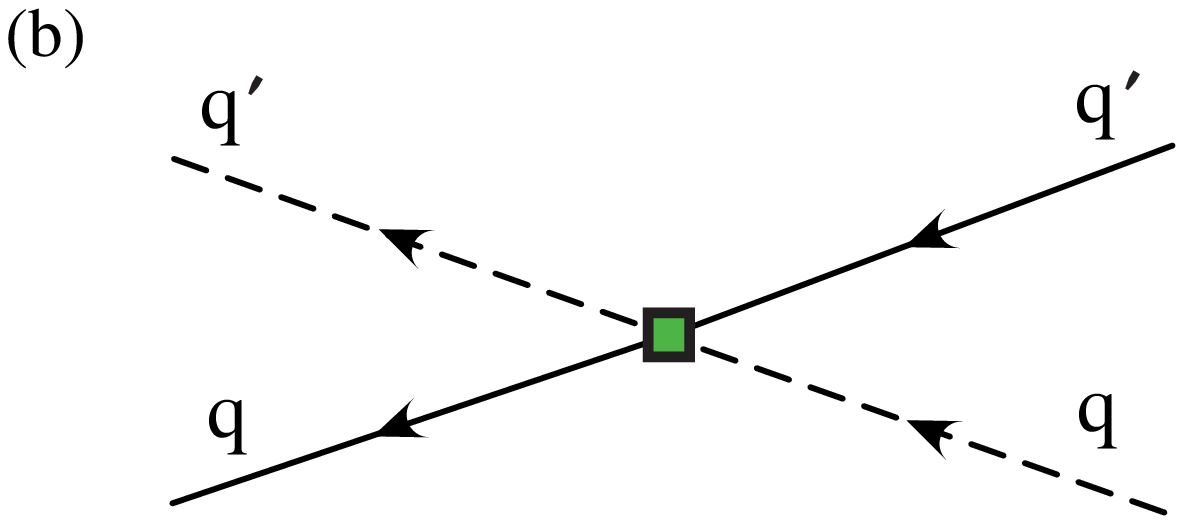}}} }
  }
\vskip-0.cm
\caption[1]{Tree level matching calculation for the $L_{L,R}^{(0,8)}$ operators,
with (a) the $T$-product in \SCETa and (b) the operator in \SCETb. Here $q,q'$ are
flavor indices and $\omega\sim \lambda^0$ are minus-momenta.}
\label{fig_scet1L} 
\vskip0cm
\end{figure}

In the \SCETb diagrams in Fig.~\ref{fig_scet1}c,d a power suppressed four quark
Lagrangian appears. It is similar to an operator introduced in Ref.~\cite{HN},
and can be obtained from  $T\{ i{\cal L}_{\xi q}^{(1)}, i{\cal L}_{\xi
  q}^{(1)} \}$ in \SCETa by a simple matching calculation \cite{bpsgi}. Summing 
over flavors $q,q'$ we find
\begin{eqnarray} \label{L4q1} 
   {\cal L}_{\xi\xi qq}^{(1)} \!\! 
  &=&\!\! \sum_{j=L,R} \sum_{\omega} \sum_{k^+}
 % \int\!\! \mbox{d}\omega\: \mbox{d}^2\!x_\perp\,  \mbox{d}k^+\, 
  \Big[ \bar J^{(0)}(\omega k^+)
  L^{(0)}_j(\omega, k^+\!\!, x) + \bar J^{(8)}(\omega k^+)
  L^{(8)}_j(\omega, k^+\!\!, x) \Big] \,,\nn\\
%\begin{eqnarray} \label{L4q2}
 L^{(0)}_{j}(\omega, k^+\!\!, x)
 &=&   \sum_{q,q'}\! %\int\!\! \frac{dx^- dx^+}{2}
   \Big[ \big(\bar \xi_n^{(q)} W\big)_{\omega} \bnslash P_{j} 
   \big(W^\dagger \xi_n^{(q')}\big)_{\omega} \Big]%(x_\perp) \nn \\
 %&& \times  
  \Big[ \big(\bar q^{\,\prime} S\big)_{k^+} \nslash P_{j}
   \big(S^\dagger q )_{k^+}  \Big](x) \, . 
\end{eqnarray}
In Eq.~(\ref{L4q1}) the soft momenta labels are defined by $(S^\dagger q)_{k^+}
= [\delta(k^+\!-\!n\mcdot P)\: S^\dagger q]$, and the positions
$(x^+,x^-,x_\perp)\sim (1/\Lambda,Q/\Lambda^{2},1/\Lambda)$. For the soft fields
the $x^-$ coordinates encode small residual plus-momenta, and for the collinear
fields the $x^+$ coordinates encode small residual minus-momenta. Thus, we used
the summation/integration notation for label/residual momenta from
Ref.~\cite{LMR}. The operator $L^{(8)}_{j}(\omega, k^+\!, x)$ has
the same form as Eq.~(\ref{L4q1}) except with color structure $T^a\otimes T^a$.
At tree level the coefficient functions are given by the calculation in
Fig.~\ref{fig_scet1L}
\begin{eqnarray} \label{L4qtree}
  &&  \overline J^{(0)}(\omega k^+) = -\frac{C_F}{2N_c}\
  \frac{4\pi\alpha_s(\mu)}{\omega k^+} \,, \qquad
  \bar J^{(8)}(\omega k^+) = \frac{1}{2N_c}\
  \frac{4\pi\alpha_s(\mu)}{\omega k^+} \,.
\end{eqnarray}
Beyond tree level they obtain contributions from loop diagrams with additional
${\cal L}_{\xi\xi}^{(0)}$ vertices.  In terms of the operator in
Eq.~(\ref{L4q1}) the \SCETb operators that contribute to $[T_j^{(i)}]_{\rm
  long}$ in the factorization theorem are
\begin{eqnarray} \label{Obar}
  \overline O_{j}^{(0,8)}(\omega_i,k^+\!, \omega,\mu) &=& 
  \int\!\! \mbox{d}^4x\: T\,
   {\cal Q}^{(0,8)}_{j}(\omega_i,x=0)\ i L^{(0,8)}(\omega, k^+ \!, x) \,.
\end{eqnarray}
\OMIT{where integrals over the momentum fractions are suppressed for convenience.}  The
operators $\overline O$ generate the diagrams (c) and (d) in
Fig.~\ref{fig_scet1}.

At any order in perturbation theory the jet functions $J$ from the $C$--topology
and $E$--topology generate one spin structure, and two color structures for the
\SCETb operators. For the six quark operators we find
\begin{eqnarray} \label{OV}
 O_{j}^{(0)}(k^+_i,\omega_k) &=& 
  \Big[ \bar h_{v'}^{(c)}  \Gamma^h_j\,  h_v^{(b)} \:
  (\bar d\,S)_{k^+_1} \nslash P_L\, (S^\dagger u)_{k^+_2} \Big]
  \Big[ (\bar \xi_n W)_{\omega_1} \Gamma_c (W^\dagger \xi_n)_{\omega_2} 
  \Big]\,, \\
 O_{j}^{(8)}(k^+_i,\omega_k) &=& 
  \Big[ (\bar h_{v'}^{(c)} S) \Gamma^h_j\, T^a\, (S^\dagger h_v^{(b)}) \:
  (\bar d\,S)_{k^+_1} \nslash P_L T^a (S^\dagger u)_{k^+_2} \Big]
  \Big[ (\bar \xi_n W)_{\omega_1} \Gamma_c (W^\dagger \xi_n)_{\omega_2} 
  \Big] \nn\,,
\end{eqnarray}
where here the $d$, $u$, $h_{v'}^{(c)}$, and $h_v^{(b)}$ fields are soft, and
the $\xi_n$ fields are collinear isospin doublets, $(\xi_n^{(u)},\xi_n^{(d)})$.
In Eq.~(\ref{OV}) $\Gamma^h_{L,R}= \nslash P_{L,R}$ as in Eq.~(\ref{QVI}), while
for the collinear isospin triplet $\Gamma_c = \tau^3 \bnslash
P_L/2$.\footnote{There are also isosinglet contributions with $\Gamma_c =
  \bnslash P_L/2$.} We do not list operators with a $T^a$ next to $\Gamma_c$ since
they will give vanishing contribution in the collinear matrix element. For light
vector mesons the spin structure $\Gamma_c$ only produces the longitudinal
polarization.  This result follows from the quark helicity symmetry of ${\cal
  L}_{\xi\xi}^{(0)}$ and is discussed in further detail in Appendix~\ref{app_J}.

\begin{figure}[!t]
\vskip-0.3cm
 \centerline{
  \mbox{\epsfxsize=5.4truecm \hbox{\epsfbox{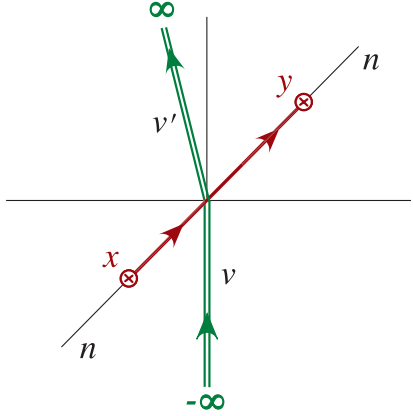}} }
  }
\vskip-0.3cm
\caption[1]{
  Non-perturbative structure of the soft operators in Eq.~(\ref{position}) which
  arise from $O_j^{(0,8)}$. 
  Wilson lines are shown for the paths $S_n(x,0)$, $S_n(0,y)$, $S_v(-\infty,0)$ and
  $S_{v'}(0,\infty)$, plus two interacting QCD quark fields inserted at the
  locations $x$ and $y$. The $S_v$ and $S_{v'}$ Wilson lines are from 
  interactions with the fields $h_v$ and $h_{v'}$ fields, respectively. The
  non-perturbative structure of soft fields in $\overline O_j^{(0,8)}$ is
  similar except that we separate the single and double Wilson lines by an amount
  $x_\perp$.}
\label{fig_wilson} 
\vskip0cm
\end{figure}

In position space the $O_j^{(i)}$ are bilocal operators, with the two soft light
quarks aligned on the $n_\mu$ light cone direction ($x^- = \frac12 n_\mu
\bn\cdot x, y^- = \frac12 n_\mu \bn\cdot y )$ passing through the point $x=0$
\begin{eqnarray}\label{position}
& &(\bar h_{v'}^{(c)} S) \Gamma_h (S^\dagger h_v^{(b)}) \:  
  (\bar d\,S)_{r^+} \Gamma_q (S^\dagger u)_{\ell^+} = \\
& &\qquad 
 \int \frac{dx^- dy^-}{(4\pi)^2}\: e^{i/2(r^+ x^- - \ell^+ y^-)}
 [\bar h_{v'}^{(c)} \Gamma_h h_v^{(b)}](0) 
  [\bar d(x^-) S_n(x^-, 0) \Gamma_q S_n(0, y^-) u(y^-)]\,.\nn
\end{eqnarray}
The gluon interactions contained in matrix elements of $O^{(0,8)}_j$ include
attachments to the light quarks $q$, to the heavy quarks $h_{v,v'}$, and to the
Wilson lines $S_n$ as shown in Fig.~\ref{fig_wilson}. The interactions with
$h_{v,v'}$ have been drawn as Wilson lines $S_{v,v'}$ along $v,v'$~\cite{KR}.
Even though we have factored the collinear and soft degrees of freedom in the
two final state hadrons, the presence of the soft Wilson lines bring in
information about the vector $n^\mu$. This allows the soft operators $O_j^{(i)}$
to be non-trivial functions of $n\cdot k_j$, $n\cdot v$, and $n\cdot v'$, and
this information gives rise to a {\em complex phase} in the soft functions
$S_{L,R}^{(0,8)}$ as shown in Appendix~\ref{app_S}.  Thus, the $S_n$ Wilson
lines are directly responsible for producing final state interactions, and the
soft fields in $O_j^{(0,8)}$ encode non-perturbative rescattering
information.\footnote{Note that in semi-inclusive processes a different
  mechanism is responsible for the phases in single-spin asymmetries which has to
  do with the boundary conditions on Wilson lines~\cite{ssa}.}  This makes good
sense given that the soft gluons in the $S_n$'s were originally generated by
integrating out attachments to the collinear quarks and gluons making up the
light energetic hadron.

The above procedure provides a {\em new} mechanism for generating
non-perturbative strong phases for exclusive decays within factorization. In the
soft $B\to D^{(*)}$ matrix elements the information about the light energetic
meson is limited to its direction of motion $n^\mu$.  Since these matrix
elements know nothing further about the nature of the light meson, these strong
phases are universal.  In particular the same strong phase $\phi$ is generated
for the decays $\bar B\to D^{(*)}\pi$ and $\bar B\to D^{(*)}\rho$.  (We caution
that this is not the isospin strong phase, but rather a different angle in the
triangle.) The same mechanism produces another universal strong phase for color
suppressed decays to $D\bar K^{(*)0}$, and a third for decays to $D_s K^{(*)-}$.
The different phases in the three classes arise in part due to the appearance of
different moments of the matrix elements of the soft operators. However, for the
kaons there are additional long distance contributions to the strong phases from
$[T]_{\rm long}$, which make the universality of the phase $\phi$ from $[T]_{\rm
  short}$ hard to test. A more complete set of phenomenological predictions is
given in Section~\ref{sect_results}, including a comparison with existing data.
Further details on the properties of the soft functions $S^{(0,8)}$ are given in
the Appendix~\ref{app_S}.

The matrix elements of the short distance operators $O_j^{(i)}$ in
Eq.~(\ref{OV}) factor into products of soft and collinear parts, respectively.
The collinear part of the matrix elements are simply given in terms of the light
cone wave function of the light meson. For the $\pi$ and $\rho$ the definitions are
[we suppress pre-factors of $\int_0^1 dx\, \delta(\omega_1-x\, \bn\mcdot
p_M)\, \delta(\omega_2+(1-x)\bn\mcdot p_M)$ on the RHS]\footnote{Our vector meson
  states are defined with an extra minus sign relative to the standard
  convention.}
\begin{eqnarray} \label{phipirho}
\langle \pi_n^0 |(\bar \xi_n W)_{\omega_1} \bnslash\gamma_5 \tau_3
  (W^\dagger \xi_n)_{\omega_2}| 0\rangle
  &=& -i\,\sqrt{2}\, f_\pi \: \bn\mcdot p_\pi\: \phi_\pi(\mu,x)\: \,,\\
\langle \rho_n^0(\varepsilon) |(\bar \xi_n W)_{\omega_1} \bnslash\tau_3
  (W^\dagger \xi_n)_{\omega_2}| 0\rangle
  &=& i\,\sqrt{2}\, f_\rho \: m_\rho\, \bn\mcdot\varepsilon^*\: \phi_\rho(\mu,x)\:
  \nn\\ 
  &=& i\,\sqrt{2}\, f_\rho \: \bn\mcdot p_\rho\: \phi_\rho(\mu,x)\: \,. \nn
\end{eqnarray}
In the last equality we have used the fact that at this order the collinear
operator only produces longitudinal $\rho$'s, for which $m_\rho \bn\mcdot
\varepsilon_L^* = \bn\mcdot p_\rho$. 

\OMIT{$\omega_1 +
  \omega_2 = \bn\mcdot p_\pi (2x-1)$ and $\theta(x)\theta(1-x)\, \delta(\omega_1
  \!-\!  \omega_2\!-\!\bn\cdot p_\pi )$ on the RHS]}

Since it no longer contains couplings to energetic gluons, the soft part of the
matrix elements of ${\cal O}^{(0,8)}_j$ can be constrained using heavy quark
symmetry. In other words, heavy quark symmetry relations can be derived for
matrix elements of soft fields.  The constraints can be implemented most
compactly using the trace formalism of the HQET~\cite{hbook}. First consider the
matrix element of the soft fields in $O_j^{(0,8)}$. For $O_j^{(0)}$ we have
\begin{eqnarray}\label{trace}
 \frac{\langle D^{(*)0}(v') | (\bar h_{v'}^{(c)} S) \Gamma\, 
  (S^\dagger h_v^{(b)}) \:(\bar d\,S)_{k^+_1} \nslash P_L\, (S^\dagger u)_{k^+_2}
 | \bar B^0(v)\rangle}{\sqrt{m_B m_D}}
  =  \mbox{Tr } [\overline{H}_{v'}^{(c)} \Gamma H_v^{(b)} X^{(0)} ]\,,
\end{eqnarray}
where $X^{(0)}=X^{(0)}(k^+_j,n,v,v')$ and we use the standard relativistic
normalization for the states (and note that the LHS is independent of $m_{b,c}$
in the heavy quark limit). An identical equation holds for $O_j^{(8)}$ with an
$X^{(8)}$. In writing the trace formula in Eq.~(\ref{trace}) we have used the
fact that the $d$ and $u$ quarks must end up in the $\bar B$ and $D^{(*)}$
states.\footnote{ The matrix element of the analogous soft operators with $(\bar
  u u)+(\bar d d)$ would contain a second term in Eq.~(\ref{trace}) of the form
  $\mbox{Tr } [\overline{H}_{v'}^{(c)} \Gamma H_v^{(b)} X] \mbox{Tr }[Y]$, which
  arises from contracting the light quarks in the operator. These types of
  traces also show up for power corrections to $\bar B^0\to D^{(*)+}M^-$ and
  $B^-\to D^{(*)0}M^-$.}  The heavy mesons $(D,D^*)$ and $(B,B^*)$ are grouped
together into superfields~\cite{hbook}, defined as
\begin{eqnarray}
  H_v = \frac{1+\vslash}{2}(P_v^{*\mu} \gamma_\mu + P_v \gamma_5)\,.
\end{eqnarray}
Now $X^{(0,8)}$ are the most general structures compatible with the symmetries
of QCD. They involve 4 functions $a_{1-4}^{(0,8)}
(k^+_1,k^+_2,v\mcdot v',n\mcdot v,n\mcdot v')$
\begin{eqnarray}  \label{Xred}
X^{(0,8)} =  a_1^{(0,8)} \nslash P_L + a_2^{(0,8)} \nslash P_R 
  + a_3^{(0,8)} P_L + a_4^{(0,8)} P_R  \,,
\end{eqnarray}
Structures proportional to $\vslash$ and $\vslash'$ can be eliminated by using
$H_v \vslash = -H_v$, etc. 

The presence of four functions in Eq.~(\ref{Xred}) would appear to restrict the
predictive power of heavy quark symmetry. However, using the properties of $H_v$
and $\bar H_{v'}$ and the fact that the two-body kinematics relates $n$ to $v$
and $v'$ via $m_B v=m_D v'+E_M n$, it is easy to see that the four functions
$a_i$ appear only in two distinct combinations. (Note that we are taking
$m_M/m_B\sim \Lambda/m_B\ll 1$.) For $\Gamma_{L,R}^h$ they give soft functions
$S_{L,R}$ defined as $S_L = (n\cdot v')(a_1 - a_3/2) - a_4/2, S_R = (n\cdot v')
(a_2 - a_4/2) - a_3/2$ and
\begin{eqnarray}\label{Sintro}
 \frac{\langle D^{0}(v') | (\bar h_{v'}^{(c)}S) \nslash P_{L,R} 
 (S^\dagger h_v^{(b)})
 (\bar d S)_{k^+_1}\nslash P_L (S^\dagger u)_{k^+_2} 
 | \bar B^0(v)\rangle}{\sqrt{m_B m_D}}
 &=& S_{L,R}^{(0)}(k^+_j) \,,\nn \\
\frac{ \langle D^{*0}(v',\varepsilon) | (\bar h_{v'}^{(c)}S) \nslash P_{L,R} 
 (S^\dagger h_v^{(b)}) (\bar d S)_{k^+_1} \nslash P_{L} (S^\dagger u)_{k^+_2} 
 | \bar B^0(v)\rangle}{\sqrt{m_B m_{D^*}}}
 &=&  \pm \frac{n\mcdot \varepsilon^*}{n\mcdot v'}\: S_{L,R}^{(0)}(k^+_j) \,,
\end{eqnarray}
where the $\pm$ for the $D^*$ refers to the choice of $P_L$ or $P_R$.  Identical
definitions hold for the matrix elements of the color-octet operators which give
$S_{L,R}^{(8)}(k^+_j)$.  We will see in section~\ref{sect_results} that the
result in Eq.~(\ref{Sintro}) relates decay amplitudes and strong phases for
$\bar B^0\to D^0 M^0$ and $\bar B^0\to D^{*0} M^0$ at leading order in the power
expansion, and up to terms suppressed by $\alpha_s(Q)/\pi$.  If one takes $n\cdot
v = 1$, then $n\mcdot v'=m_B/m_D^{(*)}$, $v\mcdot v' =(m_B^2+m_{D^{(*)}}^2)/(2
m_B m_{D^{(*)}})$. The $D,D^*$ variables are equal in the heavy quark limit.

For the long distance operators $\overline O_j^{(i)}$ the same set of arguments
in Eqs.~(\ref{trace}--\ref{Sintro}) can be applied except that now we must add
terms $a_5^{(0,8)}\xslash_\perp P_L +a_6^{(0,8)}\xslash_\perp P_R$ to
$X^{(0,8)}$, and the $a_i$'s can also depend on $x_\perp^2$.  The functions
analogous to $S_{L,R}^{(0,8)}$ are defined as
$\Phi_{L,R}^{(0,8)}(k^+,x_\perp,\varepsilon_{D^*}^*)$.  In this case the $D$ and
$D^*$ decompositions are no longer related since the matrix element involves
both $n\cdot \varepsilon^*$ and $x_\perp\cdot \varepsilon^*$ terms for the
$D^*$.  Thus, due to the long distance contributions for light vector meson we
must restrict ourselves to the longitudinal polarization in order to have
equality for the $D$ and $D^*$ amplitudes.  In the case of the $\rho$ this
restriction is not important since the long distance contributions vanish (see
Appendix~\ref{app_G}).  However this observation does have phenomenological
implications for decays to $K^*$'s.

We are now in a position to write down the most general factorized result for
the amplitude for the decays $\bar B^0\to D^{(*)0}M^0$. Combining all the
factors, this formula contains the soft functions $S^{(0,8)}(k_1^+,k_2^+)$ from
Eq.~(\ref{Sintro}), the jet functions $J^{(i)}$ from Eq.~(\ref{jets}), and the
Wilson coefficients $C^{(0,8)}_{L,R}$ from Eq.~(\ref{QVI0}). In
$J^{(i)}(\tau_i,k_\ell^+,\omega_k)$ we can pull out a factor of
$\delta(\tau_1-\tau_2-\omega_1+\omega_2)$ by momentum conservation. This leaves
the variables $\tau_1+\tau_2=2 E_M(2z-1)$ and $\omega_1+\omega_2=2 E_M(2x-1)$
unconstrained, which give convolutions with the momentum fractions $z$ and $x$
respectively. In defining $J^{(i)}(z,x,k_\ell^+)$ we multiply
$J^{(i)}(\tau_i,k_\ell^+,\omega_k)$ by $\omega_1-\omega_2=\bn\cdot p_M$.  All
together the result for the $\bar B^0\to D^{(*)0} M^0$ amplitude is
\begin{eqnarray}\label{result}
 A^{D^{(*)}}_{00} &=&  N_0^M
     \int_0^1\!\!\!dx\, dz\!\!  \int\!\! dk_1^+ dk_2^+\, 
   \Big[ C^{(i)}_{L}(z)\: 
   J^{(i)}(z,x,k_1^+,k_2^+)\: S^{(i)}_L(k_1^+,k_2^+)\:  \phi_M(x) \\
  &&\qquad\qquad\qquad\qquad\quad
   \pm  C^{(i)}_{R}(z)\: 
   J^{(i)}(z,x,k_1^+,k_2^+)\: S_R^{(i)}(k_1^+,k_2^+)\:  \phi_M(x) \Big] \nn \\
 && + A_{\rm long}^{D^{(*)}\!M} \nn \,,
\end{eqnarray}
where we sum over $i=0,8$ and the $\mu_0,\mu$ dependence is as in
Eq.~(\ref{jets}).  The $A_{\rm long}^{D^{(*)}\!M}$ in Eq.~(\ref{result}) denotes
the contributions from the matrix elements of the \SCETb time-ordered products
$[T]_{\rm long}$. Also the $\pm$ refer to $D/D^*$,
$C^{(i)}_{L,R}(z)=C^{(i)}_{L,R}(\tau_1+\tau_2 , E_M, m_{b}, m_c, \mu)$,
and
\begin{eqnarray} \label{N0}
  N_0^M &=&  \frac{G_F V_{cb}^{\phantom{*}} V_{ud}^*}{2} \: f_M 
  \:  \sqrt{m_B m_{D^{(*)}}}\,.
\end{eqnarray}
The normalization factor is common since $m_D=m_D^*$ and $n\mcdot
\varepsilon^{(D^*)}=n\mcdot v'$.  This follows since the $M$'s produced by
$O_j^{(0,8)}$ are longitudinally polarized.  

The long distance amplitudes also obey a factorization theorem which can be
derived by examining the matrix elements of the $\overline O_{L,R}^{(0,8)}$
operators in Eq.~(\ref{Obar}).  First factorize the collinear fields into the
matrix element with the $M$ and the soft fields into the matrix element with the
$B,D^{(*)}$.  The independence of the collinear propagators on the residual soft
minus-momenta leads to a $\delta(x^+)$ and the independence of the soft
propagators on the residual collinear plus-momenta leads to a $\delta(x^-)$
(somewhat similar to the calculation for $B\to X_s\gamma$ as described in
Ref.~\cite{bps2}). The result is
\begin{eqnarray} \label{Along}
  A_{\rm long}^{D^{(*)}\!M} &=& 
  % \frac{G_F}{\sqrt2} V_{cb} V^*_{ud} 
   N_0^M \int_0^1\!\!\! dz\!\!  \int\!\! dk^+ d\omega \!\! \int\!\! d^2\!x_\perp \Big[
   C^{(i)}_{L}(z)\: 
   \bar J^{(i)}(\omega k^+)\: \Phi_{L}^{(i)}(k^+\!,x_\perp,\varepsilon^*_{D^*}) 
    \Psi_M^{(i)}(z,\omega,x_\perp,\varepsilon^*_M) \nn\\
 && \qquad \pm   C^{(i)}_{R}(z)\: 
   \bar J^{(i)}(\omega k^+)\: \Phi_{R}^{(i)}(k^+\!,x_\perp,\varepsilon^*_{D^*}) 
    \Psi_M^{(i)}(z,\omega,x_\perp,\varepsilon^*_M)
   \Big] \,.
\end{eqnarray}
where the $\pm$ is for $D$ and $D^*$ and we defined the non-perturbative
functions in a way which gives the same prefactor as in Eq.~(\ref{result}). Here
$C_{L,R}^{(i)}$ are the Wilson coefficients of the weak operators in
Eq.~(\ref{QV}), and the jet functions $\overline J^{(0,8)}$ are the coefficients of
the \SCETb Lagrangian in Eq.~(\ref{L4q1}). The $\Phi_{L,R}^{(i)}$ and
$\Psi_M^{(i)}$ are soft and collinear matrix elements from the operators
$\overline O$ and are given by [with prefactor $\int_0^1 dz\,
\delta(\omega_1-z\bn\mcdot p_M)\delta(\omega_2+(1-z)\bn\mcdot p_M)$ for $\Psi_M^{(0)}$]
\begin{eqnarray} \label{PsiPhi}
  && \Big\langle M^0(p_M,\epsilon_M) \Big|  % \int\!\! dx^- 
  \Big[ (\bar\xi_n^{(d)} W)_{\omega_1}
  \bnslash P_L (W^\dagger \xi_n^{(u)} )_{\omega_2} \Big](0_\perp) % x^-\!,
  \Big[ (\bar\xi_n^{(u)} W)_{\omega}
  \bnslash P_L (W^\dagger \xi_n^{(d)} )_{\omega} \Big](x_\perp) % x^-\!, 
  \Big| 0 \Big\rangle \nn\\
  &&\qquad 
   = i f_M/\sqrt{2}\:
   \Psi_M^{(0)}(z,\omega,x_\perp,\varepsilon_M^*) 
   \nn \,,\\
 && \Big\langle D^{(*)0}(v',\epsilon_{D^*}) \Big|  % \int\!\! dx^+ 
  \Big[ (\bar h_{v'}^{(c)} S)
   \Gamma_{L,R}^h (S^\dagger h_v^{(b)} ) \Big](0_\perp) % x^+\!, ) 
   \Big[ (\bar d S)_{k^+}
  \nslash P_L (S^\dagger u )_{k^+} \Big](x_\perp) % x^+\!, 
  \Big| \bar B^0 \Big\rangle \nn\\
 && \qquad 
  = \pm \sqrt{m_B m_{D^{(*)}}}\ \Phi_{L,R}^{(0)}(k^+,x_\perp,\varepsilon_{D^*}^*)  
  \,,
\end{eqnarray}
and $\Psi_M^{(8)}$ and $\Phi_{L,R}^{(8)}$ are defined by analagous equations
with color structure $T^a\otimes T^a$. The $\pm$ is for $P_L$ and $P_R$
respectively.  In a more traditional language the $A_{\rm long}^{D^{(*)}\!M}$
contributions might be referred to as ``non-factorizable'' since they involve an
direct $x_\perp$ convolution between non-perturbative functions.
Eqs.~(\ref{result}) and (\ref{Along}) are the main results of our paper.
Additional details about the derivation of Eq.~(\ref{Along}) will be presented
in Ref.~\cite{mps2}.

Using the \SCETb power counting in $\eta=\Lambda/Q$ we can verify that the short
and long distance contributions to the factorization theorem are indeed the same
order. The coefficients $C_{L,R}^{(i)}\sim \eta^0$.  The results in
Eqs.~(\ref{L4qtree}) and (\ref{Jtree}) for the jet functions imply $J^{(i)}\sim
1/\Lambda^2$ and $\overline J^{(i)}\sim 1/(Q\Lambda)$.  Furthermore,
$\phi_M\sim \eta^0$ from the definitions in Eq.~(\ref{phipirho}).  For the soft
function in Eq.~(\ref{Sintro}) we get $(\eta^{3/2})^4$ from the fields,
$\eta^{-3}$ from the states, times $\eta^{-2}$ from the delta functions
indicated by the momentum subscripts. This gives $S(k_1^+,k_2^+)\sim\eta$, ie.
$S(k_1^+,k_2^+)\sim \Lambda$. A similar calculation for the collinear and soft
long distance matrix elements in Eq.~(\ref{PsiPhi}) gives $\Psi_{M}^{(0,8)}\sim
\Lambda^2/Q$ and $\Phi_{L,R}^{(0,8)}\sim \Lambda$.  In the factorization theorem
the measures have scaling $(dk_1^+ dk_2^+)\sim \Lambda^2$ and $(dk^+
d^2x_\perp)\sim 1/\Lambda$.  Combining all the factors for the short distance
amplitude gives $(\Lambda)(\Lambda^2)(1/\Lambda^2)(\Lambda)(\Lambda^0)=
\Lambda^2$, while for the long distance amplitude we find
$(\Lambda)(1/\Lambda)(1/\Lambda)(\Lambda)(\Lambda^2)= \Lambda^2$ also.
Therefore, both terms in $A_{00}^{D^{(*)}}$ are the same order in the power
counting. They also give the complete set of contributions at this order.

For numerical results with $M=\pi, \rho$ the $A_{\rm long}^{D^{(*)}\!M}$
contributions are very small since taking $C_{L,R}^{(i)}(z)$ independent of $z$
gives $A_{\rm long}^{D^{(*)}\!M}=0$ as shown in Appendix~\ref{app_G}. This
implies that $A_{\rm long}^{D^{(*)}\!M}/A_{00}\sim \alpha_s(Q)/\pi$, and
together with the helicity structure of the jet function discussed in
Appendix~\ref{app_J} implies that the production of transverse $\rho$ mesons is
suppressed.  In Section~\ref{sect_results} we explore further phenomenological
implications.

%%%%%%%%%%%%% Tree level results %%%%%%%%%%%%%%%%%%%%%%%%%%%%%%%%%%%%%%%%%

Next tree level results are presented for the jet functions $J^{(0,8)}$.  The
SCET$_{\rm I}$ graphs in Fig.~\ref{fig_scet1} are computed with insertions of
${\cal Q}_j^{(0,8)}$ and taking momenta $-k_1$ and $-k_2$ for the initial and
final light soft antiquarks, together with momenta $p_1$ and $p_2$ for the
collinear quark and antiquark. The diagrams in Fig.~\ref{fig_scet1}a,b with
insertions of $\{{\cal Q}_j^{(0)},{\cal Q}_j^{(8)}\}$ are
\begin{eqnarray}
  \mbox{C:}\ &&\!\! g^2\:
  \frac{\big(\bar u_{v'}^{(c)} \gamma^\nu P_L \{1,T^B\} u_v^{(b)}\big)
  \big(\bar u_n^{(d)} \gamma_\nu P_L \nslash/2  \{1,T^B\} 
   T^A \gamma_\perp^\mu v_s^{(u)}\big)
  \big(\bar v_s^{(d)} T^A \gamma_\mu^\perp v_n^{(d)}\big)}
  {[n\mcdot(k_1\!-\! k_2)\!+\!i\epsilon]
   [\bn\mcdot p_2\: n\mcdot k_1\!+\!i\epsilon]} \,,\nn\\
  \mbox{E:}\ &&\!\! -g^2\:
  \frac{\big(\bar u_{v'}^{(c)} \gamma^\nu P_L \{1,T^B\} u_v^{(b)}\big)
  \big(\bar u_n^{(u)} T^A \gamma_\perp^\mu v_s^{(u)}\big)
  \big(\bar v_s^{(d)} T^A\{1,T^B\} \gamma_\mu^\perp 
   \nslash/2 \gamma_\nu P_Lv_n^{(u)}\big)}
  {[n\mcdot(k_1\!-\! k_2)\!+\!i\epsilon]
   [-\bn\mcdot p_1\: n\mcdot k_2\!+\!i\epsilon]}\,.\quad 
\end{eqnarray}
Adding these contributions with factors of $C_L^{(0)}$ and $C_L^{(8)}$ to
distinguish the two color structures, and then Fierzing gives
\begin{eqnarray} \label{Fierz}
&&\hspace{-1cm}
   C_L^{(0)} \Big[ {\bar u}_{v'}^{(c)}  \nslash P_L  u_v^{(b)} \:  
  \bar v_s^{(d)}\nslash P_L v_{s}^{(u)} \Big] \\
& &\hspace{-0.8cm}
 \times \frac{2\pi \alpha_s C_F}{N_c}\, \left( 
   \frac{ \bar u_{n}^{(d)}  {\bnslash}{} P_L v_{n}^{(d)} }
  { [n\mcdot (k_1\!-\! k_2)+i\epsilon] [\bn\mcdot p_2 n\mcdot k_1+i\epsilon]} -
  \frac{ \bar u_{n}^{(u)} {\bnslash}{} P_L  v_{n}^{(u)} }
   {[n\mcdot (k_1\!-\! k_2)+i\epsilon] [-\bn\mcdot p_1 n\mcdot k_2+i\epsilon]}
   \right)\nn\\
&&\hspace{-1cm} -\:  C_L^{(8)} \Big[ {\bar u}_{v'}^{(c)}\nslash P_L T^a u_v^{(b)} \:  
  \bar v_s^{(d)}\nslash P_L T^a v_{s}^{(u)} \Big] \nn\\
& &\hspace{-0.8cm} \times \frac{\pi \alpha_s }{N_c^2} \left( 
 \frac{ \bar u_{n}^{(d)}  {\bnslash}{} P_L v_{n}^{(d)} }
 { [n\mcdot (k_1\!-\! k_2)+i\epsilon][\bn\mcdot p_2 n\mcdot k_1+i\epsilon]} -
 \frac{ \bar u_{n}^{(u)} {\bnslash}{} P_L  v_{n}^{(u)} }
  { [n\mcdot (k_1\!-\! k_2)+i\epsilon][ -\bn\mcdot p_1 n\mcdot k_2+i\epsilon]} 
 \right)\,.\nn
\end{eqnarray}
where $C_F=(N_c^2-1)/(2N_c)$ and we set $C_{R}^{(0,8)}=0$. The first term in
each round bracket originates from the $C-$type graph (Fig.~\ref{fig_scet1}a) and
the second term from the $E-$type graph (Fig.~\ref{fig_scet1}b).  It is convenient
to group the result into isosinglet and isotriplet terms for the collinear
spinors. Since the $\pi^0$ and $\rho^0$ have definite charge conjugation we can
freely interchange the positive momenta $\bn\cdot p_1 \leftrightarrow \bn\cdot
p_2$, so a factor of $1/\bn\cdot p_1$ can be pulled out front. For the terms in
round brackets we find
\begin{eqnarray} \label{group}
 \left( \frac12\:
 \frac{\big[\bar u_{n}^{(d)}  {\bnslash} P_L v_{n}^{(d)}-
       \bar u_{n}^{(u)}  {\bnslash} P_L v_{n}^{(u)} \big] }
 { [ n\mcdot k_1+i\epsilon][-n\mcdot k_2+i\epsilon]} -
 \frac12\: \frac{ \big[[\bar u_{n}^{(d)}  {\bnslash} P_L v_{n}^{(d)}+
         \bar u_{n}^{(u)} {\bnslash} P_L  v_{n}^{(u)} \big]\:
  n\mcdot(k_2+k_1) }
  { [n\mcdot (k_1\!-\! k_2)+i\epsilon][n\mcdot k_1+i\epsilon]
    [-n\mcdot k_2+i\epsilon]} 
 \right) \,.
\end{eqnarray}

For $\bar B^0 \to D^{(*)0} \pi^0$ and $\bar B^0 \to D^{(*)0} \rho^0$ where we
have isotriplet $M^0$'s the contributions from the \SCETb diagrams in
Figs.~\ref{fig_scet1}c,d cancel. Thus, the denominator in Eq.~(\ref{group}) 
directly gives the tree level isotriplet jet functions
\begin{eqnarray} \label{Jtree}
 J^{(0)}(z,x,k_1^+,k_2^+) &=& - \frac{4\pi\alpha_s(\mu) C_F}{N_c}\: \frac{\delta(z-x)}{x\:
  [n\mcdot k_1+i\epsilon][-n\mcdot k_2+i\epsilon]} \,,\quad \\
 J^{(8)}(z,x,k_1^+,k_2^+) &=& \frac{2\pi\alpha_s(\mu)}{N_c^2}\: \frac{\delta(z-x)}{x\:
  [n\mcdot k_1+i\epsilon][-n\mcdot k_2+i\epsilon]} \,, \nn
\end{eqnarray}
where $\bn\mcdot p_1=x\: \bn\mcdot p_M$.  These jet functions are non-singular
given that the non-perturbative soft function $S(k_1^+,k_2^+)$ vanishes for
$k_1^+=0$ or $k_2^+=0$, and that $\phi_{\pi,\rho}(x)$ vanishes at $x=0$ and
$x=1$. On the other hand for isosinglet $M^0$'s the result in Eq.~(\ref{group})
has a singular denominator $1/[n\mcdot (k_1-k_2)+i\epsilon]$. The singularity
occurs when the collinear quark propagators in Figs.~\ref{fig_scet1}a,b get too
close to their mass shells, ie. when $n\cdot (k_1-k_2)\lesssim \Lambda^2/Q$.
This singularity is exactly what is canceled by subtracting the \SCETb diagrams
in Figs.~\ref{fig_scet1}c,d, which then gives a non-singular isosinglet jet
function.

Next we consider the result for the factorization theorem for $M=\pi,\rho$
with these tree level jet functions. Taking the matrix elements of the
$O_{L}^{(0,8)}$ operators, the collinear part factors from the soft operators as
explained above. Their matrix elements are given in terms of the $M^0$ light
cone wave function, and the $S^{(0,8)}(k_+,l_+)$ functions, respectively.  This
gives the explicit result for the $\bar B^0 \to D^{(*)0} \pi^0$ and $\bar B^0
\to D^{(*)0} \rho^0$ decay amplitudes, at lowest order in the matching for $C$
and $J$
\begin{eqnarray}\label{col}
A(\bar B^0 \to D^{(*)0} \pi^0) &=& N_0^\pi \left\{
  -\frac{4\pi\alpha_s(\mu_0) C_F}{N_c}\, C_L^{(0)} \   s^{(0)}
 + \frac{2\pi\alpha_s(\mu_0)}{N_c^2}\, C_L^{(8)}
  \ s^{(8)} \right\} \langle x^{-1} \rangle_\pi \,, \nn\\
A(\bar B^0 \to D^{(*)0} \rho^0) &=&  N_0^\rho \left\{
  -\frac{4\pi\alpha_s(\mu_0) C_F}{N_c}\, C_L^{(0)} \   s^{(0)}
 + \frac{2\pi\alpha_s(\mu_0)}{N_c^2}\, C_L^{(8)}
  \ s^{(8)} \right\}  \langle x^{-1} \rangle_\rho \,.\hspace{0.5cm}
\end{eqnarray}
We choose to evaluate $C_L^{(0,8)}$, $s^{(0,8)}$, and $\langle x^{-1}\rangle$ at
the common scales $\mu=\mu_0\sim \sqrt{E_\pi\Lambda}$ since one of the hard
scales $m_c^2$ is not much different than $E_\pi\Lambda$.  In Eq.~(\ref{col})
the convolutions of the soft and collinear matrix elements are defined by
\begin{eqnarray}\label{S08}
  s^{(0,8)} &=& |s^{(0,8)}| e^{i\phi^{(0,8)}}
  = \int \mbox{d}k_1^+ \mbox{d}k_2^+\
    \frac{ S_L^{(0,8)}(k_1^+,k_2^+,\mu)}{(k_1^+ + i\epsilon)(-k_2^+ + i\epsilon)} 
   \,,\nn\\
 \langle x^{-1} \rangle_M &=& \int_0^1\!\! dx\: \frac{\phi_M(x,\mu)}{x} \,.
\end{eqnarray}

>From Eq.~(\ref{S08}) we can immediately verify the result of the power counting
for operators described earlier.  Since $\langle x^{-1} \rangle_M\sim \langle
x^0 \rangle_M \sim \lambda^0$, comparing Eqs.~(\ref{factLO},\ref{N}) and
(\ref{col}) we see that
\begin{eqnarray} \label{A00pc}
  \frac{A(\bar B^0 \to D^0\pi^0)}{A(\bar B^0\to D^+\pi^-)}  
   \sim {4\pi\alpha_s(\mu_0)\:} \frac{N_0\: s^{(0)}}{N\: E_\pi} 
   \sim {4\pi\alpha_s(\mu_0)\:} \frac{s^{(0)}}{E_\pi} 
   \sim {4\pi\alpha_s(\mu_0)\:} \frac{\Lambda_{\rm QCD}}{E_\pi} \,,
\end{eqnarray}
where we have used the standard HQET power counting for the soft matrix elements
to determine that $s^{(0,8)}\sim \Lambda_{\rm QCD}$.  Thus, the ratio of type-II
to type-I amplitudes scales as $\Lambda/Q$ just as predicted.  Due to the factor
of $4\pi$ the suppression by $\alpha_s$ does not have much effect numerically.
The $4\pi$ arises because the $\alpha_s$ is generated at tree level. It is
expected that perturbative corrections to the matching for $C$ and $J$ will be
suppressed by factors of $\alpha_s(Q)/\pi$ and
$\alpha_s(\sqrt{E_\pi\Lambda})/\pi$ respectively. In Eq.~(\ref{A00pc}) grouping
$g^2 N_c\sim 1$ gives an extra factor of $1/N_c$, so with this counting the
ratio is color suppressed as expected.

%%%%%%%%%%%%%%%%%%%%%%%%%%%%%% Kaons %%%%%%%%%%%%%%%%%%%%%%%%%%%%%%%%%%%%%%%%%%%%%%%

\section{Adding strange quarks}\label{sect_Kaon}

In this section we consider how the factorization theorem derived in
section~\ref{sect_SCET} is modified in the case of color suppressed decays
involving kaons, which include $\bar B^0\to D^{(*)}_s K^-$, $\bar B^0\to D^{(*)}_s
K^{*-}$, as well as the Cabbibo suppressed decays $\bar B^0\to D^{(*)0}K^0$ and
$\bar B^0\to D^{(*)0}K^{*0}$. 

If strange quarks are included in the final state then operators with different
flavor structure appear. In the exchange topology we can have the production of
an $s\bar s$ pair (as shown by the s-quarks in brackets in
Fig.~\ref{fig_scet1}b).  This gives \SCETb six-quark operators 
\begin{eqnarray} \label{OV2}
 O_{j}^{(0)}(k^+_i,\omega_k) &=& 
  \Big[ \bar h_{v'}^{(c)}  \Gamma^h_j\,  h_v^{(b)} \:
  (\bar d\,S)_{k^+_1} \nslash P_L\, (S^\dagger s)_{k^+_2} \Big]
  \Big[ (\bar \xi_n^{(s)} W)_{\omega_1} \Gamma_c 
  (W^\dagger \xi_n^{(u)})_{\omega_2} \Big]\,, \\
 O_{j}^{(8)}(k^+_i,\omega_k) &=& 
  \Big[ (\bar h_{v'}^{(c)} S) \Gamma^h_j\, T^a\, (S^\dagger h_v^{(b)}) \:
  (\bar d\,S)_{k^+_1} \nslash P_L T^a (S^\dagger s)_{k^+_2} \Big]
  \Big[ (\bar \xi_n^{(s)} W)_{\omega_1} \Gamma_c 
  (W^\dagger \xi_n^{(u)})_{\omega_2}  \Big] \nn\,,
\end{eqnarray}
which mediate $\bar B^0 \to D_s^{(*)} K^{(*)-}$.  For the long distance
contribution we take flavors $q'=d$ and $q=s$ in the Lagrangian in
Eq.~(\ref{L4q1}), which leads to $s$,$\bar s$ quarks replacing $u$, $\bar u$
quarks in $\overline O_{L,R}^{(i)}$.  The result for the factorization theorem
is then identical to Eqs.~(\ref{result}) and (\ref{Along}), except that only the
E-topology contributes. For this case the long distance contribution is not
suppressed, and serves to regulate the singularity when matching onto the
E-topology jet functions $J^{(0,8)} = J_E^{(0,8)}$.  Further discussion of the
singularities is left to Ref.~\cite{mps2}. The hard coefficients
$C_{L,R}^{(0,8)}$ are the same as in the previous section.

The remaining difference for $\bar B^0 \to D_s^{(*)} K^{(*)-}$ are the
non-perturbative functions.  The light-cone wavefunctions for $K^-$, $\bar
K^0$, $K^{*-}$, and $\bar K^{*0}$ are [with $q=u,d$, $\omega_1=\bn\mcdot p x_s$,
$\omega_2=-\bn\mcdot p x_q$, and a prefactor as in Eq.~(\ref{phipirho})]
\begin{eqnarray}
\langle K_n |(\bar \xi_n^{(s)} W)_{\omega_1} \bnslash\gamma_5 
  (W^\dagger \xi_n^{(q)} )_{\omega_2}| 0\rangle
  &=& -2i f_K \: \bn\mcdot p_K\: \phi_K(\mu,x_s)\: \,,\\
\langle K_n^*(\epsilon) |(\bar \xi_n^{(s)}  W)_{\omega_1} \bnslash
  (W^\dagger \xi_n^{(q)})_{\omega_2}| 0\rangle
  &=& -2i f_{K^*} \: m_{K^*}\, \bn\mcdot\epsilon^*\: \phi_{K^*}(\mu,x_s)\:
  \nn\\ 
  &=& 
  -2i f_{K^*} \: \bn\mcdot p_{K^*}\: \phi_{K^*}(\mu,x_s)\: \,. \nn
\end{eqnarray}
The collinear functions $\Psi_M^{(0,8)}$ also depend on the light meson $M$.
The non-perturbative soft functions involve strange quarks and are also
different from section~\ref{sect_SCET}, $S\to \tilde S^{(i)}_{L,R}$ and
$\Phi_{L,R}^{(0,8)}\to \tilde \Phi_{L,R}^{(0,8)}$.  The non-perturbative
functions are related to those in the previous section in the $SU(3)$ flavor
symmetry limit. However, the jet functions are not related in this limit, they
differ since different topologies contribute. This leads to different
convolutions over the non-perturbative functions.

Next consider the Cabibbo suppressed $b\to c s\bar u$ transition with the color
suppressed topology (as shown by the brackets in Fig.~\ref{fig_scet1}a). For the
six quark operators we have\footnote{Note that the flavor structure was not
  distinguished in naming the operators in Eqs.~(\ref{OV},\ref{OV2},\ref{OV3}).
  This should not cause confusion since they always contribute to different
  decays.}
\begin{eqnarray} \label{OV3}
 O_{j}^{(0)}(k^+_i,\omega_k) &=& 
  \Big[ \bar h_{v'}^{(c)}  \Gamma^h_j\,  h_v^{(b)} \:
  (\bar d\,S)_{k^+_1} \nslash P_L\, (S^\dagger u)_{k^+_2} \Big]
  \Big[ (\bar \xi_n^{(s)} W)_{\omega_1} \Gamma_c 
  (W^\dagger \xi_n^{(d)})_{\omega_2} \Big]\,, \\
 O_{j}^{(8)}(k^+_i,\omega_k) &=& 
  \Big[ (\bar h_{v'}^{(c)} S) \Gamma^h_j\, T^a\, (S^\dagger h_v^{(b)}) \:
  (\bar d\,S)_{k^+_1} \nslash P_L T^a (S^\dagger u)_{k^+_2} \Big]
  \Big[ (\bar \xi_n^{(s)} W)_{\omega_1} \Gamma_c 
  (W^\dagger \xi_n^{(d)})_{\omega_2} \Big] \nn\,,
\end{eqnarray}
which mediate the decays $\bar B^0\to D^{(*)0} \bar K^{(*)0}$. In this case the
\SCETb Lagrangian in Eq.~(\ref{L4q1}) has the same flavor structure as in
section~\ref{sect_SCET}. Since only the C-topology contributes the long distance
contribution is not suppressed in the factorization theorem, and the jet
function $J^{(0,8)} \to J_C^{(0,8)}$.  For both the short and long distance
non-perturbative functions the change of flavor appears only through the
collinear quarks in the weak operator, so the collinear functions depend on the
$K^{(*)0}$ but the soft functions $S_{L,R}^{(0,8)}$ and $\Phi_{L,R}^{(0,8)}$ are
identical to those in section~\ref{sect_SCET}. (However, now $J_C^{(0,8)}$
appears, so the moments over the soft function $S_{L,R}^{(0,8)}$ will be
different.) Finally note that if we allow a strange quark in the initial state
(for $B_s$-decays) then the E-topology can also contribute and more operators
are generated.

Due to the non-negligible long distance contributions the number of model
independent phenomenological predictions for kaons are more limited.  The main
predictions are the equality of branching fractions and strong phase shifts for
decays to $D$ versus $D^*$.  For $M=K^0, K^-$ an identical proof to the one for
$\pi^0$ and $\rho^0$ can be used.  For the vector mesons the proof can also be
used if we restrict our attention to longitudinal polarizations, so the final
states $D^{(*)}K^{*0}_\parallel$ are related, and so are $D^{(*)} K^{*-}_\parallel$.
The factorization theorem allows for transversely polarized kaons at the same
order in the power counting, but only through the long distance contribution.

%%%%   Large Nc   %%%%%%%%%%%%%%%%%%%%%%%%%%%%%%%%%%%%%%%%%%%%%%%%%%%

\section{Discussion and comparison with the large $N_c$ limit}\label{sect_lNc}

It is instructive to compare the $N_c$ scaling of the different terms in the
SCET result Eq.~(\ref{result}) (or Eq.~(\ref{col})) with that expected from QCD
before expanding in $1/Q$ given in Eq.~(\ref{fierzHw}).  Combining the matrix
elements in Eq.~(\ref{fierzHw}) written in a form similar to Eq.~(\ref{col})
gives the decay amplitude at leading order in $1/Q$ as
\begin{eqnarray}
A(\bar B^0 \to D^{0} M^0) &= & N_0^M \left(C_1 + \frac{C_2}{N_c}\right)
\left[ \frac{1}{N_c} (F_0 + 2G_1) + \cdots \right]\\
&+&  N_0^M\: C_2
\left[ F_0 + \frac{1}{N_c^2}(-F_0 + F_2 - 2G_1) + \cdots \right] \cdots \nn\,.
\end{eqnarray}
The ellipses denote power suppressed terms.  This reproduces the $1/N_c$
expansion of the SCET amplitude in Eq.~(\ref{col}) with the identification
\begin{eqnarray}\label{Nclimit}
   F_0 = 0\,,\qquad 
  \left.  G_1  = -\pi \alpha_s C_F\, s^{(0)}\right|_{N_c\to \infty} \,,\qquad 
  \left.  F_2 - 2G_1 = 2\pi \alpha_s\, s^{(8)}\right|_{N_c\to \infty} \,,
\end{eqnarray}
where $s^{(0)}\sim N_c^0$ and $s^{(8)}\sim N_c$.  This implies that the
factorizable term $F_0$ is power suppressed in the limit of an energetic pion
relative to the leading order amplitude in Eq.~(\ref{col}).

The naive factorization approach in Eq.~(\ref{naive}) keeps only the $F_0$ term,
which is expressed in terms of the $\bar B\to \pi$ form factor in the large
$N_c$ limit.  We comment here on the form of this contribution in the effective
theory.  They appear in the matching of the $(\bar db)_{V-A} (\bar cu)_{V-A}$
operator onto \SCETa $T-$products such as
\begin{eqnarray}\label{TprodF0}
  T_1^{(4)} &=& T\{ {\cal Q}^{(2)}_0\,, i{\cal L}_{q\xi}^{(2)}\}\,,\qquad
  T_2^{(4)} = T\{ {\cal Q}^{(3)}_{1a,1b}\,, i{\cal L}_{q\xi}^{(1)}\}\,,
\end{eqnarray}
where the operators ${\cal Q}^{(2,3)}$ contain one usoft light quark.  From the
leading order operators in Eq.~(\ref{QV}) they can be constructed by switching
$\xi_n \to q$ to give ${\cal Q}^{(2)}$, and adding a further $W^\dagger iD_\perp
W$ to get ${\cal Q}^{(3)}$.  Their precise form is different depending on
whether they are introduced by matching from the color-suppressed (C) or the
$W$-exchange (E) graph. Schematically
\begin{eqnarray}
\mbox{C-type: } & & {\cal Q}^{(2)}_0 = [(\bar \xi_n^{(d)} W)_\omega 
 \Gamma_c h_v^{(b)}]
 [\bar h_{v'}^{(c)} \Gamma_h u]\\
 & & {\cal Q}^{(3)}_{1a} = [(\bar \xi_n^{(d)}\frac{\bnslash}{2}
i\Dslash^\dagger_{\perp c} W)_\omega \frac{1}{\bn\mcdot {\cal P}^\dagger} 
\Gamma_c h_v^{(b)}] [\bar h_{v'}^{(c)} \Gamma_h u]\nonumber\\
 & & {\cal Q}^{(3)}_{1b} = \frac{1}{m_c}
[(\bar \xi_n^{(d)} W)_{\omega_1}\Gamma_c h_v^{(b)}]
[\bar h_{v'}^{(c)} [W^\dagger i\Dslash_{\perp c} W]_{\omega_2} \Gamma_h u]\nonumber\\
\mbox{E-type: } & & {\cal Q}^{(2)}_0 = [\bar d \Gamma_h h_v^{(b)}]
[\bar h_{v'}^{(c)} \Gamma_c (W^\dagger \xi_n^{(u)})_\omega]\\
& &{\cal Q}^{(3)}_{1a} = 
[\bar d \Gamma_h h_v^{(b)}]
[\bar h_{v'}^{(c)} \Gamma_c \frac{1}{\bn\mcdot {\cal P}} (W^\dagger 
i\Dslash_{\perp c} \frac{\bnslash}{2} \xi_n^{(u)})_\omega]\nonumber \\
& &{\cal Q}^{(3)}_{1b} = \frac{1}{m_c}
[\bar d \Gamma_h h_v^{(b)}]
[\bar h_{v'}^{(c)} [W^\dagger i\Dslash_{\perp c} W]_{\omega_1}\Gamma_c 
(W^\dagger \xi_n^{(u)})_{\omega_2}]\,.\nonumber
\end{eqnarray}

The presence of the usoft quark field $q$ in these operators introduces an
additional suppression factor of $\lambda^2$, such that the $T-$products
$T_{1,2}^{(4)}$ are $O(\lambda^4)\sim \Lambda^2/Q^2$ down relative to the
operators ${\cal Q}_{L,R}^{(0,8)}$ in Eq.~(\ref{QV}). (Note that since the form
factors enter as time ordered products we do not expect a different $\alpha_s$
suppression for $T_{1,2}^{(4)}$ relative to those in
Eq.~(\ref{Tprod})~\cite{bps4}.) This explains the absence of the $F_0$
contributions at order $\Lambda/Q$, as noted in (\ref{Nclimit}).  Although $F_0$
is part of the leading order result in the large $N_c$ limit, it is subleading
in the $1/Q$ expansion.

After soft-collinear factorization, the $T$-products (\ref{TprodF0}) match onto
factorizable operators in \SCETb. For example, the $C-$type
time-ordered product containing $Q_{1a}^{(3)}$ gives (schematically) 
\begin{eqnarray}\label{T14}
  T_2^{(4)} \to \int d\omega_1 d\omega_2\: {\cal J}(\omega_i,k_1^+)
  \Big[(\bar dS)_{k_1} \Gamma (S^\dagger h_v^{(b)})\Big][\bar h_{v'}^{(c)}
  \Gamma_h u]  [(\bar \xi_n^{(u)} W)_{\omega_1} \Gamma_c 
  (W^\dagger \xi_n^{(u)})_{\omega_2}]
\end{eqnarray}
Apart from the $(\bar cu)$ soft bilinear, this is similar to a factorizable
operator contributing to the $B\to \pi$ form factor \cite{bps4}.  The presence
of the $D$ meson in the final state implies that the matrix element of the soft
operator in Eq.~(\ref{T14}) is different from that appearing in $\bar B \to \pi$.
Therefore, naive factorization of type-II decay amplitudes, as written in
Eq.~(\ref{naive}), does not follow in general from the large energy limit.
Still, in the large $N_c$ limit, the matrix element of $T_1^{(4)}$ above can be
indeed expressed in terms of the $B\to \pi$ form factor, as required by
Eq.~(\ref{F0})

Recently an analysis of color-suppressed decays was performed using the ``pQCD''
approach working at leading order in an expansion in $m_{D^{(*)}}/m_B$ and
$\Lambda_{\rm QCD}/m_{D^{(*)}}$~\cite{Keum}.  This differs from the expansion
used here in that we do not expand in $m_{D^{(*)}}/m_B$.  The non-perturbative
functions in their proposed factorization formula include the light-cone
wavefunctions $\phi_\pi^{(p)}(x_3)$, $\phi_D(x_2)$ and a $B$ light-cone
wavefunction that depends on a transverse coordinate $\phi_B(x_1,b_1)$.  This
differs from our result which involves a $B\to D$ function $S(k_1^+,k_2^+)$ and
also has additional long distance contributions, $A_{\rm long}^{D^{(*)}M}$, at
the same order in our power counting. Our long distance contributions are
``non-factorizable'' in the sense that the non-perturbative functions
$\Phi_{L,R}^{(i)}(k^+,x_\perp)$ and $\Psi_{M}^{(i)}(z,\omega,x_\perp)$
communicate directly through their $x_\perp$ dependence without going through a
hard kernel.  In Ref.~\cite{Keum} strong phases only occur from the perturbative
$\mu_0^2\simeq E_M\Lambda$ scale, whereas we also find non-perturbative strong
phases from the $\Lambda^2$ scale (in $S(k_1^+,k_2^+)$). The non-perturbative
phases are expected to dominate in our result.  Finally, the results in
Ref.~\cite{Keum} do not manifestly predict the equality of the $D$ and $D^*$
amplitudes since at the order they are working contributions from different
$B\to M$ form factors show up.  For example their pQCD prediction $Br(\bar B^0
\to D^{*0}\rho^0)/Br(\bar B^0 \to D^{0}\rho^0)=2.7$ is much different than the
prediction of $1.0$ that we obtain in the next section using heavy quark
symmetry.

The time ordered products presented in this paper in Eq.~(\ref{Tprod}) are only
$\Lambda/Q$ down from the class-I $T$ amplitudes.  Therefore, they give the
dominant contribution to the color-suppressed and $W$-exchange amplitudes in the
limit of an energetic pion ($\Lambda/Q\ll 1$). This is a new result, not noticed
previously in the literature.  The power counting of ``factorizable'' $F_0$ type
contributions are indeed suppressed by $\Lambda^2/Q^2$ in our analysis in
agreement with the literature. However, these terms do not give the dominant
contribution. 

%%%%    Phenomenology   %%%%%%%%%%%%%%%%%%%%%%%%%%%%%%%%%%%%%%%%%%%%%%%%%%%%%%%%

\section{Phenomenological predictions}\label{sect_results}

A factorization theorem for color-suppressed $\bar B^0\to D^0 M^0$ decays was
proven in Section~\ref{sect_SCET} and extended to decays to kaons in
Section~\ref{sect_Kaon}. The amplitudes at leading order in $\Lambda_{\rm
  QCD}/Q$ with $Q=\{m_b,m_c,E_\pi\}$ have the form
\begin{eqnarray}\label{result1}
 A_{00} &=& A(\bar B^0 \to D^{(*)0}M^0) \nn\\
 &=&  N_0^M
  % \frac{G_F}{\sqrt2} V_{cb} V^*_{ud} 
   \int_0^1\!\!\!dx\, dz\!\!  \int\!\! dk_1^+ dk_2^+\,\sum_{i=0,8} 
 \left[ C^{(i)}_L(z)\: S_L^{(i)}(k_j^+) \pm C^{(i)}_R(z)\: S_R^{(i)}(k_j^+)
 \right] J^{(i)}(z,x,k_j^+)\:   \phi_M(x) \nn \\
& & + A^{D^{(*)}\!M}_{\rm long} \,,
\end{eqnarray}
where the sign $\pm$ corresponds to a $D^0$ or $D^{*0}$ meson in the final
state, respectively.  In this section the implications of Eq.~(\ref{result1}) for
the phenomenology of color suppressed decays are discussed.  One class of
predictions follow without  any assumptions about the form of $J$:
\begin{itemize}
  
\item Heavy quark symmetry relates the nonperturbative soft matrix elements
  appearing in the $\bar B^0\to D^0 M^0$ and $\bar B^0\to D^{*0} M^0$ decays
  with the same light meson at leading order in $\alpha_s(Q)/\pi$. This implies
  relations among their branching fractions and equal strong phases in their
  isospin triangles.

\end{itemize}
These relations are encoded in the ratios $R_0^M$ in Eq.~(\ref{ratios}).  A
second class of predictions depend on using a perturbative expansion of
$J$ in $\alpha_s(\mu_0)$ for $\mu_0^2 \sim E_M\Lambda$:
\begin{itemize}
  
\item Using a perturbative description of $J$ the amplitudes and strong
  phases for decays to different light mesons $M$ can be related at leading
  order in $\alpha_s(\mu_0)/\pi$.

\end{itemize}
These predictions are encoded in the ratios $R_0^{M/M'}$, $R_c$, and strong
phase $\phi$ in $R_I$, as defined in Eq.~(\ref{ratios}).  We consider the two
classes of predictions in turn.

First, consider relations between color-suppressed $\bar B\to D M$ and $\bar
B\to D^* M$ decays with the same light meson. At tree level in the matching at
the hard scale $\mu \simeq Q$, two of the Wilson coefficients vanish
$C^{(0,8)}_R=0$. Therefore both amplitudes for $D$ and $D^*$ contain only the
soft functions $S_L^{(0,8)}(k_j^+)$ appearing in the same linear combination.
This implies model-independent predictions, which can be made even in the
absence of any information about the jet functions $J^{(i)}$ and the
non-perturbative functions $S_L^{(i)}$, $\phi_M$, and without knowing $A_{\rm
  long}^{D^{(*)}\!M}$.  For $M=\pi^0, \rho^0$, we have $A_{\rm
  long}^{D^{(*)}\!M}=0$ so Eq.~(\ref{result1}) gives
\begin{eqnarray}\label{Rpi} 
  R^\pi_0 \equiv \frac{A(\bar B^0 \to D^{*0}\pi^0)}
    {A(\bar B^0 \to D^0 \pi^0)} = 1\,,\qquad
  R^\rho_0 \equiv \frac{A(\bar B^0 \to D^{*0}\rho^0)}
    {A(\bar B^0 \to D^0 \rho^0)} = 1 \,.
%\frac{n\mcdot \varepsilon^*}{n\mcdot v'} =  1\,.
\end{eqnarray}
For decays to $D_s^{(*)} K^-$, $D_s^{(*)} K^{*-}_\parallel$, $D^{(*)0} \bar
K^0$, and $D^{(*)0} \bar K_\parallel^{*0}$ it was shown that $A_{\rm
  long}^{DM}=A_{\rm long}^{D^*\!M}$ and so
\begin{eqnarray}\label{RK} 
  R^{K^-}_0 &=& \frac{A(\bar B^0 \to D^{*}_s K^-)}
    {A(\bar B^0 \to D_s K^-)} = 1\,,\qquad
  R^{K^{*-}_\parallel}_0 = \frac{A(\bar B^0 \to D^{*}_s K^{*-}_\parallel )}
    {A(\bar B^0 \to D_s K^{*-}_\parallel )} = 1 \,, \nn\\
  R^{K^0}_0 &=& \frac{A(\bar B^0 \to D^{*} \bar K^0)}
    {A(\bar B^0 \to D \bar K^0)} = 1\,,\qquad
  R^{K^{*0}_\parallel}_0 = \frac{A(\bar B^0 \to D^{*} \bar K^{*0}_\parallel )}
    {A(\bar B^0 \to D \bar K^{*0}_\parallel )} = 1 \,.
\end{eqnarray}
The ratios in Eqs.~(\ref{Rpi}) and (\ref{RK}) have calculable corrections of
order $\alpha_s(Q)/\pi$ and power corrections\footnote{Note that using the
  observed $D$ and $D^*$ masses $R_0^M=N_0^*/N_0= 1.04$. This small difference
  corresponds to keeping an incomplete set of higher order corrections.} of
order $\Lambda/Q$, which can be expected to be $\sim 20\%$.  

\begin{figure}[!t]
\vskip-0.1cm
 \centerline{
  \mbox{\epsfxsize=9truecm \hbox{\epsfbox{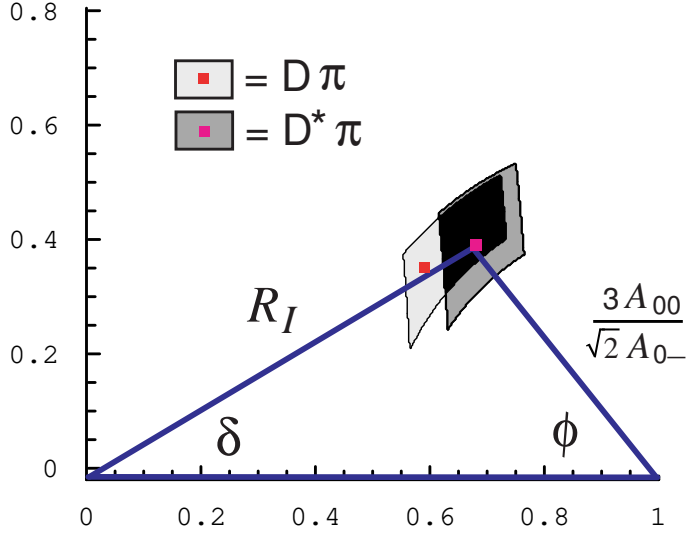}} }\hspace{1.4cm}
  %{\mbox{\epsfxsize=6truecm \hbox{\epsfbox{work/Dstrho.eps}}} }
    %  \hbox{\epsfbox{DvsDs_ai.eps}}
  } 
\vskip-0.3cm
\caption[1]{
  The ratio of isospin amplitudes $R_I = A_{1/2}/(\sqrt2 A_{3/2})$ and strong
  phases $\delta$ and $\phi$ in $\bar B\to D\pi$ and $\bar B\to D^*\pi$. The
  central values following from the $D$ and $D^*$ data in Table I are denoted by
  squares, and the shaded regions are the $1\sigma$ ranges computed from the
  branching ratios. The overlap of the $D$ and $D^*$ regions show that the two
  predictions embodied in Eq.~(\ref{Rpi}) work well. }
\label{fig_isospin} 
\vskip0cm
\end{figure}

These amplitude relations imply the equality of the branching fractions. They
also imply the equality of the non-perturbative strong phases between isospin
amplitudes, namely the phases $\delta^{D^{(*)}M}$ in the ratios $R_I^{D^{(*)}M}$
as shown in Fig.~\ref{fig_isospin}.  Thus for each of $M=\pi^0,\rho^0, K^0, 
K^{*0}_\parallel$
\begin{eqnarray} \label{p1}
  Br(\bar B^0 \to D^{*0}M^0) = Br(\bar B^0 \to D^{0}M^0)\,,\qquad\quad
  \delta^{D^{*0}M^0} = \delta^{D^{0}M^0} \,,
\end{eqnarray}
and for $M=K^-, K^{*-}_\parallel$
\begin{eqnarray} \label{p2}
  Br(\bar B^0 \to D^{*}_s M) = Br(\bar B^0 \to D_s M)\,,\qquad\quad
  \delta^{D^{*}_s M} = \delta^{D_s M} \,.
\end{eqnarray}
The predictions in Eqs.~(\ref{Rpi},\ref{p1}) agree well with the data for
$D^{(*)}\pi$ in Table~\ref{table_data}, which give
\begin{eqnarray}
 |R_0^\pi|^{\rm exp} = 0.94\pm 0.21 \,,\qquad
 \delta^{D\pi} =  30.3^{\circ\:+7.8}_{\:\ -13.8}\,, \qquad
   \delta^{D^*\pi} = 30.1^\circ \pm 6.1^\circ \,.
\end{eqnarray}
This agreement is represented graphically by the overlap of the $1$$\sigma$
regions in Fig.~\ref{fig_isospin}, with small squares indicating the central
values. The dominant contribution to the phase $\delta$ is generated by the
$(C-E)$ amplitudes which have complex phases from $J^{(i)}\, S_L^{(0,8)}$ in
Eq.~(\ref{result1}).  Since the phases in $S_L^{(0,8)}$ are non-perturbative and
can be large it is expected that they will dominate. Note that with this choice
of triangle the power suppressed side in Fig.~\ref{fig_isospin} is enlarged by a
isospin prefactor of $3/\sqrt{2}=2.1$.

For $\bar B^0$ decays to $D^{(*)0}\rho^0$, $D^{(*)0}\bar K^0$, $D^{(*)0}\bar
K^{*0}_\parallel$, $D^{(*)}_s K^-$ and $D^{(*)}_s K^{*-}_\parallel$ only upper
bounds on the branching ratios exist, so our relation between $D$ and $D^*$
triangles has not yet be tested.  For each of these channels similar triangles
to the one in Fig.~\ref{fig_isospin} can be constructed once data becomes
available.

The results in Eqs.~(\ref{Rpi}) and (\ref{RK}) can be contrasted with the
absence of a definite prediction in the large $N_c$ limit as in
Eq.~(\ref{R0lgN}).  Even when only the $F_0$ term is included (naive
factorization), $R_\pi$ is given by a ratio of $B\to \pi$ form factors, which
for generic $m_{b,c}$ are not related by heavy quark symmetry. Thus, one does
not expect a relation between the branching fractions or strong phases unless
the $1/Q$ expansion is used.

%%%%%%%%%%%%%%%%%%%%%%  jet fn dep. results  %%%%%%%%%%%%%%%%%%%%%%%%%%%%%%%%%%%%%%%%%

Next consider the second class of predictions, which follow from the
perturbative expansion of the jet function in Eq.~(\ref{result1}). We now assume
that $\alpha_s(\mu_0)$ is perturbative, and focus on $M=\pi,\rho$ since the
kaons are contaminated by contributions from $A_{\rm long}^{D^{(*)}\!M}$. The
tree level result for $J$ is given in Eq.~(\ref{Jtree}), and was used to define
the nonperturbative parameters $s^{(0,8)}$ through convolutions with the soft
distribution functions $S_L^{(0,8)}(k_i^+)$ as in Eq.~(\ref{S08}).  It is
convenient to introduce an effective moment parameter,
\begin{eqnarray} \label{seff}
  s_{\rm eff} = - s^{(0)} + \frac{1}{2 N_c C_F} \frac{C_L^{(8)}}{C_L^{(0)}} 
    s^{(8)} = |s_{\rm eff} | e^{-i\phi} \,.
\end{eqnarray}
In terms of the effective moment the result in Eq.~(\ref{result1}) at lowest
order in $\alpha_s(Q)$ and $\alpha_s(\mu_0)$ becomes
\begin{eqnarray}\label{col2eff}
 A(\bar B^0 \to D^{(*)0} M^0) &=& N_0^M\, C_L^{(0)}\:
  \frac{16\pi\alpha_s(\mu_0) }{9} \ s_{\rm eff}(\mu_0) \ 
  \langle x^{-1} \rangle_M \,,\hspace{0.5cm}
\end{eqnarray}
where $N_0^M$ is defined in Eq.~(\ref{N0}).  Since $s_{\rm eff}$ is independent 
of $M=\pi,\rho$ the same phase $\phi$ is predicted for these two light mesons. 

\begin{figure}[!t]
\vskip-0.1cm
 \centerline{ \hspace{-0.5cm}
  \mbox{\epsfxsize=8truecm \hbox{\epsfbox{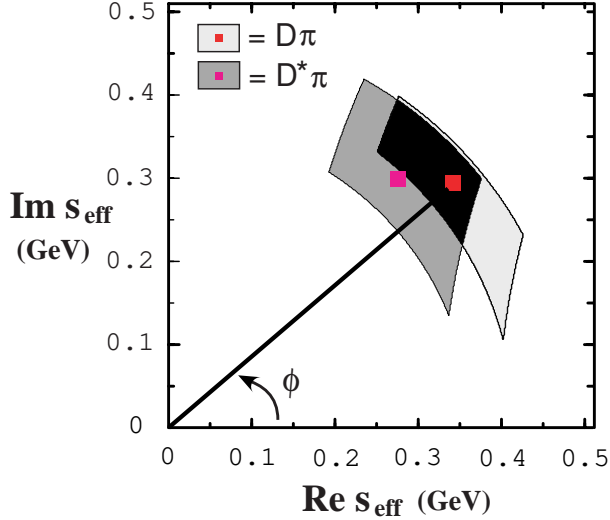}} } \hspace{1cm}
  } 
\vskip-0.3cm
\caption[1]{
  Fit to the soft parameter $s_{\rm eff}$ defined in the text, represented
  in the complex plane with the convention that $A_{0-}$ is real. The regions
  are derived by scanning the $1\sigma$ errors on the branching fractions
  (which may slightly overestimate the uncertainty). The light
  grey area gives the constraint from $\bar B \to D \pi$ and the dark grey area
  gives the constraint from $\bar B\to D^{*} \pi$.   }
\label{fig_soft} 
\end{figure}

At leading order in $1/Q$ the type-I amplitude $A_{0-} = A(B^- \to D^0 \pi^-)$
factors as in Eq.~(\ref{factLO}) giving the product of a form factor and decay
constant, both of which are real (with the usual phase conventions for the
states, and neglecting tiny $\alpha_s(m_b)$ strong phases ($\sim 2^\circ$)
generated by the coefficients $C_{L,R}^{(0)}$ at one-loop~\cite{bbns}).
Therefore the amplitude $A_{0-}$ is real at leading order in $1/Q$, up to
calculable corrections of order $\alpha_s(Q)$.  Choosing the orientation of the
triangle so that $A_{0-}$ lies on the real axis, the phase $\phi$ can be
directly extracted as one of the angles in the isospin triangle
\begin{eqnarray}\label{triangle}
  \sqrt2 A_{00} + A_{+-}  = A_{0-}  \,.
\end{eqnarray}
This is shown in Fig.~\ref{fig_soft} where we divide by $A_{0-}$ to normalize
the base.  The data on $\bar B^0\to D^0\rho^0$ is not yet sensitive enough to
test the prediction that $\phi$ is the same for $\pi^0$ and $\rho^0$.

Using Eqs.~(\ref{factLO}) and (\ref{col2eff}) it is possible to make a
prediction for the ratio $R_c$ in Eq.~(\ref{ratios}) at NLO in the power
expansion. Since $R_c=A_{+-}/A_{0-}$ contains only charged light mesons it is
easier to measure than neutral pion channels. Data is available for all four of
the $D^{(*)}\pi$ and $D^{(*)}\rho$ channels.  Using the triangle relation in
Eq.~(\ref{triangle}) one finds for the ratio of any two such modes
[$M=\pi,\rho$]
\begin{eqnarray} \label{Rc} 
  R_c^{D^{(*)}M} 
  = 1-\sqrt{2} \frac{A_{00}}{A_{0-}} = 1 - \frac{16\pi\alpha_s(\mu_0) m_{D^{(*)}}}
    {9\,E_M(m_B+m_{D^{(*)}})}\ \frac{s_{\rm eff}(\mu_0) 
     }{ \xi(w_0,\mu_0)}\ \langle x^{-1} \rangle_M \,.
  %\sqrt2\: \frac{A_{00}}{A_{0-}}\,.  
\end{eqnarray} 
It is easy to see that the ratio of amplitudes on the right-hand side is common
to final states containing a $D$ or $D^*$, and has only a mild dependence on the
light meson, introduced through the inverse moment $\langle x^{-1}\rangle_M$.
In particular we note that there is no dependence on the decay constant $f_M$ on
the RHS of Eq.~(\ref{Rc}), since it cancels in the ratio $A_{00}/A_{0-}$. This
implies that the ratios $R_c$ are comparable for all four channels $D^{(*)}\pi$
and $D^{(*)}\rho$, up to corrections introduced by $\langle
x^{-1}\rangle_\pi\neq \langle x^{-1}\rangle_\rho$. These corrections can be
smaller than the correction one might expect from the ratio of decay
constants $f_\rho/f_\pi\simeq 1.6$ (which appear in the naive $a_2$
factorization).  The experimental values of these ratios can be extracted from
Table I and are in good agreement with a quasi-universal prediction
\begin{eqnarray} \label{Rcdata}
 |R_c^{(D\pi)}| &=& \frac{|A(\bar B^0\to D^+ \pi^-)|}
    {|A(\bar B^- \to D^0\pi^-)|} =  0.77 \pm 0.05 \,, \\
 |R_c^{(D^*\pi)}| &=& \frac{|A(\bar B^0\to D^{*+} \pi^-)|}
    {|A(\bar B^- \to D^{*0}\pi^-)|} =  0.81 \pm 0.05 \,, \nn \\
 |R_c^{(D\rho)}| &=& \frac{|A(\bar B^0\to D^+ \rho^-)|}
    {|A(\bar B^- \to D^0\rho^-)|} = 0.80 \pm 0.09 \,, \nn\\
 |R_c^{(D^*_L\rho_L)}| &=& \frac{|A(\bar B^0\to D^{*+} \rho^-)|}
    {|A(\bar B^- \to D^{*0}\rho^-)|} = 0.86 \pm 0.10\,. \nn
\end{eqnarray}
This lends support to our prediction for the universality of the strong phase
$\phi$ in $\bar B\to D^{(*)}\pi$ and $\bar B\to D^{(*)}\rho$ decays from the
$s_{\rm eff}$ in Eq.~(\ref{Rc}).  The central values of $R_c\simeq 0.8$ are well
described by an $s_{\rm eff}$ of the expected size ($\sim\Lambda_{\rm QCD}$) as
discussed in the fit to the isospin triangle below. Further data on these
channels may expose other interesting questions, such as whether $R_c^{(D^{*}M)}$
is closer to $R_c^{(DM)}$ than $R_c^{(D^{(*)}\pi)}$ is to $R_c^{(D^{(*)}\rho)}$.

An alternative use of Eq.~(\ref{Rc}) and the $R_c$ amplitude ratios is to give
us a method for extracting the ratio of $\rho$ and $\pi$ moments. Using the $D
\pi$ and $D\rho$ measurements which have smaller errors than for $D^*$, we find
\begin{eqnarray} \label{xmpp}
  \frac{ \langle x^{-1} \rangle_\rho }{  \langle x^{-1} \rangle_\pi }
  = \frac{|R_c^{(D\rho)}|-1 }{ |R_c^{(D\pi)}|-1 } 
  = 0.87 \pm 0.42 \,.
\end{eqnarray}
where only the experimental uncertainty is shown.  The extraction in
Eq.~(\ref{xmpp}) is smaller, but still in agreement with the ratio extracted
from light-cone QCD sum rules. The best fit from the $\gamma^*\gamma\to \pi^0$
data performed in Ref.~\cite{bakulev} gives $\langle x^{-1}\rangle_{\pi}=3.2\pm
0.4$ in agreement with sum rule estimates of the moment. The QCD sum-rule result
$\langle x^{-1}\rangle_{\rho} = 3.48 \pm 0.27$~\cite{BaBr}, then implies
\begin{eqnarray} \label{SRrhopi}
 \frac{ \langle x^{-1}\rangle_{\rho} }{ \langle x^{-1}\rangle_{\pi} }
 \bigg|_{\rm SR} = 1.10 \pm  0.16 \,.
\end{eqnarray}
The result that this ratio is close to unity is consistent with the universality
of the data in Eq.~(\ref{Rcdata}).  This data can be contrasted with cases where
the single light meson is replaced by a multibody state such as~\cite{PDG}
\begin{eqnarray} 
 && \frac{Br(\bar B^0\to D^{*+} \pi^- \pi^-
  \pi^+\pi^0)}{Br(\bar B^- \to D^{*0}\pi^+ \pi^- \pi^-\pi^0)} 
 = 1.02\pm 0.27 \,,
\end{eqnarray}
For the four pion final state our proof of the factorization theorem does not
work, since for many events one or more of the pions will be slow.  We therefore
would expect less universality in branching ratios involving more than one light
meson. (For these decays a different type of factorization involving large $N_c$
works well for the $q^2$ spectrum~\cite{Ligeti}.)

The result in Eq.~(\ref{col2eff}) also leads to predictions for the ratios of
color-suppressed decay amplitudes to final states containing different light
mesons $M^0 = \pi^0, \rho^0$. We find
\begin{eqnarray} \label{Rrhopi}
  R_0^{\rho/\pi} \equiv \frac{|A(\bar B^0 \to D^{0} \rho^0)|}
   {|A(\bar B^0 \to D^0 \pi^0)|}
   = \frac{f_\rho}{f_\pi} \ \frac{\langle x^{-1}\rangle_{\rho}}
   {\langle x^{-1}\rangle_{\pi}} = 1.40 \pm 0.77\,,
\end{eqnarray}
where we used $f_{\pi^\pm} = 130.7 \pm 0.4$ MeV, and $f_{\rho^\pm} = 210\pm 10$
MeV, and inserted the result in Eq.~(\ref{xmpp}) for the moments. This can be
compared with the experimental result $(R_{\rho/\pi})^{\rm exp}= 1.02\pm 0.21$.
The large uncertainty in the ratio of moments in Eq.~(\ref{xmpp}) dominates the
error in Eq.~(\ref{Rrhopi}). With the QCD sum rule result in Eq.~(\ref{SRrhopi})
we find $R_{\rho/\pi} = 1.64\pm 0.35$, a result whose central value is farther
from the experimental data, but still consistent with it.

In contrast to the first class of predictions, the predictions for the ratios in
Eqs.~(\ref{Rc}), (\ref{xmpp}), and (\ref{Rrhopi}) and the prediction for the
universality of $\phi$ can receive corrections from neglected
$[\alpha_s(\mu_0)^2/\pi]$ terms in $J$.  The dominant theoretical corrections to
this extraction are expected to come again from these perturbative corrections
to $J$ or from power corrections, which we estimate may be at the $\sim 30\%$
level. A future study of the perturbative corrections is possible within the
framework of our factorization theorem and SCET.

The result in Eq.~(\ref{col2eff}) and the data on $B\to D\pi$ and $B\to D^*\pi$
decays can be used to extract values of the moment parameters $|s_{\rm eff}|$
and strong phase $\phi$.  We present in Fig.~\ref{fig_soft} the constraints on
the parameter $s_{\rm eff}$ in the complex plane, obtained from $D\pi$ (light
shaded region) and $D^*\pi$ data (darker shaded area). We used in this
determination $\mu_0=E_\pi=2.31\,{\rm GeV}$, and leading order running which
gives $\alpha_s(\mu_0) = 0.25$, $C_1(\mu=\mu_0)=1.15$, and
$C_2(\mu=\mu_0)=-0.32$. The good agreement between the $D\pi$ and $D^*\pi$
$1\sigma$ regions marks a quantitative success of our factorization relation in
Eq.~(\ref{result1}).  Averaging over the $D\pi$ and $D^*\pi$ results, we find
the following values for the soft parameters at $\mu=\mu_0$
\begin{eqnarray} \label{snum}
  |s_{\rm eff}| &=& \Big( 428 \pm 48 \pm 100 \mbox{ MeV}\Big)\ 
   \bigg( \frac{0.26}{C_L(\mu_0)\alpha_s(\mu_0)}\bigg)\ 
   \bigg( \frac{3.2}{\langle x^{-1} \rangle_{\pi} } \bigg) \,,\nn\\
  \phi &=& 44.0^\circ \pm 6.5 ^\circ\,.
\end{eqnarray}
In this determination the inverse moment of the pion wave function was taken
from the best fit to the $\gamma^*\gamma\to \pi^0$ data~\cite{bakulev}, $\langle
x^{-1}\rangle_\pi = 3.2\pm 0.4$. For $|s_{\rm eff}|$ the first error is
experimental, while the second is our estimate of the theoretical uncertainty in
the extraction from varying $\mu_0$ from $E_\pi/2$ to $2E_\pi$. At the order we
are working the extraction of the phase $\phi$ is independent of the scale,
since the prefactor $\alpha_s(\mu_0) \langle x^{-1}\rangle_\pi$ drops out. The
result in Eq.~(\ref{snum}) agrees well with the dimensional analysis estimates
$s_{\rm eff}\sim s^{(0,8)}\sim \Lambda_{\rm QCD}$. Since $\phi$ is
non-perturbative its value is unconstrained, and a large value of this phase is
allowed.

The recent $\bar B^0\to D^0\rho^0$ data from Belle allows us to extract $|s_{\rm
  eff}|$ and $\phi$ in a manner independent of the above determination.  Keeping
only experimental errors we find
\begin{eqnarray} \label{srho}
  |s_{\rm eff}| &=& \Big( 259 \pm 124 \mbox{ MeV}\Big) 
   \bigg( \frac{0.26}{C_L(\mu_0)\alpha_s(\mu_0)}\bigg)\ 
   \bigg( \frac{3.5}{\langle x^{-1} \rangle_{\rho} } \bigg)\,,\nn\\
    \phi &=& 17^\circ \pm 70^\circ\,.
\end{eqnarray}
The results agree with Eq.~(\ref{snum}) within $1\sigma$, but currently have
errors that are too large to significantly test the factorization prediction of
equality on the $20$-$30\%$ level of the parameters extracted from $D\rho$ and
$D\pi$.

%%%%%%%%%%%%%%%%%%%%%%  B -> D K, B -> Ds K-  %%%%%%%%%%%%%%%%%%%%%%%%%%%%%%%%

The $\bar B^0 \to D_s^{(*)+} K^-$ channels proceed exclusively through the
$W$-exchange graph and have been the object of recent theoretical work
\cite{DsKtheory}.  For the result analogous to Eq.~(\ref{col2eff}) we would
have [$M=K, K^*$]
\begin{eqnarray}\label{col2effDsK}
 A(\bar B^0 \to D^{(*)}_s M) &=& \sqrt{2} N_0^M\, C_L^{(0)}\:
  \frac{16 \pi\alpha_s(\mu_0) }{9} \ s^E_{\rm eff}(\mu_0) \ 
  \langle x_s^{-1} \rangle_M + A_{\rm long}^{D^{(*)}_s\!M} \,.\hspace{0.5cm}
\end{eqnarray}
Both the $\bar B^0 \to D_s^{(*)+} K^-$ modes and the Cabibbo-suppressed decays
$\bar B \rightarrow D^{(*)} \bar K^{(*)}$ receive this additional contribution
from $A_{\rm long}^{D^{(*)}\!M}$. This makes the factorization theorem less
predictive, and so we do not attempt an analysis of ratios $R_c^{D^{(*)}K^{(*)}}$,
$R_0^{M/M'}$, or the universal phases $\phi_E$ and $\phi_C$ that are analogous to
the $\phi$ in Eq.~(\ref{seff}).

On the experimental side both the Babar and Belle Collaborations \cite{DsK} recently
observed the $\bar B^0 \to D_s^+ K^-$ decay, and set an upper limit on the
branching ratio of $\bar B^0 \to D_s^{*+} K^-$
\begin{eqnarray}\label{DsKdata}
 B(\bar B^0 \to D_s^+ K^-) &=& [3.2 \pm 1.0\mbox{ (stat) } \pm 1.0 \mbox{ (sys) }]
 \times 10^{-5}\qquad \mbox{ (Babar)}\\
&=& [4.6^{+1.2}_{-1.1}\mbox{ (stat) } \pm 1.3 \mbox{ (sys) }]\times 10^{-5}\qquad
 \quad\mbox{ (Belle)}\nn\\
 B(\bar B^0 \to D_s^{*+} K^-) &\leq & 2.5 \times 10^{-5}\, (90\% \mbox{CL} )
 \hspace{3.9cm} \mbox{ (Babar)}\,.\nn
\end{eqnarray}
The branching fraction for $\bar B^0 \to D_s^+ K^-$ is an order of magnitude
smaller than that for $\bar B^0 \to D^0\pi^0$. This indicates that the
$W$-exchange amplitude $E^{D_s K^-}$ is suppressed relative to $(C-E)^{D\pi}$
and $(V_{ud}/\sqrt{2}V_{us})\: C^{D^0\bar K^0}$.  In SCET the SU(3) breaking
between $\phi_\pi(x)$ and $\phi_K(x)$ is generated by masses in the collinear
quark Lagrangian~\cite{Leib}. This causes an asymmetry in the light-cone kaon
wavefunction.  This SU(3) violation can be expected to be at most a canonical
$\sim 20$-$30\%$ effect, which would not account for the observed suppression.

However, there is one important source of potentially larger SU(3) breaking from
an enhancement in moments of the light-cone kaon wavefunction which appear in
the short distance amplitude. This may account for the observed suppression.
Basically strange quark mass effects imply a larger SU(3) violation for inverse
moments than expected for $\phi_\pi$ versus $\phi_K$ alone, and implies that
$\langle x_s^{-1} \rangle_K < \langle x_d^{-1} \rangle_K$.  Using the result
from QCD sum rules the ratio of moments~\cite{BaBr} is ${\langle x_d^{-1}
  \rangle_K}/{\langle x_s^{-1} \rangle_K} \sim 1.4$.  Furthermore, we anticipate
a similar large effect from the moments that appear in the soft matrix elements
which again differ by factors of $(k^+_d)^{-1}$ versus $(k^+_s)^{-1}$, and
appear in a way that suppresses $D_sK^-$.  The combination of these two
suppression factors might accommodate the observed factor of three suppression
in the $D_s K^-$ amplitudes.\footnote{In general this argument gives a dynamic
  explanation for the suppression of $s\bar s$-popping at large energies which
  could be tested elsewhere. The production of an $s\bar s$ pair which end up in
  different strange hadrons is likely to be accompanied by a suppression from
  inverse momentum fractions that arise from the gluon propagator that produced
  these quarks.  This enhances the SU(3) violation in a well defined direction
  so that less $s\bar s$ pairs are produced. A factor of $3$ suppression of
  $s\bar s$ popping is implemented in JETSET~\cite{jetset}. }  The long distance
amplitude also involves two inverse momentum fractions through $\overline
J^{(0,8)}$ in Eq.~(\ref{L4qtree}), although admittedly much less is known about
the non-perturbative functions $\Psi_{M}^{(0,8)}$ and $\Phi_{L,R}^{(0,8)}$.
Thus, we find that the suppression of $E^{D_sK^-}$ may not imply much about the
relative size of $C^{D\pi}$ and $E^{D\pi}$.  Finally, we note that the
suppression mechanism for $s\bar s$ creation that we have identified is
particular to problems involving large energies where light-cone wavefunctions
arise.

Further information on the relative size of the short and long distance
contributions to the kaon factorization theorem is clearly desirable. In
section~\ref{sect_Kaon} it was noted that in type-II decays transverse $K^*$'s
are produced only by the long distance contribution at this order in
$\Lambda_{\rm QCD}/Q$. Therefore, measuring the polarization of the $K^*$ in
both the $\bar B^0\to D_s^{*} K^{*-}$ and $\bar B^0 \to D^{*0} K^{*0}$ decays
can give us a direct handle on whether there might be additional dynamical
suppression of either the long or short distance contributions, or whether they
are similar in size as one might expect apriori from the power counting.

%%%%%%%%%%%%%%%%%%%%%%%%%%%%%%%%%%%%%%%%%%%%%%%%%%%%%%%%%%%%%%%%%%%%%%%%%%%%%%%

\section{Summary and Conclusions}\label{sect_concl}

We presented in this paper the first model-independent analysis of
color-suppressed $\bar B^0 \to D^{0(*)} M^0$ decays, in the limit of an
energetic light meson $M^0$. The soft-collinear effective theory (SCET) was used
to prove a factorization theorem for these decay amplitudes at leading order in
$\Lambda_{\rm QCD}/Q$, where $Q=\{m_b,m_c,E_M\}$.  Compared with decays into a
charged pion these decays are suppressed by a factor $\Lambda_{\rm QCD}/Q$.
Therefore, in the effective theory they are produced exclusively by subleading
operators.\footnote{In type-I decays, other subleading operators can compete
  with the time ordered products we have identified at the same order in
  $\Lambda/Q$. This makes a complete analysis of power corrections to type-I
  decays more complicated than our analysis of type-II decays.}

We have identified the complete set of subleading operators which contribute to
$\bar B^0 \to D^{0(*)}M^0$ decays with $M=\pi,\rho,K,K^*$, as well as for the
decays $\bar B^0 \to D_s^{(*)}K^{(*)-}$.  After hard-soft-collinear
factorization, their matrix elements are given by i) a short distance
contribution involving a jet function convoluted with nonperturbative soft
distribution functions, and the non-perturbative light-cone meson wave function,
and ii) a long distance contribution involving another jet function and
additional $x_\perp$ dependent nonperturbative functions for the soft $B, D$ and
collinear $M$. The long distance contributions were shown to vanish for
$M=\pi,\rho$ at lowest order in $\alpha_s(Q)/\pi$.

The factorization formula is given in Eqs.~(\ref{result}) and (\ref{Along}). It
may seem surprising that the type-II decays factor into a pion light-cone wave
function and a $B\rightarrow D^{(*)}$ soft distribution function rather than
being like the naive $a_2$ factorization in Eq.~(\ref{naive}). Our results
indicate that factorization for type-II decays is similar to factorization for
type-I decays (albeit with new non-perturbative soft functions and additional
long distance contributions for kaons).  To derive Eq.~(\ref{result}), QCD was
first matched onto \SCETa at the scale $\mu ^2=Q^2$.  In \SCETa it is still
possible for gluons to redistribute the quarks. This intermediate theory
provides a mechanism for connecting the soft spectator quark in the $B$ to a
quark in the pion, and for connecting the energetic quark produced by the
four-quark operator with the soft spectator in the $D$ (see
Fig.~\ref{fig_scet1}).  This process is achieved by the power suppressed time
ordered products given in Eq.~(\ref{Tprod}).  \SCETa is then matched onto \SCETb
at a scale $\mu_0^2=E_M\Lambda$.  In \SCETb the collinear quarks and gluons are
non-perturbative and bind together to make the light meson $M$. This second
stage of matching introduces a new coefficient function (jet functions) as in
Eq.~(\ref{jets}). The jet function $J$ contains the information about the SCET I
graphs that move the spectator quarks into the pion. The physics at various
scales is neatly encoded in Eq.~(\ref{result}). The Wilson coefficient $C(z)$
from matching QCD onto SCET I depends on physics at the scale $Q^2 $, the jet
functions $J,\overline J$ from matching SCET I onto SCET II depends on $Q
\Lambda$ physics which is where quark redistribution occurs, and finally the
soft distribution functions $S,\Phi$ and the pion light cone wavefunction
$\phi_M,\Psi_M$ depend on non-perturbative physics at $\Lambda^2$ which is where
the binding of hadrons occur.

The soft functions $S$ are complex, and encode information about strong rescattering
phases. This information is introduced through Wilson lines along the light
meson direction of motion, which exchange soft gluons with the final state meson
$D^{(*)}$. They provide a new mechanism which generates non-perturbative strong
phases.  In the literature other mechanisms which generate perturbative strong
phases have been proposed. In particular in Ref.~\cite{bss,bbns} a method for
identifying perturbative strong phases with an expansion in $\alpha_s(Q^2)$ was
developed. In Ref.~\cite{LiSanda,Keum} it was pointed out that strong phases can
also be generated perturbatively at the intermediate scale $\alpha_s(E_M\Lambda)$.
In the language of our factorization theorem in Eq.~(\ref{result}) these phases
roughly correspond to imaginary parts in the hard coefficients $C_{L,R}^{(0,8)}$
and jet functions $J$ respectively. These phases exist, but for the $B\to D\pi$
channels they only show up at next-to-leading order in the $\alpha_s(m_b)$ or
$\alpha_s(\mu_0)$ expansion. (In type-I $B\to D^{(*)}\pi$ decays the hard strong
phase is very small, $\sim 2^\circ$~\cite{bbns}). In contrast, our new source of
strong phases is entirely non-perturbative in origin and can produce
unconstrained phases. For the case of $B\to D^{(*)}M$ these phases show up in
the power suppressed class-II amplitudes.

The factorization theorem proven in this paper leads to predictions which were
tested against existing experimental data on color-suppressed decays.  We
derived two model independent relations, which related
\begin{itemize}
 \item the $\bar B^0\to D^0 M^0$ and $\bar B^0\to
D^{*0} M^0$ decay branching fractions and
\item the $\bar B\to D M$ and $\bar B\to D^{*} M$ strong phases.
\end{itemize}
Here $M=\pi,\rho,K,K^*_\parallel$, and these relations are true to all orders in
the strong coupling at the collinear scale. The same predictions are also
obtained for $\bar B^0 \to D_s^{(*)} K^{-}$ and $\bar B^0 \to D_s^{(*)}
K^{*-}_\parallel$.  The good numerical agreement observed between the strong
phases and branching fractions in the $D\pi$ and $D^*\pi$ channels gives strong
backing to our results.  This prediction can be tested further since the
equality of the strong phases for the $\rho$, $K$, and $K^*_\parallel$ channels
have not yet been tested experimentally.

Additional predictions followed from the factorization theorem by using a
perturbative expansion for the jet function, including $[M=\pi,\rho]$
\begin{itemize}
 \item the ratios $|R_c|=|A(\bar B^0\to D^{(*)+} M^-)/ A( B^-\to D^{(*)0}
M^-)|$ to subleading order
\item the ratios $|R_0^{\rho/\pi}|=|A(\bar B^0\to D^{(*)0} \rho^0)/ A( \bar B^0\to
  D^{(*)0} \pi^{0})|$ to subleading order
\item universal parameters $\{|s_{\rm eff}|, \phi\}$ which appear for both
  $D^{(*)}\pi$ and $D^{(*)}\rho$, and
\item a mechanism for enhanced $SU(3)$ violation in $s\bar s$ production for the
  short distance amplitude which might explain the suppression of the $\bar B^0
  \to D_s^{(*)} K^-$ rates relative to $\bar B^0 \to D^0\pi^0$.
\end{itemize}
For $|R_c|$ taking different values of $M$ with the same isospin the power
corrections only differ by the moments $\langle x^{-1} \rangle_M$, giving an
explanation for the observed quasi-universality of these ratios.  The isospin
triangles for these $M$'s are predicted to involve a universal angle $\phi$.
The ratio of neutral modes $|R_0^{\rho/\pi}|$ are determined by inverse moments
of the light-cone wavefunctions and decay constants.  Finally extractions of the
non-perturbative soft moment parameter $s_{\rm eff}$ agrees with the $\sim
\Lambda_{\rm QCD}$ size estimated by dimensional analysis.

In the case of $\bar B^0\to D_s^{(*)} K^{(*)-}$ an additional suppression
mechanism was identified, which arises from enhanced SU(3) violation due to the
asymmetry of non-perturbative distributions involving strange versus down
quarks. The inverse moments that appear in the factorization theorem enhances
this difference, and can lead to a dynamic suppression of $s\bar
s$-popping. Further information on the size of the short and long distance
amplitudes would help in clarifying this observation.

A more detailed experimental study of the channels in Tables~I and II is crucial
to further test the accuracy of the factorization theorem and improve our
understanding of the structure of power corrections. Work on extending these
results to decays to isosinglet mesons is in progress.  It should be evident
that $B_s$ decays could also be considered although we have not done so here.

\acknowledgments We would like to thank Z.~Ligeti and I.~Rothstein for
discussion and comments on the manuscript.  This work was supported in part by
the U.S.  Department of Energy (DOE) under the cooperative research agreement
DF-FC02-94ER40818, by the U.S.  NSF Grant PHY-9970781, and by the Research
Corporation through grant No.CS0362.

%%%%     Appendix A     %%%%%%%%%%%%%%%%%%%%%%%%%%%%%%%%%%%%%%%%%%%%%%%%%%%%%%

\appendix

\section{Long Distance contributions for $\pi$ and $\rho$}
\label{app_G}

The factorization theorem derived in Sec.~\ref{sect_SCET} for the
color-suppressed $B^0\to D^0 M^0$ amplitude contains both short- and
long-distance contributions.  In this Appendix we show that, working at lowest
order in the Wilson coefficients at the hard scale $Q$, the long-distance
amplitude vanishes for the case of an isotriplet light mesons $M = \pi, \rho$.

We start by recalling the factorized form of the long-distance amplitude, which
is given by \SCETb time ordered products $\bar T_{L,R}^{(0,8)}$
\begin{eqnarray} \label{Along2}
  A_{\rm long}^{D^{(*)}\!M} &=& 
  % \frac{G_F}{\sqrt2} V_{cb} V^*_{ud} 
   \int_0^1\!\!\! dz\!\!  \int\!\! dk^+ d\omega \!\! \int\!\! d^2\!x_\perp \Big[
   C^{(i)}_{L}(z)\: 
   \bar J^{(i)}(\omega k^+)\: \Phi_{L}^{(i)}(k^+\!,x_\perp,\varepsilon^*_{D^*}) 
    \Psi_M^{(i)}(z,\omega,x_\perp,\varepsilon^*_M) \nn\\
 && \qquad \pm   C^{(i)}_{R}(z)\: 
   \bar J^{(i)}(\omega k^+)\: \Phi_{R}^{(i)}(k^+\!,x_\perp,\varepsilon^*_{D^*}) 
    \Psi_M^{(i)}(z,\omega,x_\perp,\varepsilon^*_M)
   \Big] \,.
\end{eqnarray}
The functions $\Psi_M^{(i)}$ and $\Phi_{L,R}^{(i)}$ are \SCETb matrix elements of
collinear and soft fields, respectively, and their precise definitions are given
in Eqs.~(\ref{PsiPhi}). The jet functions $\bar J^{(i)}(\omega k^+)$ appear in the
definition of the subleading soft-collinear Lagrangian ${\cal L}^{(1)}_{\xi\xi qq}$
and their lowest order expressions are given in Eq.~(\ref{L4qtree}).

In the following we derive a few general properties of the functions
$\Psi_M^{(i)}$ and $\overline J^{(0,8)}$ following from isospin, charge
conjugation, parity and time-reversal.  The collinear function
$\Psi_{M}^{(i)}(z,\omega,x_\perp,\varepsilon_M^*)$ is defined as the matrix element
\begin{eqnarray} \label{A2}
  \Big\langle M^0(\varepsilon) \Big|  \Big[ (\bar\xi_n^{(d)} W)_{\tau_1}
  \bnslash P_L (W^\dagger \xi_n^{(u)} )_{\tau_2} \Big](0_\perp) %\\
 % && \qquad\quad \times 
  \Big[ (\bar\xi_n^{(u)} W)_{\omega}
  \bnslash P_L (W^\dagger \xi_n^{(d)} )_{\omega} \Big](x_\perp) 
  \Big| 0 \Big\rangle \,.
\end{eqnarray}
We will prove that $\Psi_{M=\pi,\rho}$ is even under $\omega\to -\omega$ and
$z\to 1-z$. As motivation consider the first bilinear in Eq.~(\ref{A2}), which
creates a $d\bar u$ collinear quark pair. The second bilinear in Eq.~(\ref{A2})
must act at some point along the collinear quark lines: it either takes a $d\to
u$ (for $\omega >0$) or takes a $\bar u \to \bar d$ (for $\omega <0$).
Examination of lowest order graphs contributing to $\Psi_M$ shows that these two
types of contributions always appear in pairs, such that the projection of
$\Psi_M$ onto an isotriplet state is even under $\omega\to -\omega$. This
suggests the existence of a symmetry argument, valid to all orders in
perturbation theory.

We will prove that $\Psi_{M}^{(0,8)}$ is even, as a consequence of G-parity.
This is defined as usual by $G=C \exp(-i\pi I_2)$ where $C$ is charge
conjugation and $I_2$ is the isospin generator, and is a symmetry of the
collinear Lagrangian in the limit $m_{u,d}\ll \Lambda_{\rm QCD}$.  Its action on
the collinear operators in Eq.~(\ref{A2}) can be worked out from that of its
components $C$ and $I_2$ (cf. Ref.~\cite{bfprs}) and is given by
\begin{eqnarray}
G\: (\bar\xi_n^{(d)} W)_{\tau_1}
  \bnslash P_L (W^\dagger \xi_n^{(u)} )_{\tau_2} \: G^\dagger &=& 
(\bar\xi_n^{(d)} W)_{-\tau_2}
  \bnslash P_R (W^\dagger \xi_n^{(u)} )_{-\tau_1} \,, \\
G\: (\bar\xi_n^{(u)} W)_{\omega}
  \bnslash P_L (W^\dagger \xi_n^{(d)} )_{\omega} \:  G^\dagger &=& 
(\bar\xi_n^{(u)} W)_{-\omega}
  \bnslash P_R (W^\dagger \xi_n^{(d)} )_{-\omega}\,.\nonumber
\end{eqnarray}
Taking into account the G-parity of the states, Eq.~(\ref{A2}) is equal to
\begin{eqnarray} \label{A3}
 \pm \Big\langle M^0(\varepsilon) \Big|   \Big[ (\bar\xi_n^{(d)} W)_{-\tau_2}
  \bnslash P_R (W^\dagger \xi_n^{(u)} )_{-\tau_1} \Big](0_\perp) %\\
 % && \qquad\quad \times 
  \Big[ (\bar\xi_n^{(u)} W)_{-\omega}
  \bnslash P_R (W^\dagger \xi_n^{(d)} )_{-\omega} \Big](x_\perp) 
  \Big| 0 \Big\rangle \,, 
\end{eqnarray}
where the $\pm$ refer to the $\rho^0$ and $\pi^0$ respectively. Next we apply
parity in the matrix element followed by switching our basis vectors
$n\leftrightarrow \bn$. Acting on Eq.~(\ref{A3}) this gives
\begin{eqnarray} \label{A4}
  \Big\langle M^0(\varepsilon^*_P) \Big|   \Big[ (\bar\xi_n^{(d)} W)_{-\tau_2}
  \bnslash P_L (W^\dagger \xi_n^{(u)} )_{-\tau_1} \Big](0_\perp) %\\
 % && \qquad\quad \times 
  \Big[ (\bar\xi_n^{(u)} W)_{-\omega}
  \bnslash P_L (W^\dagger \xi_n^{(d)} )_{-\omega} \Big](-x_\perp) 
  \Big| 0 \Big\rangle \,, 
\end{eqnarray}
where the overall sign is now the same for $M=\rho,\pi$. Now since
$\Psi_{M}^{(0,8)}$ is a scalar function the only allowed perpendicular dot
products are $(-x_\perp)^2=x_\perp^2$ and $-x_\perp\mcdot \varepsilon_P^*= x_\perp
\mcdot \varepsilon^*$. Finally we note that the change in $\tau_{1,2}$ from
Eqs.~(\ref{A2}) to (\ref{A4}) is equivalent $z\to 1-z$.  Thus the invariance of
\SCETb under G-parity and regular parity has allowed us to prove that
\begin{eqnarray}
  \Psi_{\pi,\rho}^{(i)}(z,\omega,x_\perp,\varepsilon^*) = 
     \Psi_{\pi,\rho}^{(i)}(1-z,-\omega,x_\perp,\varepsilon^*) \,.
\end{eqnarray}

Next we prove that $\overline J^{(0,8)}(\omega k^+)$ is odd under $\omega\to
-\omega$.  By reparameterization invariance type-III~\cite{mmps} only the
product $\omega k^+$ will appear. Consider applying time reversal plus the
interchange $(n\leftrightarrow \bn)$ to the \SCETb Lagrangian. Since this
Lagrangian does not have coefficients that encode decays to highly virtual
offshell states it should be invariant under this transformation. Acting on
Eq.~(\ref{L4q1}) this implies that $\overline J^{(0,8)}$ must be real,
\begin{eqnarray}
  [ \overline J^{(0,8)}(\omega k^+) ]^* = \overline J^{(0,8)}(\omega k^+).
\end{eqnarray}
At tree level this implies that we should drop the $i\epsilon$ in the collinear
gluon propagator in matching onto this operator. This was done in arriving at
the odd functions $J^{(0,8)}\propto 1/(\omega k^+)$ in Eq.~(\ref{L4qtree}). The
imaginary part would give a $\delta(\omega k^+)$ and corresponds to cases where the
\SCETa $T$-product is reproduced by a purely collinear \SCETb $T$-product $(k^+=0)$,
or a purely soft \SCETb $T$-product ($\omega=0$). Thus dropping the $i\epsilon$
also saves us from double counting.

Now consider what functions can be generated by computing loop corrections to
$\overline J^{(0,8)}$. By dimensional analysis $\overline J^{(0,8)}$ must be
proportional to $1/(\omega k^+)$ times a dimensionless function of $\omega
k^+/\mu^2$.  Since at any order in perturbation theory the matching calculation
will involve only massless quarks we can only generate logarithms.  Therefore,
we must study functions of the form
\begin{eqnarray}
  \frac{1}{\omega k^+}  \ln^n\Big( \frac{\omega k^+ \pm i\epsilon}{\mu^2} \Big)\,.
\end{eqnarray} 
To demand that only the real part of these functions match onto $\overline
J^{(0,8)}$ we average them with their conjugates. It is straightforward to check
that only terms odd in $\omega\to -\omega$ survive.  Thus, all the terms that
can correct the form of $\overline J^{(0,8)}$ at higher orders in $\alpha_s$ are
odd under $\omega\to -\omega$.

Now in Eq.~(\ref{Along2}) the integration over $\omega$ is from $-\infty$ to
$\infty$, while $z$ varies from $0$ to $1$.  Consider the change of variable
$\omega\to -\omega$ and $z\to 1-z$. If $C_{L,R}^{(i)}(z)=C_{L,R}^{(i)}(1-z)$
then under this interchange one of the functions in the integrand is odd
($\overline J$) and the other two are even ($C_{L,R}^{(i)}$ and
$\Psi_{\pi,\rho}^{(i)}$), so the integral would vanish.

Now if $C_{L,R}^{(i)}(z)$ are kept only to leading order then they are
independent of $z$ and thus unchanged under $z\to 1-z$. So at this order in the
$\alpha_s(Q)/\pi$ expansion of $C_{L,R}^{(i)}(z)$ we find $A_{\rm
  long}^{D^{(*)}\!M}=0$.  This completes the proof of the assertion about the
vanishing of the long distance contributions for $M=\pi,\rho$.

%%%%%%%%     Appendix B     %%%%%%%%%%%%%%%%%%%%%%%%%%%%%%%%%%%%%%%%%%%%%%%%%%%%

\section{Helicity Symmetry and Jet functions}
\label{app_J}

In this appendix we discuss the general structure of the jet functions
$J^{(0,8)}(z,x,k_j^+)$ in Eq.~(\ref{result}), which are generated by matching
\SCETa and \SCETb at any order in $\alpha_s(\mu_0)$. In Fig.~\ref{fig_scet1}a,b
this means adding additional collinear gluons which generate loops by attaching
to the collinear lines already present (as well as vacuum polarization type
collinear quark, gluon, and ghost loops). Additional collinear loops should also
be added to Figs.~\ref{fig_scet1}c,d,e, and the difference at lowest order in
$\lambda$ gives $J$. Throughout this appendix we continue to drop isosinglet
combinations of $\bar\xi_n \cdots \xi_n$. These will also have additional
contributions from topologies where the outgoing collinear quarks are replaced
by outgoing gluons (through $B_{\perp}^\mu$ operators).

The leading order collinear Lagrangian has a U(1) helicity spin symmetry for the
quarks, see reference 2 in~\cite{SCET}. It is defined by a generator $h_n$ 
that has the quark spin projection along the $n$ direction, which is different
from usual definition of helicity as the projection of the spin along its
momentum.  Unlike QCD, the collinear fields in SCET only allow quarks and
antiquarks that move in the $n$ direction.  For $h_n$ we have
\begin{eqnarray}
   h_n = \frac{1}{4}\ \epsilon_{\perp}^{\mu\nu}  \sigma_{\mu\nu} \,,\qquad
   h_n^2 =1 \,,\qquad
   [h_n,\nslash]=[h_n,\bnslash] = 0\,,\qquad
   \{ h_n, \gamma_\perp^\sigma\} = 0 \,.
\end{eqnarray}
After making a field redefinition~\cite{bps2} to decouple ultrasoft gluons the
leading order collinear quark Lagrangian is
\begin{eqnarray}\label{Lc0}
{\cal L}_{\xi\xi}^{(0)} = \bar \xi_{n,p'} \left\{in\cdot D_c \!+\!
 i\DSppP \frac{1}{i\bn\!\cdot D_c} i\DSppP \right\} 
 \frac{\bnslash}{2}\: \xi_{n,p} ,
\end{eqnarray}
where $iD_c^{\mu}$ contains only collinear $A_{n,q}^\mu$ gluons.  ${\cal
  L}_{\xi\xi}^{(0)}$ is invariant under the transformation $\xi_n \to
\exp(i\theta h_n)\xi_n$, $\bar \xi_n \to \bar \xi_n\exp(-i\theta h_n)$.  This
means that any number of leading order collinear quark interactions preserve the
quark helicity $h_n$. The collinear gluon interactions take $u_n(\uparrow)\to
u_n(\uparrow)$, ${u_n(\downarrow)}$$\to u_n(\downarrow)$, $v_n(\uparrow)\to
v_n(\uparrow)$, $v_n(\downarrow)\to v_n(\downarrow)$, and can also produce or
annihilate the quark-antiquark combinations $u_n(\uparrow)\, v_n(\downarrow)$ or
$u_n(\downarrow)\, v_n(\uparrow)$ (the arrows refer to the helicity of the
antiparticles themselves rather than their spinors). For this reason we refer to
${\cal L}_{\xi\xi}^{(0)}$ as a $\Delta h_n=0$ operator.

The leading order \SCETa operators in Eq.~(\ref{QVI}) are also unchanged by the
$h_n$-transformation and therefore does not change collinear quark helicity.  In
contrast the operators ${\cal L}_{\xi q}^{(1)}$ do generate or annihilate a
collinear quark giving $\Delta h_n=\pm 1/2$. However, at tree level we showed in
Section~\ref{sect_SCET} that the two graphs in Figs.~\ref{fig_scet1}a,b match
onto an overall $\Delta h_n=0$ operator in \SCETb as given in Eqs.~(\ref{OV}).
Since at higher orders the ${\cal L}_{\xi\xi}^{(0)}$ will not cause a change in
the helicity they also match onto these same operators, so the structure
$\bnslash \gamma_\perp^\nu$ will not occur. At tree level only the structure
$\nslash P_L\otimes \bnslash P_L$ appeared in Eq.~(\ref{OV}). To rule out the
appearance of $P_R$ beyond tree level we note that the weak operator projects
onto left handed collinear fermions, and for the jet function the conservation
of helicity in ${\cal L}_{\xi\xi}^{(0)}$ implies a conservation of chirality.
This leaves us with the desired result.

It is perhaps illustrative to see this more explicitly by looking at the spin
structure of the loop graphs. We begin by noting that the spin and color
structure in $\bar h_{v'}^{(c)}\cdots h_v^{(b)}$ is unaffected by this second
stage of matching.  Adding additional collinear attachments only can affect the
spin and color structure generated in putting the collinear quark fields and
light ultrasoft quark fields together.

Consider how additional gluon attachments effect the spin structures that appear
in Figs.~\ref{fig_scet1}a,b.  The leading order collinear quark Lagrangian is
${\cal L}_{\xi\xi}^{(0)}$ in Eq.~(\ref{Lc0}). Each attachment of a collinear
gluon to a collinear quark lines in the figures generates a $\bnslash/2$ from
the vertex and a $\nslash/2$ from the quark propagator. These combine to a
projector which can be eliminated by commuting them to the right or left to act
on the collinear quark spinors, via $(\nslash\bnslash)/4\ \xi_n = \xi_n$.
Therefore, at most we have additional pairs of $\gamma_\perp$'s that appear
between the light quark spinors. The aim is to show that just like the tree
level calculation in Eq.~(\ref{Jtree}) the resulting operators have spin
structure $(\bar d \nslash P_L u)\:(\bar\xi_n \bnslash P_L \xi_n)$.

For the contraction of $T_{j}^{(0,8)}$ which gives the $C$ topology the spin
structure is
\begin{eqnarray} \label{string1}
 && \Big[ \bar u_n^{(d)} \gamma_\perp^{\mu_1} \gamma_\perp^{\mu_2} \cdots
  \gamma_\perp^{\mu_{2k-1}} \gamma_\perp^{\mu_{2k}} \Big( \frac{\bnslash}{2} P_L
  \Big) \gamma_\perp^{\nu_1} \cdots \gamma_\perp^{\nu_{2\ell}} 
  \Big(\frac{\nslash}{2} \gamma_\perp^\alpha\Big)
  u_s^{(u)}  \Big] \  %\nn\\
 %&& \times 
 \Big[ \bar u_s^{(d)} \Big(\gamma_\perp^\beta\Big) 
  \gamma_\perp^{\lambda_1} \cdots \gamma_\perp^{\lambda_{2j}} u_n^{(d)} \Big]
  \nn\\
 &&=  \big[ \bar u_n^{(d)} \gamma_\perp^{\mu_1}  \cdots
  \gamma_\perp^{\mu_{2k'}}  \gamma_\perp^\alpha P_L u_s^{(u)} \big] \  
  \big[ \bar u_s^{(d)}\gamma_\perp^\beta
  \gamma_\perp^{\lambda_1} \cdots \gamma_\perp^{\lambda_{2j}} u_n^{(d)} 
  \big] \,.
\end{eqnarray}
In the first line $(\frac{\bnslash}{2} P_L)$ comes from ${\cal
  Q}^{(0,8)}_{L,R}$, the $\gamma_\perp^\alpha$ and $\gamma_\perp^\beta$ are
terms generated by the ${\cal L}_{\xi q}^{(1)}$ insertions and the $\nslash/2$
is from the extra collinear quark propagator. In the second line the $P_L$
projector was moved next to $u_s^{(u)}$ without a change of sign (for
anticommuting $\gamma_5$), and the remaining $\bnslash$ and $\nslash$ were then
moved next to the $\bar u_n^{(d)}$ and canceled.  The remaining free $\perp$
indices in the second line are contracted with each other in some manner.
Fierzing the set of $\gamma$ matrices in Eq.~(\ref{string1}) by inserting
$1\otimes 1$ next to the collinear spinors gives
\begin{eqnarray} \label{string2}
  && \big[ \bar u_n^{(d)}\: \Gamma_1\: u_n^{(d)}  \big] \  
  \big[ \bar u_s^{(d)}\gamma_\perp^\beta
  \gamma_\perp^{\lambda_1} \cdots \gamma_\perp^{\lambda_{2j}}\: \Gamma_1'\:
  \gamma_\perp^{\mu_1}  \cdots \gamma_\perp^{\mu_{2k'}}  
  \gamma_\perp^\alpha P_L u_s^{(u)} \big]\,,
\end{eqnarray}
where
\begin{eqnarray} \label{G1G1}
  \Gamma_1\otimes \Gamma_1' &=& 
          \frac{\bnslash}{2}\otimes \nslash 
        - \frac{\bnslash\gamma_5}{2} \otimes \nslash\gamma_5 
        - \frac{\bnslash\gamma_\perp^\nu}{2} \otimes \nslash\gamma^\perp_\nu
        \nn\\
 &\to &  \frac{\bnslash}{2}(1-\gamma_5) \otimes \nslash 
        - \frac{\bnslash\gamma_\perp^\nu}{2} \otimes \nslash\gamma^\perp_\nu \,.
\end{eqnarray}
In the second line of Eq.~(\ref{G1G1}) we have used the fact that the $\gamma_5$
in the bracket with soft quark spinors can be eliminated by moving it next to
the $P_L$. To eliminate the $\bnslash\gamma_\perp^\nu$ Dirac structure we note
that between the soft spinors in Eq.~(\ref{string2}) there are an odd number of
$\gamma_\perp$'s to the left and right of $\nslash\gamma_\nu^\perp$, and so at
least one set of indices are contracted between the sets
$\{\beta,\lambda_1,\ldots, \lambda_{2j}\}$ and
$\{\mu_1,\ldots,\mu_{2k'},\alpha\}$. The identity $\{\gamma_\perp^\sigma,
\gamma_\perp^\tau\}=2g_\perp^{\sigma\tau}$ can be used to move these matrices so
that they sandwich $\gamma_\nu^\perp$, and this gives the product
$\gamma_\perp^\mu \gamma_\nu^\perp \gamma^\perp_\mu =0$.  After these
manipulations only the spin structure $(\bar d \nslash P_L
u)\:(\bar\xi_n \bnslash P_L \xi_n)$ remains. A similar argument can be applied
to the E-topology with the same result.

In several places in the above argument we made use of Dirac algebra that is
particular to $4$-dimensions (anticommuting $\gamma_5$ and setting
$\gamma_\perp^\mu \gamma_\nu^\perp \gamma^\perp_\mu =0$). If the $\gamma_\perp$'s
are taken in in full dimensional regulation then it is not apriori clear if the
manipulations survive regulation. However, the original helicity symmetry 
argument shows that as long as the theory can be regulated in a way that
preserves this symmetry this will indeed be the case. 

\OMIT{ Since for the jet functions
and \SCETa diagrams we can work order by order in $\alpha_s(\mu_0)$ it is not
necessary to consider non-perturbative effects which might break the chiral
symmetry. 
}

%%%%%     Appendix C    %%%%%%%%%%%%%%%%%%%%%%%%%%%%%%%%%%%%%%%%%%%%%%%%%%%%%%%%%%

\section{Properties of Soft Distribution Functions}
\label{app_S}

In this appendix we derive some useful properties of the soft functions
$S^{(0,8)}$. In particular we show that these functions are complex. The
imaginary parts have a direct interpretation as non-perturbative contributions
to final state rescattering between the $D^{(*)}$ and final energetic meson as
discussed in section~\ref{sect_SCET}.

To be definite we consider the function $S_L^{(0)}$, and suppress the index $L$.
The manipulations for the remaining soft functions $S_R^{(0)}$ and
$S_{L,R}^{(8)}$ are identical.  The definition in Eq.~(\ref{Sintro}) is
\begin{eqnarray} \label{Sstart}
 {\langle D^{0}(v') | (\bar h_{v'}^{(c)}S) \nslash P_L 
  (S^\dagger h_v^{(b)}) (\bar d S)_{k^+_1}\nslash P_L (S^\dagger u)_{k^+_2} 
  | \bar B^0(v)\rangle}
 &=& S^{(0)} \,,
\end{eqnarray}
where the Wilson lines are defined as
\begin{eqnarray} \label{WS}
  W &=&  \Big[ \sum_{\rm perms} \exp\Big( -\frac{g}{\bnP}\: 
  \bn\mcdot  A_{n,q}(x) \ \Big) \Big] \,,\quad %\nn\\
  S = \Big[ \sum_{\rm perms}
  %\lower7pt \hbox{ $\stackrel{\mbox{\small$\sum$}}{\mbox{\small perms}}$ }
  %\!\! 
  \exp\Big(-\!g\,\frac{1}{n\mcdot\cP}\ n\mcdot A_{s,q} \Big) \Big] \,.
\end{eqnarray}
In general $S^{(0)}$ is a dimensionless function of $v\mcdot v'$, $n\mcdot v$,
$n\mcdot v'$, $n\mcdot k_1$, $n\mcdot k_2$, $\Lambda_{\rm QCD}$, and $\mu$.  Since
$(S^\dagger q)_{k_2^+}=\delta(k_2^+-n\mcdot \cP) (S^\dagger q)$ the LHS is
invariant under a type-III reparameterization transformation~\cite{mmps}
($n\to e^\alpha n$, $\bn \to e^{-\alpha}\bn$). Therefore the RHS can only
be a function of $w$, $t=n\mcdot v/n\mcdot v'$, $z=n\mcdot k_1/n\mcdot k_2$,
$K/\mu =[n\mcdot k_1\: n\mcdot k_2/(n\mcdot v\: n\mcdot v'\mu^2)]^{1/2}$, and
$\Lambda_{\rm QCD}/\mu$.

Rather than study the matrix element in Eq.~(\ref{Sstart}) directly it is useful
to instead consider
\begin{eqnarray} \label{newS}
 && {\langle H_i(v') | (\bar h_{v'} S) \nslash P_{L} 
  (S^\dagger h_v) (\bar q S)_{k^+_1}\nslash P_L \tau^a (S^\dagger q)_{k^+_2} 
  |  H_j(v)\rangle} \nn\\
 &&\qquad =   S^{(0)}\Big(t,z, v\mcdot v', \frac{K}{\mu},
  \frac{\Lambda_{\rm QCD}}{\mu} \Big) (\tau^a)_{ij}  \,,
\end{eqnarray}
where $h_v$ are doublet fields under heavy quark flavor symmetry, and $q$ and $|
H_{i=1,2}(v)\rangle $ are isospin doublets of $(u,d)$.  The last three
variables in Eq.~(\ref{newS}) will not play a crucial role so we will suppress
this dependence.  Taking the complex conjugate of Eq.~(\ref{newS}) gives
\begin{eqnarray} \label{S1}
 {\langle H_j(v) | (\bar h_{v} S) \nslash P_{L} 
  (S^\dagger h_{v'}) (\bar q S)_{k^+_2}\nslash P_L \tau^a (S^\dagger q)_{k^+_1} 
  |  H_i(v')\rangle}
 &=&   [S^{(0)}(t,z)]^*  (\tau^a)_{ji}\nn\\
 &=& S^{(0)}\Big(\frac{1}{t},\frac{1}{z}\Big) (\tau^a)_{ji}\,.
\end{eqnarray}
The dependence on $w$ and $K$ is unchanged since they are even under the
interchange $v\leftrightarrow v'$, $n\mcdot k_1\leftrightarrow n\mcdot k_2$.
Next, decompose the functions $S^{(0)}$ in terms of even and odd functions
under $t\to 1/t$, $z\to 1/z$:
\begin{eqnarray}
  S^{(0)} &=&  S_E^{(0)} +  S_O^{(0)} \,,
\end{eqnarray}
where $S_{E,O}^{(0)}=[S^{(0)}(t,z)\pm S^{(0)}(1/t,1/z)]/2$. Now
Eq.~(\ref{S1}) implies that
\begin{eqnarray}
 [ S^{(0)}_E(t,z) ]^* = S^{(0)}_E(t,z) \,,\qquad
 [ S^{(0)}_O(t,z) ]^* = - S^{(0)}_O(t,z)
\end{eqnarray}
so $S_E^{(0)}$ is real and $S_O^{(0)}$ is imaginary.  An identical argument for
$S^{(8)}$ implies that it too is a complex function.

For the above analysis it is important to note that $n\cdot v' = m_B/m_D$ is not
$1$ in the heavy quark limit where we have new spin and flavor symmetries. These
symmetries arise from taking $m_B \gg \Lambda_{\rm QCD}$ and $m_D \gg
\Lambda_{\rm QCD}$, not from having $m_B=m_D$.


\begin{thebibliography}{99}

\bibitem{BSW}
M.~Bauer, B.~Stech and M.~Wirbel,
%``Exclusive Nonleptonic Decays Of D, D(S), And B Mesons,''
Z.\ Phys.\ C {\bf 34}, 103 (1987).
%%CITATION = ZEPYA,C34,103;%%

\bibitem{DG}
M.~J.~Dugan and B.~Grinstein,
%``QCD Basis For Factorization In Decays Of Heavy Mesons,''
Phys.\ Lett.\ B {\bf 255}, 583 (1991).
%%CITATION = PHLTA,B255,583;%%

\bibitem{phen}
M.~Neubert and B.~Stech,
%``Non-leptonic weak decays of B mesons,''
Adv.\ Ser.\ Direct.\ High Energy Phys.\  {\bf 15}, 294 (1998)
[arXiv:hep-ph/9705292];
%%CITATION = HEP-PH 9705292;%%
A.~Ali, G.~Kramer and C.~D.~Lu,
%``Experimental tests of factorization in charmless non-leptonic two-body  B decays,''
Phys.\ Rev.\ D {\bf 58}, 094009 (1998)
[arXiv:hep-ph/9804363];
%%CITATION = HEP-PH 9804363;%%
H.~Y.~Cheng and B.~Tseng,
%``Nonfactorizable effects in spectator and penguin amplitudes of hadronic  
%charmless B decays,''
Phys.\ Rev.\ D {\bf 58}, 094005 (1998)
[arXiv:hep-ph/9803457].
%%CITATION = HEP-PH 9803457;%%

\bibitem{BlSh}
B.~Blok and M.~A.~Shifman,
%``Nonfactorizable amplitudes in weak nonleptonic decays of heavy mesons,''
Nucl.\ Phys.\ B {\bf 389}, 534 (1993)
[arXiv:hep-ph/9205221];
%%CITATION = HEP-PH 9205221;%%
I.~E.~Halperin,
%``Soft gluon suppression of 1/N(c) contributions in color suppressed heavy
% meson decays,''
Phys.\ Lett.\ B {\bf 349}, 548 (1995)
[arXiv:hep-ph/9411422].
%%CITATION = HEP-PH 9411422;%%

\bibitem{BuSi}
A.~J.~Buras and L.~Silvestrini,
%``Non-leptonic two-body B decays beyond factorization,''
Nucl.\ Phys.\ B {\bf 569}, 3 (2000)
[arXiv:hep-ph/9812392].
%%CITATION = HEP-PH 9812392;%%

\bibitem{PW}
H.~D.~Politzer and M.~B.~Wise,
%``Perturbative Corrections To Factorization In Anti-B Decay,''
Phys.\ Lett.\ B {\bf 257}, 399 (1991).
%%CITATION = PHLTA,B257,399;%%

\bibitem{bbns}
M.~Beneke, G.~Buchalla, M.~Neubert and C.~T.~Sachrajda,
Nucl.\ Phys.\ B {\bf 591}, 313 (2000)
[arXiv:hep-ph/0006124].
%%CITATION = HEP-PH 0006124;%%


\bibitem{Ligeti}
Z.~Ligeti, M.~E.~Luke and M.~B.~Wise,
%``Comment on studying the corrections to factorization in B $\to$ D(*) X,''
Phys.\ Lett.\ B {\bf 507}, 142 (2001)
[arXiv:hep-ph/0103020].
%%CITATION = HEP-PH 0103020;%%

\bibitem{bps}
C.~W.~Bauer, D.~Pirjol and I.~W.~Stewart,
%``A proof of factorization for B $\to$ D pi,''
Phys.\ Rev.\ Lett.\  {\bf 87}, 201806 (2001)
[arXiv:hep-ph/0107002].
%%CITATION = HEP-PH 0107002;%%

\bibitem{xing0}
Z.~z.~Xing,
%``Determining the factorization parameter and strong phase differences in  B $\to$ D(*) pi decays,''
arXiv:hep-ph/0107257.
%%CITATION = HEP-PH 0107257;%%

\bibitem{NePe}
M.~Neubert and A.~A.~Petrov,
%``Comments on color suppressed hadronic B decays,''
Phys.\ Lett.\ B {\bf 519}, 50 (2001)
[arXiv:hep-ph/0108103].
%%CITATION = HEP-PH 0108103;%%

\bibitem{Ben}
C.~W.~Bauer, B.~Grinstein, D.~Pirjol and I.~W.~Stewart,
%``Testing factorization in B $\to$ D(*) X decays,''
Phys.\ Rev.\ D {\bf 67}, 014010 (2003)
[arXiv:hep-ph/0208034].
%%CITATION = HEP-PH 0208034;%%

\bibitem{Rosner}
C.~W.~Chiang and J.~L.~Rosner,
%``Final-state phases in B $\to$ D pi, D* pi, and D rho decays,''
Phys.\ Rev.\ D {\bf 67}, 074013 (2003)
[arXiv:hep-ph/0212274].
%%CITATION = HEP-PH 0212274;%%

\bibitem{xing}
Z.~z.~Xing,
%``Final-state rescattering and SU(3) symmetry breaking in B $\to$ D K and  B $\to$ D K* decays,''
Eur.\ Phys.\ J.\ C {\bf 28}, 63 (2003)
[arXiv:hep-ph/0301024].
%%CITATION = HEP-PH 0301024;%%

\bibitem{Li}
H.~n.~Li,
%``QCD aspects of exclusive B meson decays,''
arXiv:hep-ph/0303116.
%%CITATION = HEP-PH 0303116;%%

\bibitem{Buras} 
G.~Buchalla, A.~J.~Buras and M.~E.~Lautenbacher,
%``Weak Decays Beyond Leading Logarithms,''
Rev.\ Mod.\ Phys.\  {\bf 68}, 1125 (1996)
[arXiv:hep-ph/9512380].
%%CITATION = HEP-PH 9512380;%%

%%%%%%%%%%%%%%%%%% start Expt papers %%%%%%%%%%%%%%%%%%%%%%%%%%%%%

\bibitem{PDG}
K.~Hagiwara {\it et al.}  [Particle Data Group Collaboration],
%``Review Of Particle Physics,''
Phys.\ Rev.\ D {\bf 66}, 010001 (2002).
%%CITATION = PHRVA,D66,010001;%%

\bibitem{CLEO} S.~Ahmed {\it et al.}  [CLEO Collaboration],
%``Measurement of B(B- $\to$ D0 pi-) and B(anti-B0 $\to$ D+ pi-) and isospin  
%analysis of B $\to$ D pi decays,''
Phys.\ Rev.\ D {\bf 66}, 031101 (2002)
[arXiv:hep-ex/0206030].
%%CITATION = HEP-EX 0206030;%%


\bibitem{CLEOdata}
T.~E.~Coan {\it et al.}  [CLEO Collaboration],
%``Observation of anti-B0 $\to$ D0 pi0 and anti-B0 $\to$ D*0 pi0,''
Phys.\ Rev.\ Lett.\  {\bf 88}, 062001 (2002)
[arXiv:hep-ex/0110055].
%%CITATION = HEP-EX 0110055;%%

\bibitem{BELLEdata}
K.~Abe {\it et al.}  [BELLE Collaboration],
%``Observation of color-suppressed anti-B0 $\to$ D0 pi0, D*0 pi0, D0 eta  and D0 omega decays,''
Phys.\ Rev.\ Lett.\  {\bf 88}, 052002 (2002)
[arXiv:hep-ex/0109021].
%%CITATION = HEP-EX 0109021;%%

\bibitem{CLEOhelicity} 
S.~E.~Csorna {\it et al.}  [CLEO Collaboration],
%``Measurements of the branching fractions and helicity amplitudes in  B $\to$ D* rho decays,''
arXiv:hep-ex/0301028.
%%CITATION = HEP-EX 0301028;%%

\bibitem{Belle}
A.~Satpathy {\it et al.}  [Belle Collaboration],
%``Study of anti-B0 $\to$ D(*)0 pi+ pi- decays,''
Phys.\ Lett.\ B {\bf 553}, 159 (2003)
[arXiv:hep-ex/0211022].
%%CITATION = HEP-EX 0211022;%%

%%%%%%%%%%%%%%%%%%

\bibitem{color}
G.~Bertsch, S.~J.~Brodsky, A.~S.~Goldhaber and J.~F.~Gunion,
%``Diffractive Excitation In QCD,''
Phys.\ Rev.\ Lett.\  {\bf 47}, 297 (1981).
%%CITATION = PRLTA,47,297;%%


\bibitem{BL}
G.~P.~Lepage and S.~J.~Brodsky,
%``Exclusive Processes In Perturbative Quantum Chromodynamics,''
Phys.\ Rev.\ D {\bf 22}, 2157 (1980);
%%CITATION = PHRVA,D22,2157;%%
%\bibitem{Efremov:1978rn}
A.~V.~Efremov and A.~V.~Radyushkin,
%``Asymptotical Behavior Of Pion Electromagnetic Form-Factor In QCD,''
Theor.\ Math.\ Phys.\  {\bf 42}, 97 (1980)
[Teor.\ Mat.\ Fiz.\  {\bf 42}, 147 (1980)].
%%CITATION = TMPHA,42,97;%%

\bibitem{SCET}
C.~W.~Bauer, S.~Fleming and M.~Luke,
%``Summing Sudakov logarithms in B $\to$ X/s gamma in effective field theory,''
Phys.\ Rev.\ D {\bf 63}, 014006 (2001)
[hep-ph/0005275].
%%CITATION = HEP-PH 0005275;%%
%\bibitem{bfps}
C.~W.~Bauer, S.~Fleming, D.~Pirjol and I.~W.~Stewart,
%``An effective field theory for collinear and soft gluons: Heavy to light
% decays,''
Phys.\ Rev.\ D {\bf 63}, 114020 (2001).
[arXiv:hep-ph/0011336].
%%CITATION = HEP-PH 0011336;%%
C.~W.~Bauer and I.~W.~Stewart,
%``Invariant operators in collinear effective theory,''
Phys.\ Lett.\ B {\bf 516}, 134 (2001)
[arXiv:hep-ph/0107001].
%%CITATION = HEP-PH 0107001;%%

\bibitem{bps2}
C.~W.~Bauer, D.~Pirjol and I.~W.~Stewart,
%``Soft-collinear factorization in effective field theory,''
Phys.\ Rev.\ D {\bf 65}, 054022 (2002)
[arXiv:hep-ph/0109045].
%%CITATION = HEP-PH 0109045;%%

\bibitem{D0K0} 
P.~Krokovny {\it et al.}  [Belle Collaboration],
%``Observation of anti-B0 $\to$ D0 anti-K0 and anti-B0 $\to$ D0 anti-K*0  decays,''
Phys.\ Rev.\ Lett.\  {\bf 90}, 141802 (2003)
[arXiv:hep-ex/0212066].
%%CITATION = HEP-EX 0212066;%%

\bibitem{bfprs}
C.~W.~Bauer {\it et al.},
%``Hard scattering factorization from effective field theory,''
Phys.\ Rev.\ D {\bf 66}, 014017 (2002)
[arXiv:hep-ph/0202088].
%CITATION = HEP-PH 0202088;%%

\bibitem{bps4}
C.~W.~Bauer, D.~Pirjol and I.~W.~Stewart,
%``Factorization and endpoint singularities in heavy-to-light decays,''
arXiv:hep-ph/0211069;
%%CITATION = HEP-PH 0211069;%%
D.~Pirjol and I.~W.~Stewart,
%``A complete basis for power suppressed collinear-ultrasoft operators,''
arXiv:hep-ph/0211251.
%%CITATION = HEP-PH 0211251;%%

\bibitem{bcdf}
M.~Beneke {\it et al.}, 
%``Soft-collinear effective theory and heavy-to-light currents beyond  leading power,''
Nucl.\ Phys.\ B {\bf 643}, 431 (2002)
[arXiv:hep-ph/0206152].
%%CITATION = HEP-PH 0206152;%%

\bibitem{HN}
R.~J.~Hill and M.~Neubert,
%``Spectator interactions in soft-collinear effective theory,''
arXiv:hep-ph/0211018.

\bibitem{bpsgi}
%%CITATION = HEP-PH 0211018;%%
C.~W.~Bauer, D.~Pirjol and I.~W.~Stewart,
%``On power suppressed operators and gauge invariance in SCET,''
arXiv:hep-ph/0303156.
%%CITATION = HEP-PH 0303156;%%

\bibitem{LMR}
M.~E.~Luke, A.~V.~Manohar and I.~Z.~Rothstein,
%``Renormalization group scaling in nonrelativistic QCD,''
Phys.\ Rev.\ D {\bf 61}, 074025 (2000)
[arXiv:hep-ph/9910209].
%%CITATION = HEP-PH 9910209;%%

\bibitem{KR}
G.~P.~Korchemsky and A.~V.~Radyushkin,
%``Infrared factorization, Wilson lines and the heavy quark limit,''
Phys.\ Lett.\ B {\bf 279}, 359 (1992) [hep-ph/9203222].
%%CITATION = HEP-PH 9203222;%%

\bibitem{ssa}
S.~J.~Brodsky, D.~S.~Hwang and I.~Schmidt,
%``Final-state interactions and single-spin asymmetries in semi-inclusive  deep inelastic scattering,''
Phys.\ Lett.\ B {\bf 530}, 99 (2002)
[arXiv:hep-ph/0201296];
%%CITATION = HEP-PH 0201296;%%,
J.~C.~Collins,
%``Leading-twist single-transverse-spin asymmetries: Drell-Yan and  deep-inelastic scattering,''
Phys.\ Lett.\ B {\bf 536}, 43 (2002)
[arXiv:hep-ph/0204004].
%%CITATION = HEP-PH 0204004;%%

\bibitem{hbook}
A.~V.~Manohar and M.~B.~Wise,
%``Heavy Quark Physics,''
Cambridge Monogr.\ Part.\ Phys.\ Nucl.\ Phys.\ Cosmol.\  {\bf 10}, 1 (2000).
%%CITATION = 00315,10,1;%%

\bibitem{mps2}
S.~Mantry, D.~Pirjol and I.~W.~Stewart,
in preparation.

\bibitem{Keum}
Y.~Y.~Keum, T.~Kurimoto, H.~N.~Li, C.~D.~Lu and A.~I.~Sanda,
%``Nonfactorizable contributions to B $\to$ D(*) M decays,''
arXiv:hep-ph/0305335.
%%CITATION = HEP-PH 0305335;%%

\bibitem{bakulev}
A.~P.~Bakulev, S.~V.~Mikhailov and N.~G.~Stefanis,
%``CLEO and E791 data: A smoking gun for the pion distribution amplitude?,''
arXiv:hep-ph/0303039.
%%CITATION = HEP-PH 0303039;%%

\bibitem{BaBr} P.~Ball and V.~M.~Braun,
%``Handbook of higher twist distribution amplitudes of vector mesons in  QCD,''
arXiv:hep-ph/9808229;
%%CITATION = HEP-PH 9808229;%%
%\bibitem{SR2}
V.~L.~Chernyak and A.~R.~Zhitnitsky,
%``Asymptotic Behavior Of Exclusive Processes In QCD,''
Phys.\ Rept.\  {\bf 112}, 173 (1984);
%%CITATION = PRPLC,112,173;%%
A.~Khodjamirian, R.~Ruckl, S.~Weinzierl, C.~W.~Winhart and O.~I.~Yakovlev,
%``Predictions on B $\to$ pi anti-l nu/l, D $\to$ pi anti-l nu/l and D $\to$ K  anti-l nu/l from QCD light-cone sum rules,''
Phys.\ Rev.\ D {\bf 62}, 114002 (2000)
[arXiv:hep-ph/0001297].
%%CITATION = HEP-PH 0001297;%%

\bibitem{DsKtheory}
C.~D.~Lu and K.~Ukai,
%``Branching ratios of B $\to$ D/s K decays in perturbative QCD approach,''
arXiv:hep-ph/0210206;
%%CITATION = HEP-PH 0210206;%%
Y.~Li and C.~D.~Lu,
%``Study of pure annihilation type decays $B \to D_s^{*} K$,''
arXiv:hep-ph/0304288;
%%CITATION = HEP-PH 0304288;%%
C.~K.~Chua, W.~S.~Hou and K.~C.~Yang,
%``Final state rescattering and color-suppressed anti-B0 $\to$ D(*)0 h0
% decays,''
Phys.\ Rev.\ D {\bf 65}, 096007 (2002)
[arXiv:hep-ph/0112148].
%%CITATION = HEP-PH 0112148;%%

\bibitem{DsK} B.~Aubert {\it et al.}  [BABAR Collaboration],
%``A study of the rare decays B0 $\to$ D/s(*)+ pi- and B0 $\to$ D/s(*)- K+,''
arXiv:hep-ex/0211053;
%%CITATION = HEP-EX 0211053;%%
P.~Krokovny {\it et al.}  [Belle Collaboration],
%``Observation of D/s+ K- and evidence for D/s+ pi- final states in neutral  
% B decays,''
Phys.\ Rev.\ Lett.\  {\bf 89}, 231804 (2002)
[arXiv:hep-ex/0207077].
%%CITATION = HEP-EX 0207077;%%

\bibitem{Leib}
A.~K.~Leibovich, Z.~Ligeti and M.~B.~Wise,
%``Comment on quark masses in SCET,''
arXiv:hep-ph/0303099;
%%CITATION = HEP-PH 0303099;%%
see also I.~Z.~Rothstein,
%``Factorization, power corrections, and the pion form factor,''
arXiv:hep-ph/0301240.
%%CITATION = HEP-PH 0301240;%%

\bibitem{jetset}
T.~Sjostrand,
%``High-energy physics event generation with PYTHIA 5.7 and JETSET 7.4,''
Comput.\ Phys.\ Commun.\  {\bf 82}, 74 (1994).
%%CITATION = CPHCB,82,74;%%

\bibitem{bss}
M.~Bander, D.~Silverman and A.~Soni,
%``CP Noninvariance In The Decays Of Heavy Charged Quark Systems,''
Phys.\ Rev.\ Lett.\  {\bf 43}, 242 (1979).
%%CITATION = PRLTA,43,242;%%

\bibitem{LiSanda}
Y.~Y.~Keum, H.~n.~Li and A.~I.~Sanda,
%``Fat penguins and imaginary penguins in perturbative QCD,''
Phys.\ Lett.\ B {\bf 504}, 6 (2001)
[arXiv:hep-ph/0004004].
%%CITATION = HEP-PH 0004004;%%

\bibitem{mmps}
J.~Chay and C.~Kim,
%``Collinear effective theory at subleading order and its application to  heavy-
% light currents,''
Phys.\ Rev.\ D {\bf 65}, 114016 (2002)
[arXiv:hep-ph/0201197];
%%CITATION = HEP-PH 0201197;%%
A.~V.~Manohar, T.~Mehen, D.~Pirjol and I.~W.~Stewart, 
%``Reparameterization invariance for collinear operators,''
Phys.\ Lett.\ B {\bf 539}, 59 (2002)
[arXiv:hep-ph/0204229].
%%CITATION = HEP-PH 0204229;%%


\end{thebibliography}
\end{document}